%% file: 2comp_paper_arxiv_fin.tex
\documentclass[aps,floats,preprintnumbers,nofootinbib,showkeys]{revtex4}
\pdfoutput=1
\usepackage{amsmath,amssymb}
\usepackage[pdftex,colorlinks=false,urlcolor=blue,linkcolor=blue]{hyperref}

\usepackage{graphicx}
\usepackage{color}
\usepackage{slashed}
\usepackage[normalem]{ulem}

\input mylatex

\def\jw#1{{\color{red}{\bf #1}}}

\def\vp{{\varphi}}
\def\lx{\lambda_x}

\def\mvp{m_\varphi}
\def\mh{m_h}
\def\mdm{m_\nu}

\def\lsp{\quad}
\def\non{\nonumber}

\def\ndm {{\nu}}

\def\yuk{g_\nu}

\def\gt{\zeta}

\begin{document}

\begin{flushright}
 UCRHEP-T537
\end{flushright}
\title{Two-Component Dark Matter}

\author{Subhadittya Bhattacharya$^1$}
\email[]{subhaditya.bhattacharya@ucr.edu }
\author{Aleksandra Drozd$^2$}
\email[]{aleksandra.drozd@fuw.edu.pl}
\author{Bohdan Grzadkowski$^2$}
\email[]{bohdan.grzadkowski@fuw.edu.pl}
\author{Jose Wudka$^1$}
\email[]{jose.wudka@ucr.edu}

\affiliation{(1) \,Department of Physics, University of California, Riverside, CA  92521, USA}
\affiliation{(2) \,Faculty of Physics, University of Warsaw, 00-681 Warsaw, Poland}

\begin{abstract}
We study an extension of the Standard Model (SM) with two interacting cold Dark Matter
(DM) candidates: 
a neutral Majorana fermion ($\nu$) and a neutral scalar singlet ($\vp$). 
The scalar $\vp$ interacts with the SM through the ``Higgs portal'' coupling while 
$\nu$ at the tree level interacts only with $\vp$ through Yukawa interactions. 
The relic abundance of $\nu$ and $\vp$ is found by solving the Boltzmann equations numerically;
for the case $\mdm > \mvp$ we also derive a reliable approximate analytical solution. 
Effects of the interaction between the two DM components are discussed.
A scan over the parameter space is performed to determine the regions 
consistent with the  WMAP data 
for DM relic abundance, and with the XENON100 direct detection limits for the
DM-nucleus cross section. We find that although
a large region of the parameter space is allowed by  the WMAP constraints,
the XENON100 data severely restricts the parameter space. Taking into account
only amplitudes generated at the tree level one finds three allowed regions
for the scalar mass: $m_\vp \sim 62.5\gev$ (corresponding to the
vicinity of the Higgs boson resonance responsible for $\vp\vp$ annihilation into 
SM particles), $\mvp\simeq 130-140\gev$ and $\mvp \gesim 3 \tev$.
1-loop induced $\nu$-nucleon scattering has been also calculated and discussed.
A possibility of DM direct detection by the CREST-II experiment was considered.

\end{abstract}


\maketitle

\section{Introduction}

Dark matter (DM) was first postulated by Oort in 1932 to account for the orbital
velocities of stars in the Milky Way, and then adopted by Zwicky in 1933 to
explain the orbital velocities of galaxies in clusters. The existence of DM is
by now well established \cite{review1} though compelling astronomical
observations, including recent ones involving Bullet cluster (1E0657-558) \cite{Markevitch:2003at}.
It is also know that DM plays a central role in cosmology, affecting both the
evolution of the early universe and structure
formation \cite{struct-form}. Understanding the properties of DM is one of the
great current problems in modern cosmology.

Despite a wealth of observations and many experimental efforts, the nature and
composition of DM remains unknown.
Since the early 80's there have been continuous attempts to determine whether DM
might be associated with one or more elementary particles, an idea that can be
probed using both collider experiments as well as cosmological observations. The
most promising possibility within this scenario is for DM to be composed
of cold non-baryonic particles; in this case current measurements of the 
anisotropy of the cosmic microwave background (CMB) 
can be used to estimate the non-baryonic DM density at \cite{Hinshaw:2012aka} 
\beq
\Omega_{DM}h^2 = 0.1138  \pm  0.0045
\label{eq:wmap-omega}
\eeq 
where the $\Omega_{DM}= \rho_{DM}/\rho_{\rm crit}$ is the ratio of the DM density
over the critical density that corresponds to flat universe, and $ h$ is the Hubble
constant in units of 100 km/(s.Mpc) (in contrast, the density of visible baryonic matter
is much much smaller: $ \Omega_{b}h^2=0.02264 \pm 0.00050$) \cite{Hinshaw:2012aka}. 

Unfortunately, all Standard Model (SM) particles are excluded as relevant
components of DM \cite{Bergstrom:2000pn}, so one has to look for extensions of the SM that provide stable (or with a decay time longer than the present age of the Universe), massive, neutral particles that might play this role. An enormous amount of work has been done
by theoreticians in this direction, considering many types of models, most
of which contain a single particle beyond the SM that is stable and might 
be considered as a DM candidate. 

This, however, may not be the case, so that
DM could have a multi-component structure (one should remember that the rich variety of SM matter is 
responsible only for a fiftieth of the matter density in the Universe),
and there have already been some studies of multi-component 
DM in the literature (see for example, \cite{Profumo:2009tb,Drozd:2011aa,Feldman:2010wy, Huh:2008vj, Aoki:2013gzs, Gu:2013iy, Cirelli:2010nh, Medvedev:2013vsa}. 
Here we would like to investigate a scenario where DM consists of two species --
a singlet scalar ($\vp$) and a singlet neutral {\em Majorana} fermion ($\nu$)
(that we will
refer to as a ``neutrino''). The scalar DM field in this model interacts with the 
SM through the Higgs field, while the fermionic DM does not couple directly to the SM. 
However, the fermionic and scalars DM components do interact, so the model provides
a simple `laboratory' where the  interesting issue of interactions 
between DM components can be studied. 

The paper is organized as follows. In sec.~\ref{model} we
introduce our specific 2-component DM model and discuss its general properties.
Then, in sec.~\ref{BE} we derive the Boltzmann equations that govern
cosmological evolution of the DM components and we obtain and discuss numerical 
and approximate analytical solutions.
In the subsequent section, sec.~\ref{relicabundance}, we calculate the 
present DM density and find the regions in parameter space for which it
is consistent with (\ref{eq:wmap-omega}). In sec.~\ref{dirdet} we derive
the constraints on our model derived from the direct detection experiments. 
Sec.~\ref{conc}
contains our conclusions. In the appendix \ref{vpvpXsec} we collect formulae
related to scalar and fermion pair annihilation.

\section{Minimal scalar-fermion model of DM}
\label{model}

Our model contains three new particles, all SM singlets:
a real scalar $ \vp $, and two {\em majorana} fermions $\nu_h$ and $\nu$
(two fermions are required in order to generate non-trivial
interactions between the DM components), only {\em one} of  the fermions will
contribute to the DM relic density.
Though the DM sector can contain particles
of any spin, the
simplest possibilities correspond to the presence of fermions and scalars.

Since all DM particles are singlets under the SM gauge group, their interaction
with the SM will be through terms of the form $ \ocal_{DM} \ocal_{SM} $, where
$ \ocal_{SM} $ is  gauge invariant operator composed of SM fields; 
of all such terms we expect those with the lowest dimension to be the most relevant.
Within the SM the lowest-dimensional scalar gauge invariant operator is $ H^\dagger H $,
where $H$ denotes the scalar isodoublet. Restricting ourselves to renormalizable 
interactions, and assuming that all DM particles transform non-trivially under a
symmetry group, fixes the leading $ \ocal_{DM} $ to be of the form $ \varphi^2 $.
Therefore, the $ \varphi $ interacts directly with the SM through the usual Higgs
portal term, while (at tree-level) the fermionic dark fields
communicate with the SM indirectly, through their interactions with $ \varphi $.

\subsection{The model}
\label{minmodel}

In order to ensure stability of DM candidates we will assume that the dark sector
is invariant under some global symmetry group  $ \gcal $ under which all the extra 
fields transform non-trivially, while all 
SM particles are $ \gcal $-singlets. 
For simplicity we choose $ \gcal=\zBB_2 \times \zBB_2 $ and, as mentioned previously,
 assume that the DM sector is composed of two majorana fermions, $\nu_h$ and $\nu$, and one 
real scalar $ \vp $, which under  $ \gcal $,
\beq
\nu_h \sim [-,+] \quad
\nu   \sim [+,-] \quad
\vp    \sim [-,-]
\eeq
We introduce the $\zBB_2 \times \zBB_2 $ symmetry to stabilize both the DM components; models with
more complicated discrete symmetries will require additional particles.

The most general, gauge- and $\gcal$-symmetric and renormalizable potential reads:
\begin{eqnarray}
V(H, \varphi) = 
- \mu_{H}^2 H^{\dagger} H 
+ \lambda_{H} (H^{\dagger} H)^2 
+ \frac{1}{2}  \mu_{\varphi}^2 \varphi^2
+ \frac{1}{4!} \lambda_{\varphi} \left(\varphi^2 \right)^2
 +  \lambda_{x} H^{\dagger} H \varphi^2 \,,
\label{pot}
\end{eqnarray}
where $H$ is the SM $SU(2)$  Higgs isodoublet and $\lambda_x$ parametrizes the `Higgs-portal'
interaction discussed above.
The Lagrangian density for the scalar sector is then given by:
\begin{eqnarray}
\lcal_{\rm scal} = \frac{1}{2} \partial_{\mu} \varphi \partial^{\mu} \varphi + 
D_{\mu}H^{\dagger} D^{\mu} H 
- V(H, \varphi)\,.
\end{eqnarray}

As usual, we require that the potential breaks spontaneously the electroweak symmetry via 
non-zero vacuum expectation value of the Higgs doublet $\left<H\right> = (0, v/\sqrt{2})$, 
$v = 246 \gev$. Since we also require the $\gcal$ symmetry 
to remain unbroken, we assume that $\mu_\varphi^2 > 0$, so $\left<\vp\right> = 0$. 
Note that $\left<\vp\right> = 0$ implies there is no mass-mixing between 
$\vp$ and $H$, so that the existing collider limits on the Higgs properties 
are not modified.
After the symmetry breaking, the physical scalars have masses 
${m_H}^2 = - \mu_{H}^2 + 3 \lambda_{H} v^2 =2 \mu_{H}^2$ 
and $m_{\varphi}^2 = \mu_{\varphi}^2 + \lambda_{x} v^2 $. 

The part of the DM Lagrangian involving fermions reads
\beq
\lcal = 
\half \overline{\nu_h} \, i \!\! \not\!\partial \, \nu_h + 
\half \overline{\nu} \, i \!\! \not\!\partial \, \nu -
\frac12 \nu_h^T C \nu_h M_h - \frac12 \nu^T C  \nu m_\nu +
\yuk \vp \, \overline{\nu_h} \nu.
\label{lag2}
\eeq
Note that the interaction between the SM and DM and the DM self-interactions
are generated by just two terms:
\beq
\lcal_{\rm int} = - \lambda_x H^\dagger H \vp^2  + g_\nu \vp \,\overline{\nu_h} \nu.
\eeq

Although this model can describe a 3 component DM sector we will introduce a further simplification by assuming
that $ M_h > m_\nu + m_\vp $, which allows the fast decay $ \nu_h \to \vp \nu $.
In this case only $ \nu $ and $ \vp $ are stable and therefore can
serve as realistic DM candidates. The  reactions that are relevant
for the evolution of DM are
$ \vp\vp \leftrightarrow {\rm SM,SM}$ and $ \vp\vp \leftrightarrow \nu\nu $ (we will
ignore the process $ \nu\nu \to {\rm SM,SM}$ that occurs at one loop).
We will  investigate this model as a simple realization of
a 2-component scalar-fermion dark sector, using it as a laboratory where the
interplay of the various dark components can be studied.

It is worth noticing that the dark sector has no 
conserved Noetherian charges, so that all the corresponding
chemical potentials vanish. This can be altered in a simple way by introducing additional 
fermions that can serve as Dirac partners of
$ \nu_h $ and $ \nu $, 
in which case the ``dark'' fermion number could be conserved; we have not
done so to simplify the discussion.
It is also worth mentioning that the scalar singlet could be used to tame the
little hierarchy problem by canceling top-quark loop induced quadratic divergences
in radiative corrections to the Higgs boson mass \cite{Drozd:2011aa, Kundu:1994bs}.

In the following we will fix $M_h$ at the smallest value that ensures 
the fast decay of $\nu_h$, so
we will effectively deal with only four parameters: $\mvp, m_\nu, \lx$ and $g_\nu$.
Our goal is to constrain the parameters taking into account available restrictions: 
theoretical (vacuum stability, unitarity/perturbativity, triviality of the scalar sector) 
and experimental (DM relic abundance, direct detection experiments).

\subsection{Theoretical constraints}
\label{Sec:stab-uni}

In order to stabilize the vacuum we require that the scalar potential in eq.~(\ref{pot}) 
is bounded from below. At the tree level it implies the following conditions  \cite{Gonderinger:2009jp}
\beq
\lambda_{\varphi} > 0  \,; \quad \lambda_x > - \sqrt{\frac{\lambda_{\varphi}\lambda_{H}}{6}}  = 
 -\frac{\mh}{2v} \sqrt{ \frac{\lambda_{\varphi}}{3}} \,,
\label{stab_con}
\eeq
where $m_h$ denotes the Higgs mass.
Amplitudes for all possible scalar-scalar scatterings will satisfy the tree-level unitarity constraints provided \cite{Cynolter:2004cq} 
\beq
\lambda_{\varphi} < 8 \pi , \lsp
|\lambda_x| < 4 \pi\,.
\label{unit_con}
\eeq

Finally, it is sufficient to require $\mu_{\varphi}^2 > 0$ for the global $\gcal$ symmetry to remain unbroken, which leads to the  very useful inequality
\beq
m_{\varphi}^2 > \lambda_{x}v^2 \,;
\label{mphi_consistency}
\eeq
as a consequence, light scalars ($\mvp \ll v$) must couple very weakly to the SM
($\lx \ll 1 $) whenever $\lx>0$.

We also impose the following perturbativity limits on $\lambda_\vp$, $\lx$ and $g_\nu$.
\beq
\lambda_{\varphi} < 4\pi , \lsp
|\lambda_x| < 4 \pi, \lsp |g_\nu| < 4\pi
\label{pert_con}
\eeq

Separating positive and negative values of $\lx$, the above constraints imply that the following regions
are allowed:
\bea
0 < &\lx& < {\rm min}\left[ \left(\frac{\mvp}{v}\right)^2, 4\pi \right] 
\label{lxregions1}
\\
-0.74 < -\frac{m_h}{2 v}\sqrt{\frac{\lambda_\vp}{3}} < &\lx& < 0,
\label{lxregions2}
\eea
where we have adopted in eq.~\ref{lxregions2} the Higgs mass $m_H=125\gev$ and 
the maximal value of $\lambda_\vp$ consistent with unitarity (\ref{unit_con}).

\section{Dark Matter Density and The Boltzmann Equation}
\label{BE}

In the following we will focus on the minimal model specified in sec.~\ref{minmodel}.
Our goal is to determine the DM relic density and test this model against the
relic density constraint derived from WMAP and the available data on direct DM detection. 

We start with formulating and solving the two Boltzmann equations (BEQ) that govern the cosmological 
evolution of our DM candidates, the DM neutrinos ($\nu$) and scalar singlets ($\vp$). 
Ignoring loop corrections the relevant reactions are  
$\nu \nu \leftrightarrow \vp  \vp $ and $\vp  \vp  \leftrightarrow $ SM SM,
where the last one occurs through the Higgs portal interaction $\lx H^\dagger H \vp^2$. 
Therefore (at tree level) for the $\nu$ to interact with the SM, they must be first converted
into $\vp$ pairs through Yukawa interactions $\propto \yuk$.
The BEQs then read: 
\bea
\dot{n}_{\vp}+3H n_{\vp} &=& - 
\int \frac{\gt_\vp d^3 p}{(2 \pi)^3 2E_p} \, \frac{\gt_\vp d^3 p'}{(2 \pi)^3 2E_p'}
\frac{\gt_{SM} d^3 q}{(2 \pi)^3 2E_q} \, \frac{\gt_{SM} d^3 q'}{(2 \pi)^3 2E_q'}
\delta^4 (p + p' - q - q')  
|M_{\vp \vp \to SM SM}|^2  \left( \tilde{f}_\vp  \tilde{f}_\vp - \tilde{f}_\vp^{EQ} \tilde{f}_\vp^{EQ} \right)
 \nonumber\\
&& - \int \frac{\gt_\vp d^3 p}{(2 \pi)^3 2E_p} \, \frac{\gt_\vp d^3 p'}{(2 \pi)^3 2E_p'}
\frac{\gt_{\nu} d^3 q}{(2 \pi)^3 2E_q} \, \frac{\gt_{\nu} d^3 q'}{(2 \pi)^3 2E_q'}
\delta^4 (p + p' - q - q')  
|M_{\vp \vp \to \nu \nu}|^2  \left( \tilde{f}_\vp  \tilde{f}_\vp - \tilde{f}_\nu \tilde{f}_\nu \right)
 \nonumber\\
&&\cr
\dot{n}_{\nu}+3H n_{\nu} &=& - \int \frac{\gt_\vp d^3 p}{(2 \pi)^3 2E_p} \, \frac{\gt_\vp d^3 p'}{(2 \pi)^3 2E_p'}
\frac{\gt_{\nu} d^3 q}{(2 \pi)^3 2E_q} \, \frac{\gt_{\nu} d^3 q'}{(2 \pi)^3 2E_q'}
\delta^4 (p + p' - q - q')  
|M_{\vp \vp \to \nu \nu}|^2  \left( \tilde{f}_\nu  \tilde{f}_\nu - \tilde{f}_\vp \tilde{f}_\vp \right)
\label{be1}
\eea
where $n_X$ denote the number density of $X=\nu, \vp$, and $n_X^{EQ}$  the corresponding equilibrium densities; a dot denotes a time derivative, $M_{i\to f}$ is the amplitude for the process $i\to f$ 
(note that $M_{\vp \vp \to \nu \nu} = M_{\nu \nu \to \vp \vp}$); $\gt_i,~
i=\vp,\,\nu, {SM}$ are the numbers of internal degrees of 
freedom ($\gt_\vp = 1$ and $\gt_{\nu}=2$, since the $\nu$ are Majorana particles), and the matrix element squared $|M|^2$ 
contains an average over the initial and final spins together with the corresponding $1/n!$ factors for $n$ identical 
particles in the initial and final states; $H$ denotes the Hubble parameter.
The phase space density $\tilde{f}_X$ and an equilibrium density $\tilde{f}_X^{EQ}$
are related to corresponding number densities as follows: 
\beq
n_X =  \int \frac{\gt_X d^3 p}{(2 \pi)^3 2E}  \tilde{f}_X,  \hspace{.4 cm}
n_X^{EQ} =  \int \frac{\gt_X d^3 p}{(2 \pi)^3 2E}  \tilde{f}_X^{EQ},    \hspace{.4 cm}   \tilde{f}_X^{EQ}  = \frac{1}{e^{E/T}\pm1}, \hspace{.4 cm} X=\vp,\nu
\eeq
where, as mentioned above, the chemical potential vanishes, and $\pm$ refers to fermions and bosons, respectively.
To simplify BEQs we will use the thermally averaged cross section $\langle \sigma_{ X X \to Y Y} v \rangle$, 
defined as:
\bea
\langle \sigma_{ X X \to Y Y} v \rangle &\equiv & \frac{1}{\left(n_X^{EQ}\right)^2}  
\int \frac{\gt_X d^3 p}{(2 \pi)^3 2E_p} \, \frac{\gt_X d^3 p'}{(2 \pi)^3 2E_p'}
\frac{\gt_{Y} d^3 q}{(2 \pi)^3 2E_q} \, \frac{\gt_{Y} d^3 q'}{(2 \pi)^3 2E_q'} \times\nonumber\\&& 
\delta^4 (p + p' - q - q') |M_{XX \to YY}|^2 e^{-(E_p+E_p')/T}
\eea
Assuming kinetic equilibrium and neglecting possible effects of quantum statistics 
the BEQs in eq.~ \ref{be1} simplify considerably:
\bea
&& \dot{n}_{\vp}+3H n_{\vp} =
 -\langle \sigma_{ \vp \vp \to SM \, SM} v \rangle \left(n_{\vp}^2- n_{\vp}^{EQ} {}^2\right) 
- \left(
\langle \sigma_{ \vp \vp \to \nu \nu} v \rangle 
n_{\vp}^2- 
\langle \sigma_{\nu \nu \to \vp \vp} v \rangle n_{\nu}^2
\right) \non \\
&& \dot{n}_{\nu}+3H n_{\nu} = 
- \left(
\langle \sigma_{\nu \nu \to \vp \vp} v \rangle n_{\nu}^2 -
\langle \sigma_{ \vp \vp \to \nu \nu} v \rangle 
n_{\vp}^2
\right)
\label{beq2}
\eea
where it is important to remember that
\beq
\langle \sigma_{\nu \nu \to \vp \vp} v \rangle = \left(\frac{n_{\vp}^{EQ}}{n_{\nu}^{EQ}}\right)^2 \langle \sigma_{ \vp \vp \to \nu \nu} v \rangle
\eeq

The above relation restates that there are just two independent cross sections that influence the dynamics of DM 
density evolution: 
$\langle \sigma_{\vp \vp \to SM \, SM} v \rangle$ and $\langle \sigma_{ \vp \vp \to \nu \nu} v \rangle$;
the first one is well known (see e.g. \cite{Drozd:2011aa}) nevertheless it is included in the appendix~\ref{vpvpXsec} 
for completeness. The
Feynman diagram and the corresponding cross section for the process  
$ \vp \vp \to \nu \nu$ are also shown in the appendix. The interactions between $\vp$ and $\nu$ 
involve an exchange of a virtual heavy neutrino $\nu_h$; if the corresponding mass $M_h$ is very large
$\langle \sigma_{ \vp \vp \to \nu \nu} v \rangle$ is strongly suppressed, which leads to an over abundance of $ \nu $. 
To remedy this  we will assume $M_h$ as small as allowed by the requirement of $\nu_h$ being unstable: we adopt $M_h = \mvp + \mdm + \Delta M_\nu$, with fixed $\Delta M_\nu=10 \gev$. 
Then the cross sections are parameterized by four parameters: 
$\mvp$, $\mdm$, the Yukawa coupling $\yuk$ and the Higgs portal coupling $\lx$.



\subsection{Solving BEQ}
Instead of a number density ($n_X$) it is more convenient to use
the number density normalized to $T^3$, so in the following we
adopt $f_X(T)\equiv n_X(T)/T^3$ (not to be confused with the
phase-space density $ \tilde f$ introduced previously).
The initial conditions are fixed
at large temperature $T_{\rm ini}= {\rm
max}(\mvp, m_\nu)$; we assume that
the couplings $\lx$ and $\yuk$ are large enough so that at
$T_{\rm ini}$ both DM
components are in equilibrium with the SM
(the SM is assumed to be in
equilibrium); hence, $f_X(T_{\rm ini}) = n_X^{EQ}(T_{\rm
ini})/T_{\rm ini}^3$.  As the Universe cools the DM components
eventually decouple from the SM
when their rate of
interaction becomes smaller than the rate of expansion of the
universe. Since here we
are looking for cold DM (CDM) candidates, we will consider only cases
where this decoupling occurs when both $ \nu $ an $ \vp $ are
non-relativistic. In the following, we will
solve the BEQs (\ref{beq2}) and determine the present,
i.e. at $T=T_{\rm CMB}=2.37 \cdot 10^{-13}\gev$, DM abundance.

The solutions can be classified according to the mass hierarchy in the dark sector:
\ben
\item[] Case A: $\mdm > \mvp$
\item[] Case B: $\mdm < \mvp$
\een
The dynamics of the DM number density evolution turns out to be very different for these two cases, as we will see. 

%
\begin{figure}[h]
\centering
\includegraphics[height = 4.8 cm]{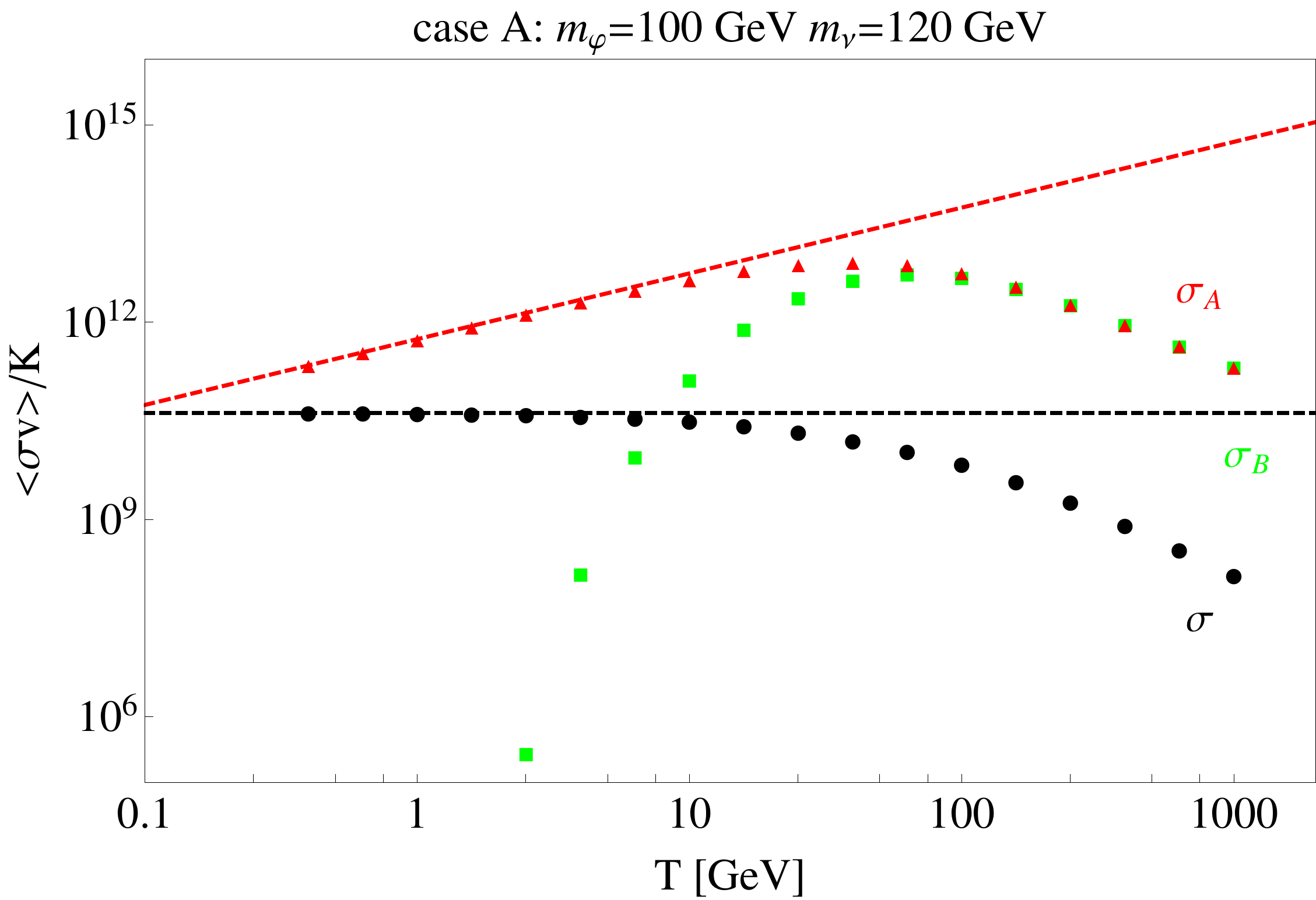}
\includegraphics[height = 4.8 cm]{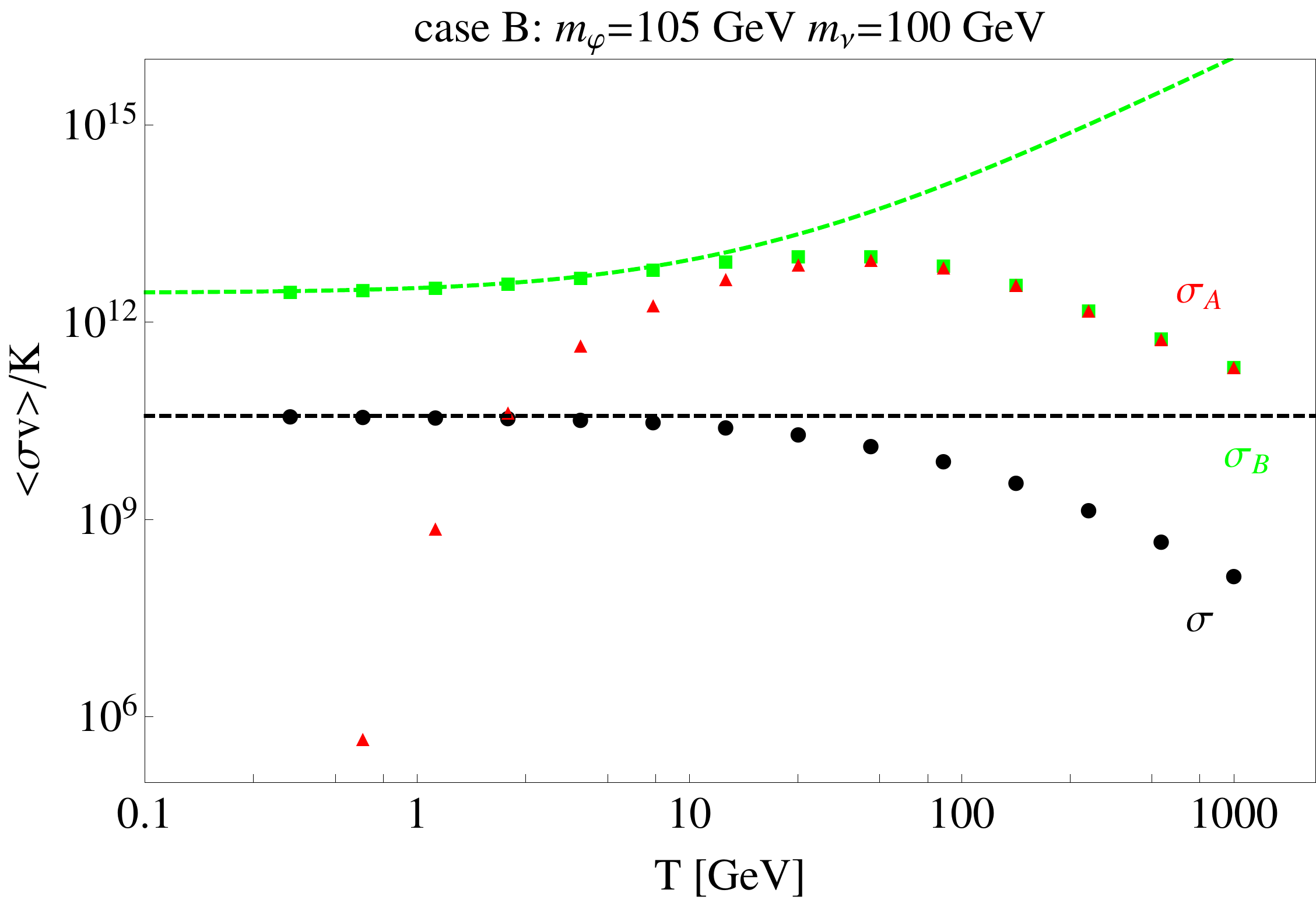}
\caption{Thermally averaged cross sections 
$\sigma \equiv $ $\langle \sigma_{ \vp \vp \to SM \, SM} v \rangle/K$ (black points);
$\sigma_B \equiv $$ \langle \sigma_{ \vp \vp \to \nu \nu} v \rangle/K$ (green points);
$\sigma_A \equiv $$\langle \sigma_{ \nu \nu \to \vp \vp} v \rangle/K$ (red points),
as a functions of $T$ (in GeV),
for $\lx = .1$ and $g_\nu = 2.5$.
In the left panel: $\mvp = 100$ GeV, $\mdm = 120$ GeV (case A); 
in the right panel: $\mvp = 120$ GeV, $\mdm = 100$ GeV (case B).
The factor $K$ is defined in (\ref{eq:defofK})}
\label{fig_cs_AB}
\end{figure}

If $\mdm > \mvp$ (Case A), there is a temperature
range where the $\vp$ do not have enough energy to create $\nu $ pairs, 
so the thermally averaged cross section
$\langle \sigma_{ \vp \vp \to \nu \nu} v \rangle\to 0$ below that temperature;
on the other hand, neutrinos still have enough energy to
maintain a high rate of annihilation
$\nu \nu \to \vp \vp$. 
This is illustrated in the left panel of fig.~\ref{fig_cs_AB} where
$\langle \sigma_{ \vp \vp \to \nu \nu} v \rangle/K$
is seen to drop precipitously below $10\gev$ while 
$\langle \sigma_{ \vp \vp\to SM \, SM} v \rangle/K$
approaches a constant value already at $T\sim 5\gev$.
$\langle \sigma_{ \nu \nu\to\vp \vp  } v \rangle/K $
is vanishing at $T \to 0$ as will be discussed below.
The $K$ factor is defined as follows
\beq
K \equiv \sqrt{\frac{4\pi^3
g(T)}{45 m_{Pl}^2}}
\label{eq:defofK}
\eeq
where $g(T)$ is the number of relativistic
degrees of freedom, and $m_{Pl}$ the Planck mass.
$K$ appears in the  BEQs for the
normalized number densities $f_X(T)\equiv n_X(T)/T^3$.
In contrast, for $\mvp > \mdm$ (Case B), it is
$\langle \sigma_{ \nu \nu\to\vp \vp  } v \rangle/K $
that becomes very small at small temperatures
(right panel of fig.~\ref{fig_cs_AB}), while $\langle \sigma_{ \vp \vp \to SM\, SM} v \rangle/K $
and $\langle \sigma_{\vp\vp \to \nu \nu} v \rangle/K $ tend to a constant value, we will
return to this issue in sec.~\ref{caseB}.

\subsubsection{Case A.  ($m_\nu > m_\varphi$) - numerical solutions}

In terms of the normalized number densities $f_X(T)$ defined earlier 
the BEQs (\ref{beq2}) for case A  become
\bea
f_{\vp}^\prime &=&
\sigma\left[f_{\vp}^2- {f_{\vp}^{EQ} }^2\right] + 
\sigma_A \left[\left(\frac{f_{\nu}^{EQ}}{f_{\vp}^{EQ}}\right)^2 f_{\vp}^2 -  f_{\nu}^2 \right] 
\label{BEAvp} \\
f_{\nu}^\prime &=& 
\sigma_A \left[f_{\nu}^2 - \left(\frac{f_{\nu}^{EQ}}{f_{\vp}^{EQ}}\right)^2 f_{\vp}^2\right],
\label{BEAnu}
\eea
where $f_X^\prime$ denotes a derivative with respect to $T$ and 
$\sigma \equiv \langle \sigma_{ \vp \vp \to SM\,SM} v \rangle /K$,
$\sigma_A \equiv \langle \sigma_{\nu \nu \to \vp \vp} v \rangle/K$; note that $\sigma,\sigma_A$ 
have dimensions of mass$^{-1}$.
Since in the non-relativistic limit $\sigma_A$ is vanishing (as implied by angular momentum and parity conservation)
therefore solving numerically the BEQs for the case A 
we have approximated $\sigma_A$ by keeping only linear 
terms in the expansion of $\sigma_A$ in powers of $x_\vp^{-1}$ where $x_\vp\equiv \mvp/T$, see \cite{Srednicki:1988ce}. 
It was assumed that $\sigma$ is T-independent. The quality of this approximation \jw{can} be estimated 
from the left panel of fig.~\ref{fig_cs_AB}. 
We have also verified this approximation for a number of points in the parameter space
by comparing results for $f_\vp(T_{\rm CMB})$ and  $f_\nu(T_{\rm CMB})$  obtained through exact numerical
solution with the one obtained adopting expansion of $\sigma_A$, relative errors obtained for the case A are: 
$\delta^A_\vp \simeq 2.3\% $, $\delta^A_\nu \simeq 1.4\% $.

\begin{figure}
\includegraphics[height = 5 cm]{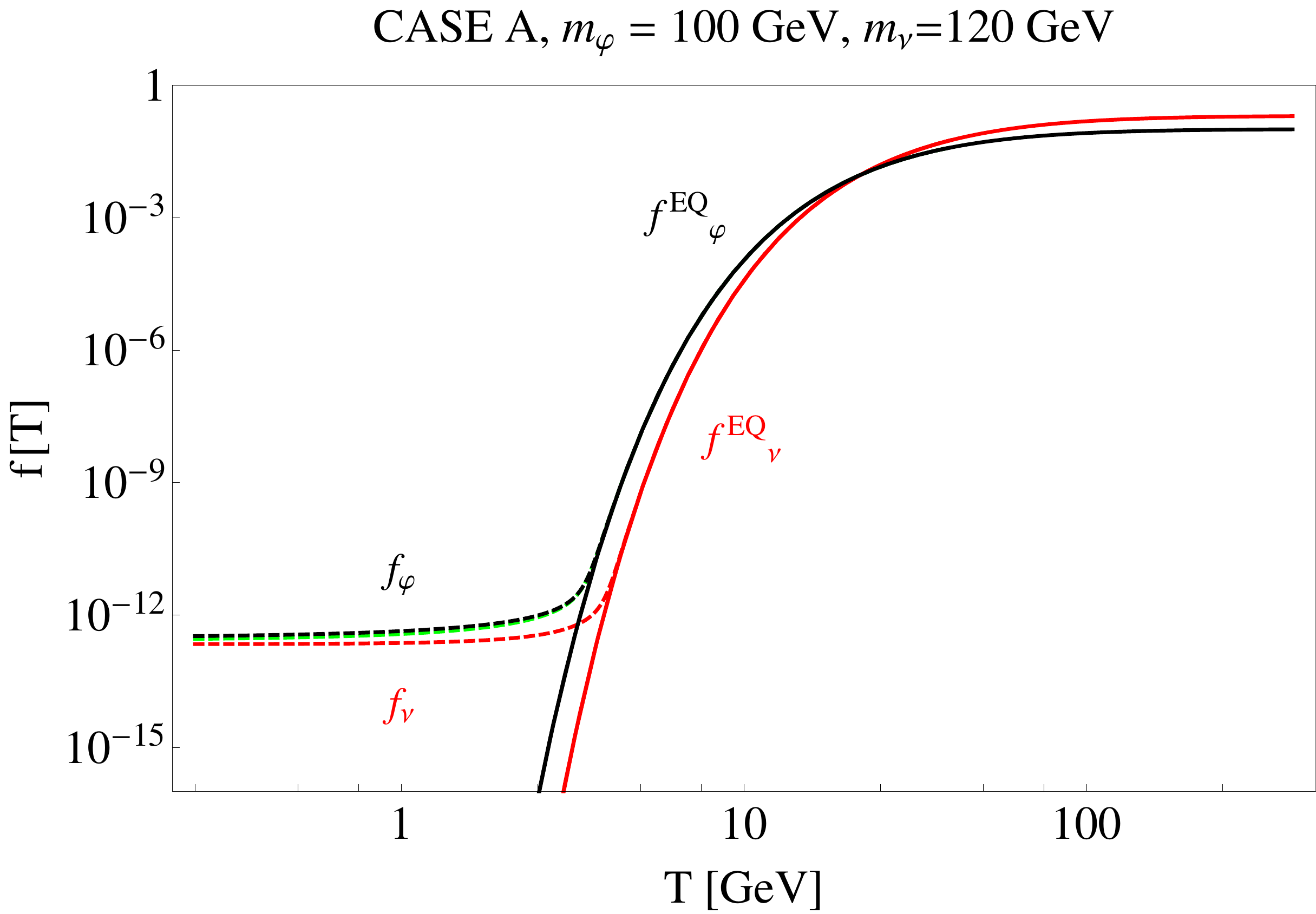} \,\,
\includegraphics[height = 5 cm]{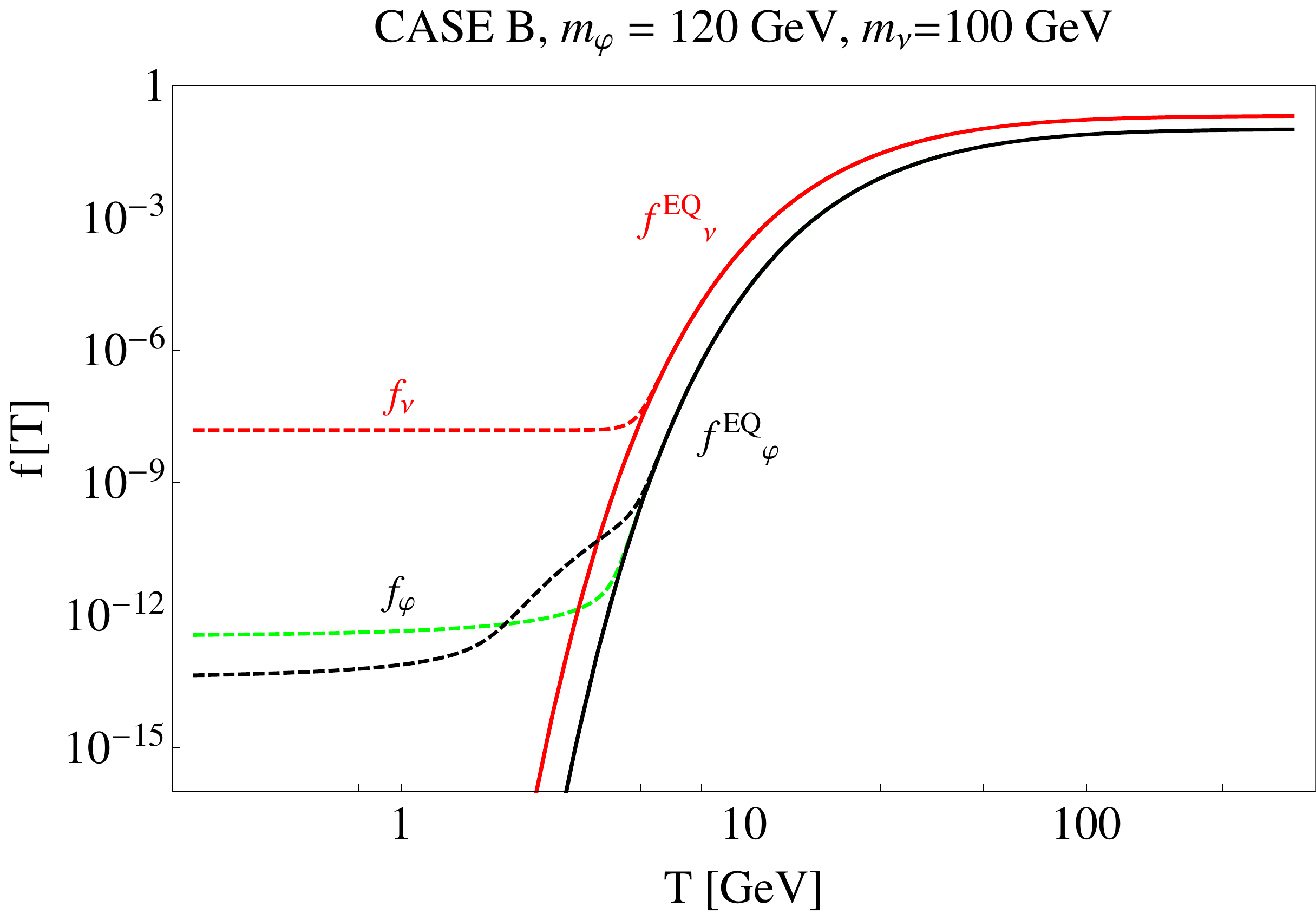}
\includegraphics[height = 5 cm]{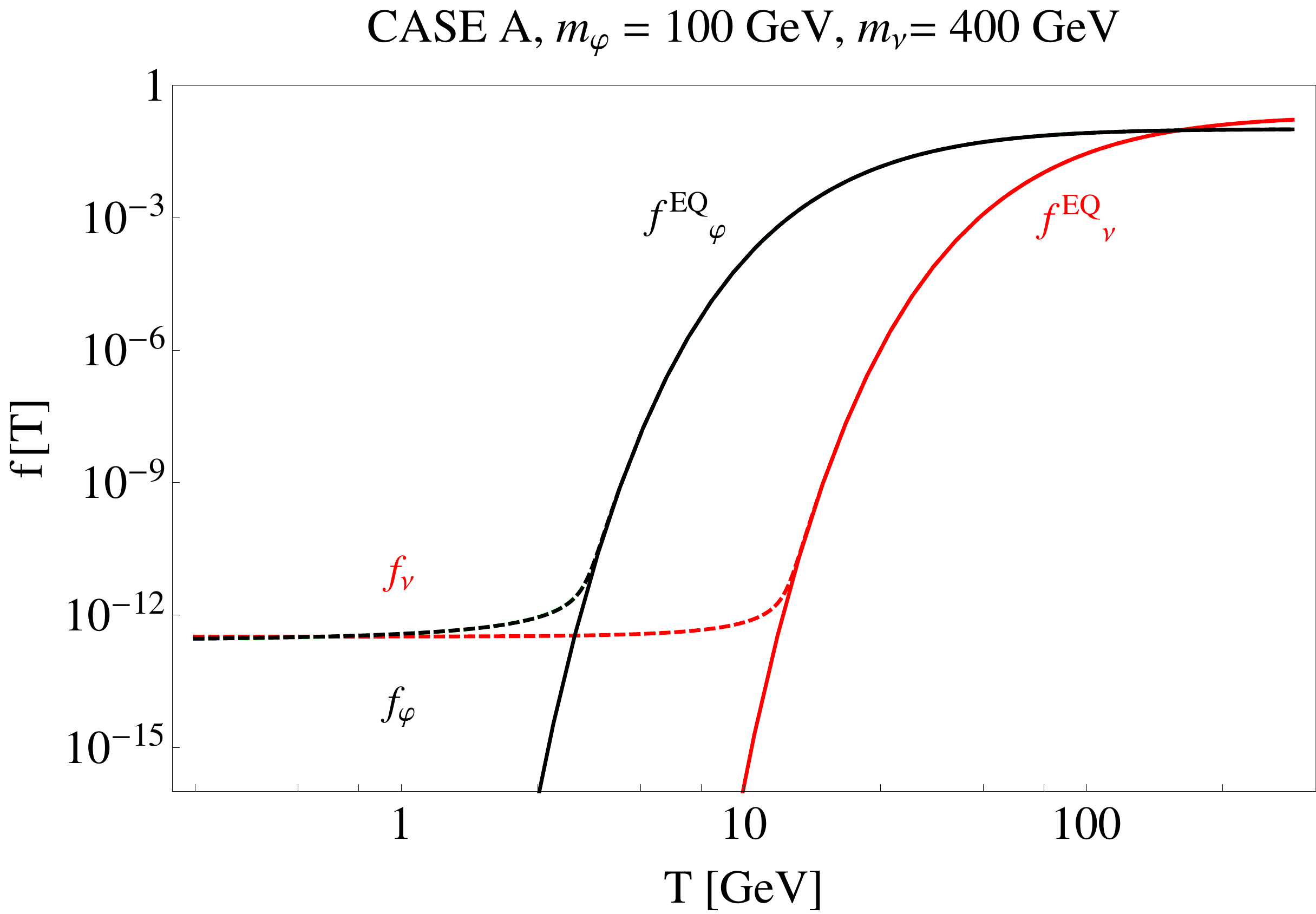} \,\,
\includegraphics[height = 5 cm]{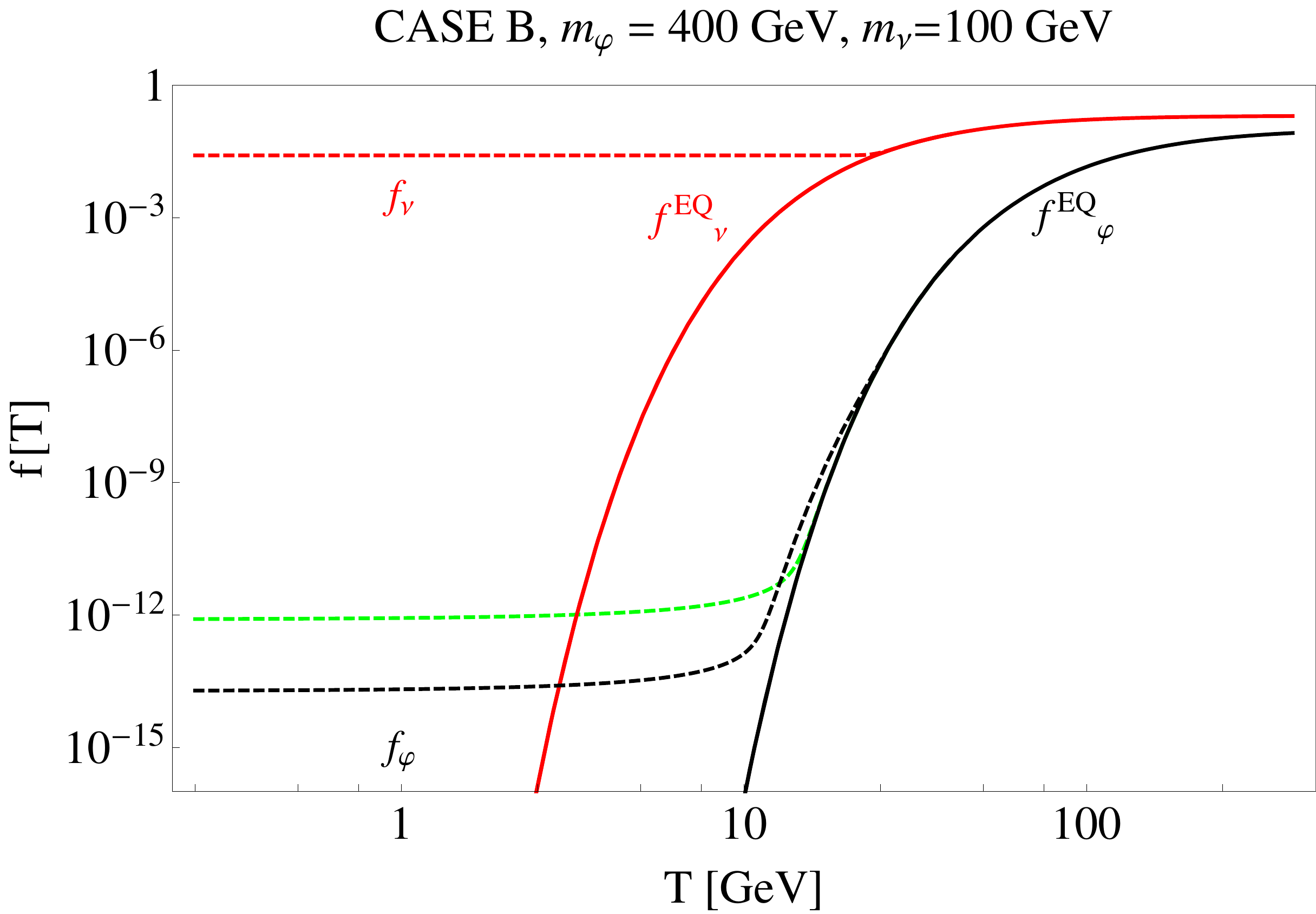}
\includegraphics[height = 5 cm]{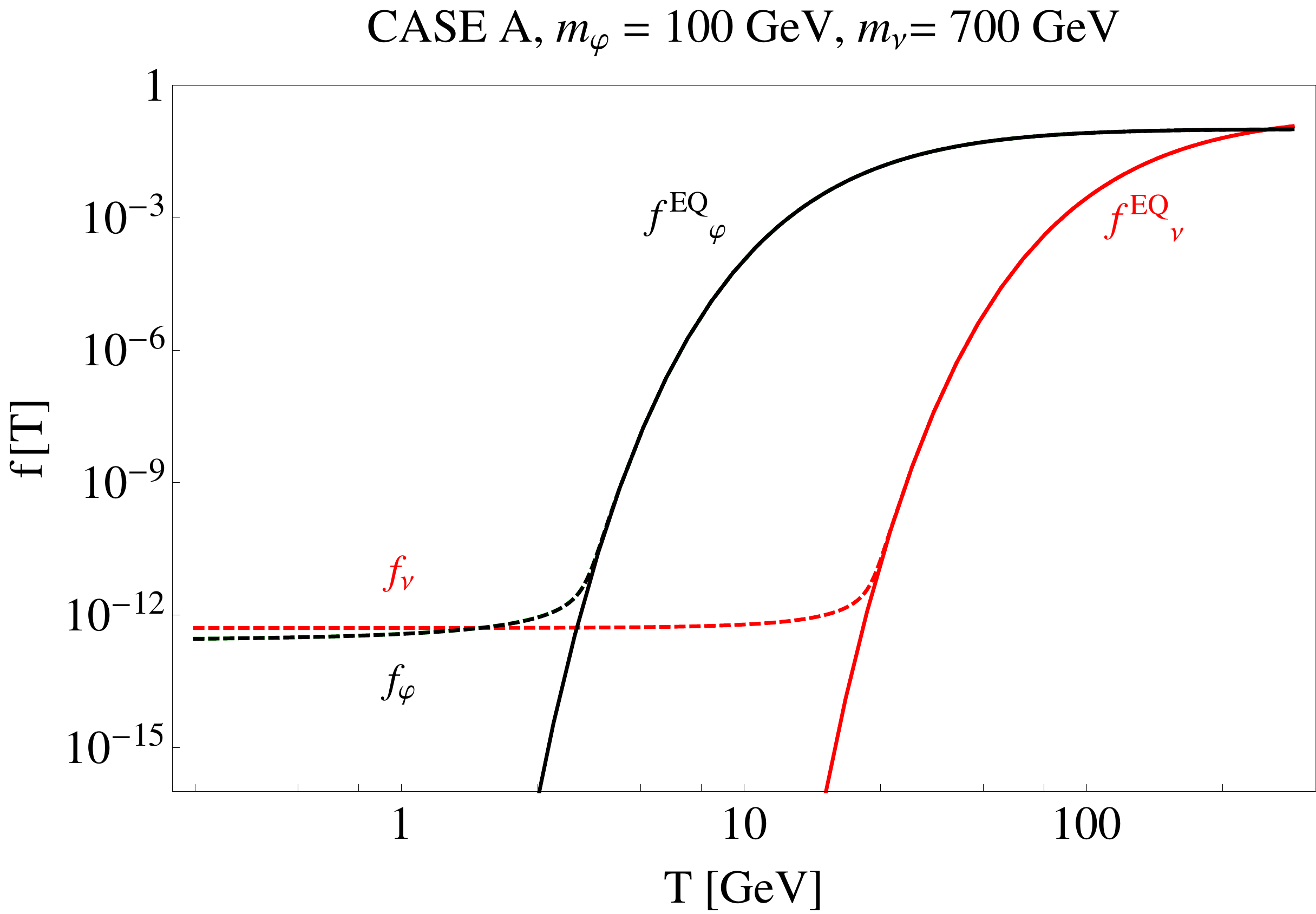} \,\,
\includegraphics[height = 5 cm]{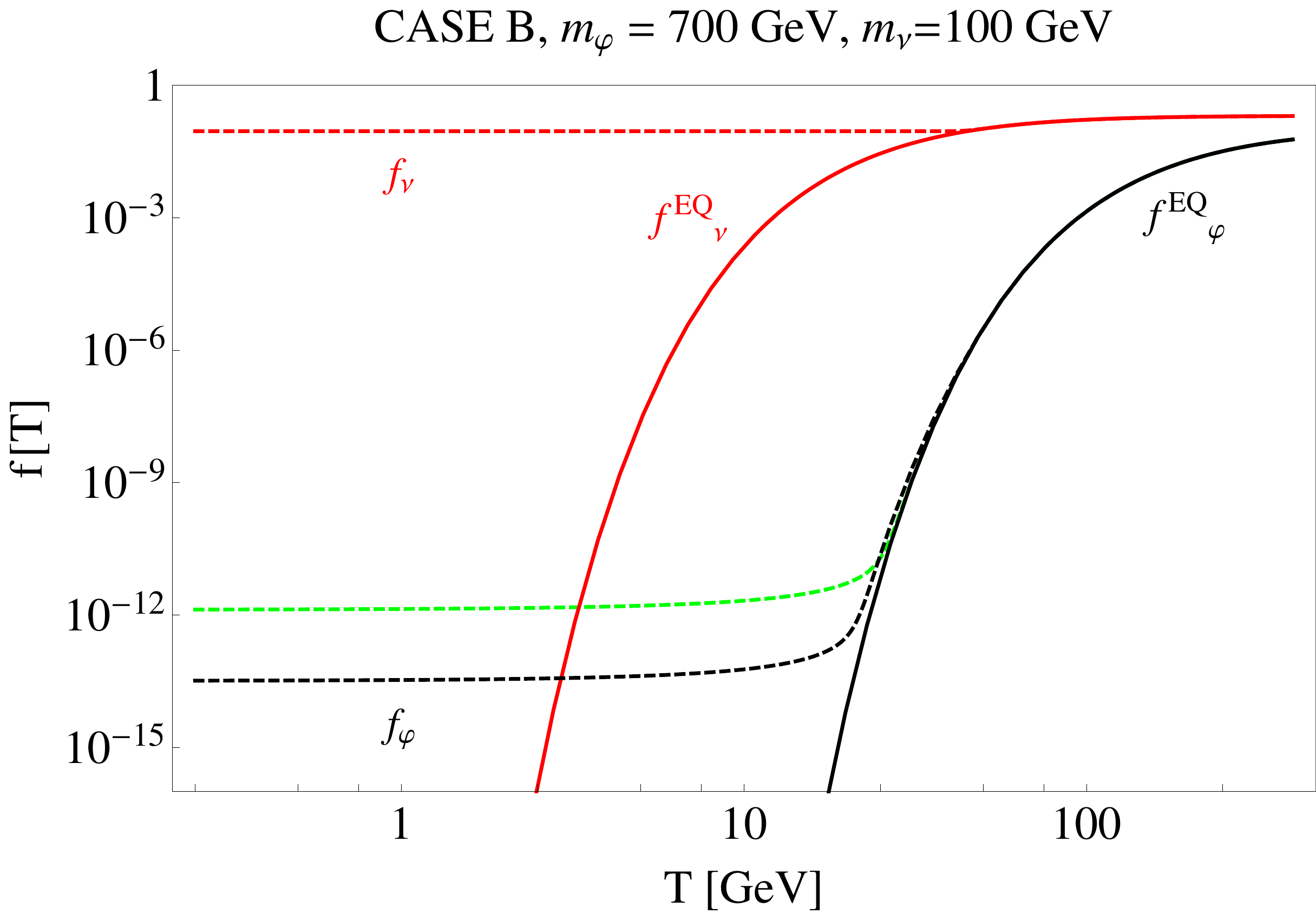}
\caption{
Solutions to the BEQs for  case A (left panels) and  case
B (right panels) for $\lx = 0.1$ and $g_\nu = 2.5$. Scalar and
neutrino DM masses are specified above each panel.
Solid black (red) lines correspond to the equilibrium distributions,
$f_{\vp}^{EQ}$ ($f_{\nu}^{EQ}$) for scalars (neutrinos), dashed
lines are the corresponding numerical solutions of the BEQs.
Green dashed lines show numerical solutions of a single BEQ for
scalars without neutrinos present in the theory. }
\label{fig_case_AB}
\end{figure}

Examples of numerical solutions of BEQs (\ref{BEAvp}-\ref{BEAnu}) 
for various illustrative parameter choices 
are shown in fig.~\ref{fig_case_AB}. The plots on the \lhs panels
correspond to case A, while case B examples are presented on the
\rhs.

For case A  we see that the $ \nu $ (red dashed line),
which are heavier, decouple from equilibrium
(solid red line) before (i.e. at a higher temperature)
the $ \vp $ (black dashed line); after
decoupling from the scalars the $\nu$ quickly freeze-out.
Sometime later (at a lower temperature) the $ \vp $ decouple from the SM,
and since there is no communication
between dark neutrinos and scalars, the latter immediately
freeze-out. 

It is seen from left panels of fig.~\ref{fig_case_AB}, the resulting low-temperature
densities for $\nu$ and $\vp$ are similar (note the logarithmic scale), which is a signal 
that both components decouple form equilibrium roughly at the same $x$ ($\sim 20-30$) 
as is typical for the standard cold DM scenario. 
Note also that
for fixed $\mvp$, the scalar decoupling temperature $T_f^\vp$ 
and the scalar DM relic density are insensitive to $\mdm$,
as a consequence of the early decoupling of the $ \nu $. Again this is an
indication that both components evolve roughly independently.
The dark neutrino decoupling temperature, $T_f^\nu$ grows
with  $m_\nu$ (since $m_\nu/T_f^\nu$ is roughly constant). 

The green line in fig.~\ref{fig_case_AB} refers
to solutions for scalar DM density when the fermionic DM
component is absent. One can see that in case A
(the left panels) the decoupling temperature of the scalar DM 
in the two component scenario is roughly the 
same as in the one component scenario with the same
$ \mvp$ and $ \lx $, though the relic density is usually (depending on parameters chosen)
smaller in the single component case.

%

\subsubsection{Case A. ($\mdm > \mvp$) - approximate analytical solutions}
\label{caseA}

In the standard case of a single cold DM candidate, it is easy
to find an approximate analytical solution of the BEQs that
allows to determine the abundance of DM at low temperatures
(see for example \cite{kolb},\cite{Gondolo:1990dk}). The
solution is often sufficiently accurate, so that one can avoid
obtaining the numerical solutions of the BEQs. In this subsection we will
derive an analogous approximate solution within our model of
two-component DM for case A.

We begin by defining  $\Delta_\vp \equiv f_\vp - f_\vp^{EQ}$, $\Delta_\nu
\equiv f_\nu - f_\nu^{EQ}$, which
parameterize the deviation from equilibrium in the solutions.
Then we can rewrite the BEQs (\ref{BEAvp}-\ref{BEAnu}) as:
\bea
 \Delta_\vp' &=& \sigma \Delta_\vp  \left[\Delta_\vp + 2 f_{\vp}^{EQ}\right] +
 \sigma_A \left[\left(\frac{f_{\nu}^{EQ}}{f_{\vp}^{EQ}}\right)^2  \Delta_\vp  (\Delta_\vp + 2 f_{\vp}^{EQ})
 - \Delta_\nu (\Delta_\nu + 2 f_{\nu}^{EQ}) \right] - f_\vp^{EQ\;\prime}
\label{a_beq_vp}
\\
 \Delta_\nu' &=&
\sigma_A \left[ \Delta_\nu  (\Delta_\nu + 2 f_{\nu}^{EQ}) - \left(\frac{f_{\nu}^{EQ}}{f_{\vp}^{EQ}}\right)^2
 \Delta_\vp  (\Delta_\vp + 2 f_{\vp}^{EQ}) \right] - f_\nu^{EQ\;\prime}
\label{a_beq_nu}
\eea
where the primes denote temperature derivatives.
Let's consider first the high temperature region - before
decoupling of the DM candidates. At these temperatures $f_\vp,\,f_\nu$ track $f_\vp^{EQ},\,f_\nu^{EQ}$
very closely, so that $\Delta_{\vp,\nu}$ and
$\Delta'_{\vp,\nu}$ are very small. The corresponding solution to 
(\ref{a_beq_nu}) is obtained by neglecting $\Delta'_{\vp,\nu}$ as
well as all terms proportional to
$(f_{\nu}^{EQ}/f_{\vp}^{EQ})^2\propto e^{-2(\mdm-\mvp)/T}$ (since
in this case $\mdm>\mvp$):
\bea
\Delta_\vp (T) &\simeq &\frac{1}{ \sigma (\Delta_\vp + 2f_\vp^{EQ})}\left({f_\vp^{EQ}}' +{f_\nu^{EQ}}' \right)
\label{delvp}\\
\Delta_\nu (T)  &\simeq &
\frac{{f_\nu^{EQ}}'}{ \sigma_A(\Delta_\nu + 2f_\nu^{EQ})}
\label{delnu}
\eea
We define the decoupling temperature (freeze-out
temperature)~\footnote{
In the case A, the freeze-out happens immediately after
decoupling,
therefore the decoupling temperature and the freeze-out
temperature are identical. As we will show shortly, this
is not true in the case B.} for scalars ($T_f^{\vp}$) and
neutrinos ($T_f^{\nu}$) as
the temperatures at which $\Delta_\vp(T_f^{\vp}) = c_\vp
f_{\vp}^{EQ}(T_f^{\vp})$ and
$\Delta_\nu(T_f^{\nu}) = c_\nu f_{\nu}^{EQ}(T_f^{\nu})$
with $c_{\vp,\nu} = O(1)$.
This means that at decoupling temperature the
number density differs from the corresponding equilibrium density
roughly by a factor of few. We will later assume $c_\vp(c_\vp+2)=c_\nu(c_\nu+2)=1$,
because this choice of $ c_\vp, c_\nu$ will provide good
agreement with numerical solutions and 
simplifies the analytical expressions. The freeze-out temperatures $T_f^{\vp}$
and $T_f^{\nu}$are then determined by
\bea
f_\vp^{EQ} (T_f^\vp) &\simeq & \frac{ 1 }{c_{\vp} (2+c_\vp) \sigma }
\left[ \frac{\mvp}{{T_f^\vp}^2} +
\frac{f_\nu^{EQ}(T_f^\vp)}{ f_\vp^{EQ}  (T_f^\vp) } 
\frac{\mdm}{{T_f^\vp}^2} 
\right]
\simeq  \frac{\mvp}{\sigma{T_f^\vp}^2} 
\label{Sol_Tvp_A}
\\
f_{\nu}^{EQ}(T_f^\nu) &\simeq &  \frac{m_\nu}{\sigma_A(T_f^\nu) {T_f^\nu}^2}
\label{Sol_Tnu_A}
\eea
where we have substituted out choice $ c_{\nu,\vp} = \sqrt{2}-1$.
In obtaining this we have assumed, consistent with the cold dark matter requirement, that 
the parameters are such that $\mdm,\mvp \gesim T_f^{\nu,\vp}$, and kept only the leading terms. 
Once the freeze-out temperatures $T_f^{\vp,\nu}$ are obtained by solving 
(\ref{Sol_Tvp_A} - \ref{Sol_Tnu_A}), $\Delta_\vp(T_f^\vp)$ and 
$\Delta_\nu(T_f^\nu)$ can be calculated using eqs. (\ref{delvp}-\ref{delnu}).
It turns out that for the choice $ c_{\nu,\vp} = \sqrt{2}-1$ our approximate
equations for $T_f^{\nu, \vp}$ reproduce the exact ones (found numerically)
very well, typical errors calculated from 20 random points are $0.9\%$ for 
$T_f^{\nu}$ and $1.2\%$ for $T_f^{\vp}$.

After freeze-out the number densities remain much larger than their equilibrium counterparts,
so that $ \Delta_{\nu,\vp} \simeq f_{\nu,\vp} $ and we can neglect all terms containing
$f_{\vp,\nu}^{EQ}$ and ${f_{\vp,\nu}^{EQ}}'$ as well as all
terms proportional to $(f_{\nu}^{EQ}/f_{\vp}^{EQ})^2\propto e^{-2(\mdm-\mvp)/T}$
In this case (\ref{a_beq_nu}) simplifies to
$ \Delta_\nu' = \sigma_A \Delta_\nu^2 $ with solutions
\beq
\Delta_\nu(T) = \frac{ \Delta_\nu(T_f^\nu)}{1- \Delta_\nu(T_f^\nu)\int_{T_f^\nu}^T \sigma_A(T') dT'}
\then
\Delta_\nu(T_{\rm CMB}) \simeq \frac{ \Delta_\nu(T_f^\nu)}{1+  \sigma_A(T_f^\nu) T_f^\nu \Delta_\nu(T_f^\nu)/2}
\eeq
where in this we assumed $\sigma_A \propto T $, as discussed above and illustrated in fig.~\ref{fig_cs_AB}. 
Using now (\ref{Sol_Tnu_A}) we find
that $  \sigma_A(T_f^\nu) T_f^\nu \Delta_\nu(T_f^\nu) > 1 $
so we obtain
\beq
f_\nu(T_{CMB}) \simeq
\Delta_\nu(T_{\rm CMB}) \simeq \frac{2}{ \sigma_A(T_f^\nu) T_f^\nu }
\label{delnu1}
\eeq

After freeze-out the evolution equation for $\vp $ becomes
\beq
\Delta_\vp' \simeq
 \sigma \Delta_\vp^2 - \sigma_A \Delta_\nu^2 
\eeq
with initial condition
$ \Delta_\vp(T_f^\vp) \simeq  c_\vp m_\vp/(\sigma T_f^{\vp\,2}) $
derived from (\ref{Sol_Tvp_A}). 
In solving this equation we will approximate $ \Delta_\nu$
by its value at $T_{\rm CMB}$ and $ \sigma_A $ by its
value at $T_f^\nu $ (we have verified the accuracy of
these assumptions by comparing the analytic results
with the exact numerical
results in a set of randomly selected paramter points).
Using these approximations the solution is easy to find: 
\beq
\Delta_\vp(T) = \frac{r_f}{\sigma T_f^\vp} \frac{ u + \tanh[r_f(1-T/T_f^\vp)]}{1+ u \tanh[r_f(1-T/T_f^\vp)]}; \quad 
r_f = 2 \; \frac{T_f^\vp}{T_f^\nu}\sqrt{\frac\sigma{\sigma_A(T_f^\vp)}},~u=\frac{c_\vp m_\vp}{ r_f T_f^\vp}
\label{delvpsol}
\eeq
Note that 
\beq
r_f \propto 2 \frac{\mvp}{m_\nu} \sqrt{\frac{\sigma}{\sigma_A(T_f^\nu)}}
\eeq
therefore in the case A, its value is typically small. Expanding (\ref{delvpsol}) around $r_f=0$
one obtains in the leading order
\beq
f_\vp(T) \simeq \Delta_\vp(T) \simeq \Delta_\vp(T_{\rm CMB}) \simeq \frac{1}{\sigma T_f^\vp}\,,
\quad 
\label{delvpsol2}
\eeq
The above expression shows that the resulting low-temperature $\vp$ density is roughly
the same as it would be in the case without neutrinos at all. That is also seen 
in the left panels of fig.~\ref{fig_case_AB} where dashed green lines (no neutrinos) coincides 
with black ones (the full system). Since $x_f$ for $\nu$ and $\vp$ are similar
therefore so are the densities. 

\begin{figure}
\centering
\includegraphics[height = 4.5 cm]{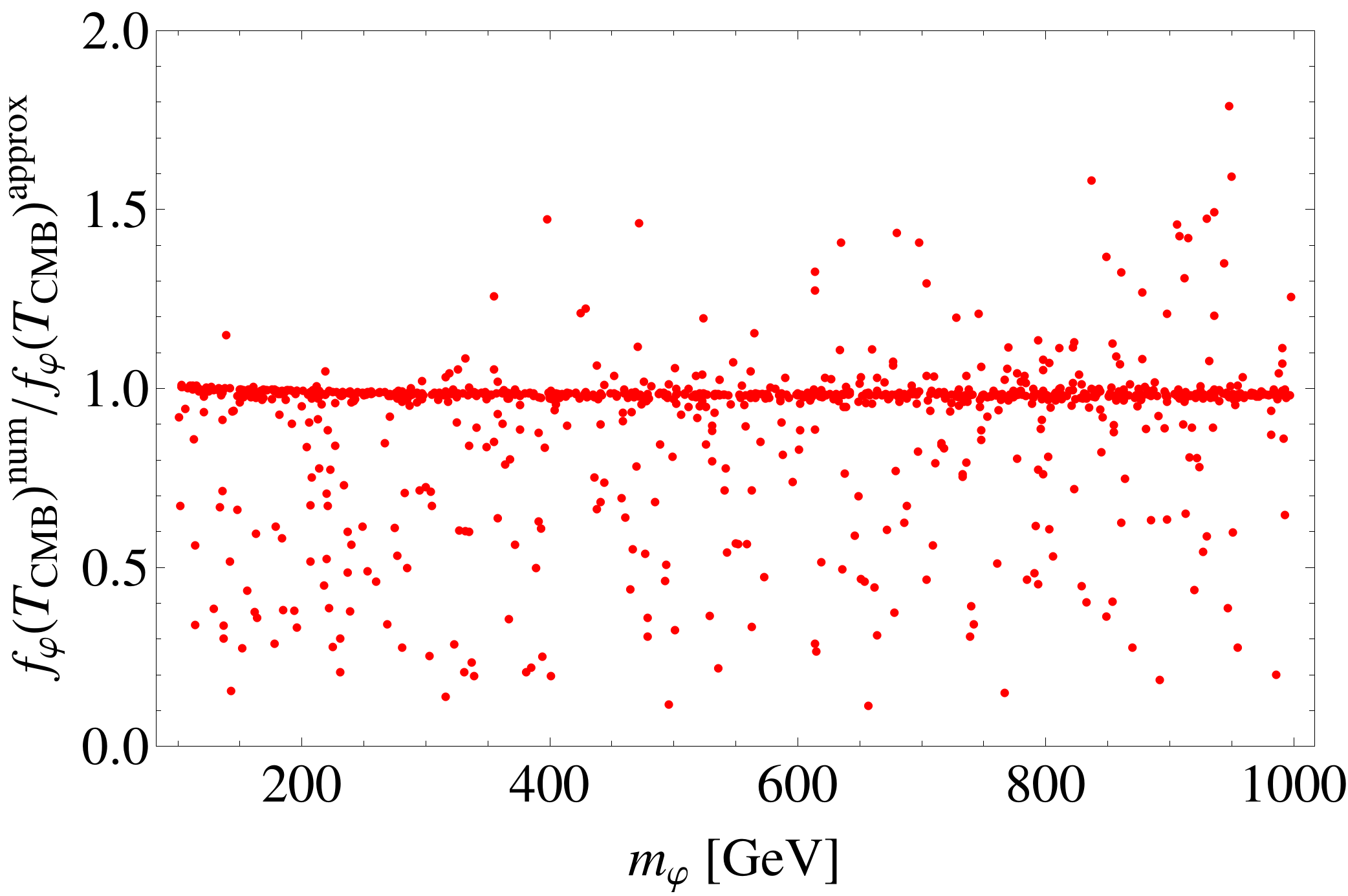}
\includegraphics[height = 4.5 cm]{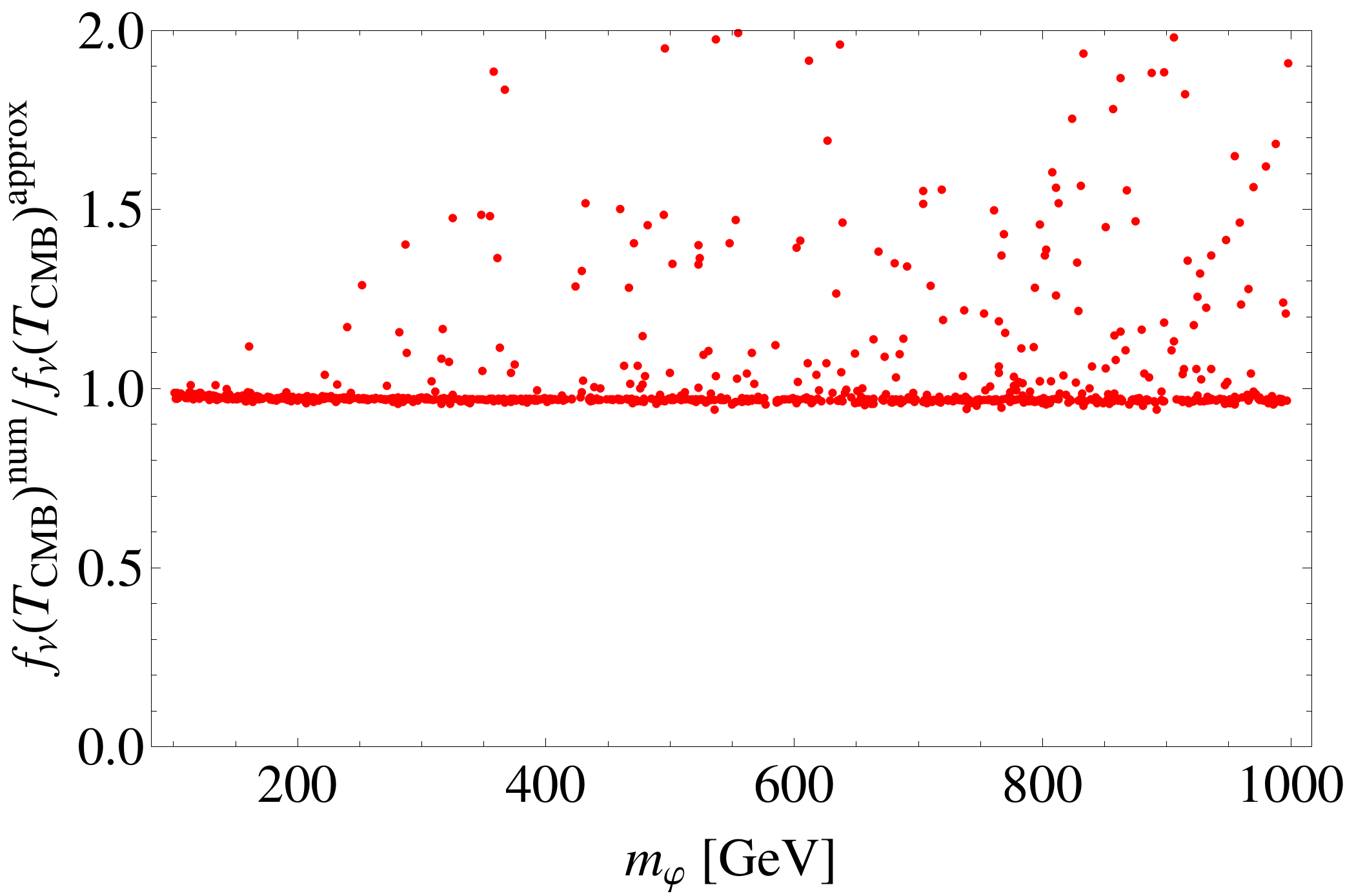}
\caption{The ratio $f_X(T_{\rm CMB})^{\rm num}/f_X(T_{\rm CMB})^{\rm approx}$ for case A for scalars (left panel) and neutrinos (right panel) as a function of scalar DM mass.
500 parameter points $(\mvp, \mdm, \lx, g_\nu)$ were chosen randomly within the ranges
$10 \gev < \mvp, \mdm < 1\tev$, $0.001 < \lx < 4\pi$ and $0.1 < g_\nu < 4 \pi$.  }
\label{APPROX_case_a}
\end{figure}

The accuracy of the above results can be gauged
by calculating the ratio of $f_X^{\rm num}$, the
numerical solution, over the corresponding analytical approximate solution, 
$f_X^{\rm approx}$, at $T=T_{\rm CMB}$, the present Universe temperature; 
the results are presented in fig.~\ref{APPROX_case_a}.
As one can see, the approximations are often satisfactory for the
chosen parameter space. In general, the result
for $f_\nu^{\rm approx}$ are more reliable
and become more accurate as the splitting between the $ \vp$
and $\nu$ masses increases (which is
natural as we are neglecting terms containing
$(f_{\nu}^{EQ}/f_{\vp}^{EQ})^2\propto e^{-2(\mdm-\mvp)/T}$).
The quality of the approximation
seems to be independent of $\lx$, both for $f_\vp$
and $f_\nu$. 

\subsubsection{Case B. ($\mdm < \mvp$) 
}
\label{caseB}

When $ \mvp>\mdm$ we again assume equilibrium at high
temperatures. As the temperature drops,
DM particles become non-relativistic and the 
neutrinos will no longer have enough energy to create pairs of
the heavier scalars through 
annihilation $\nu \nu  \to  \vp \vp$,
so that $\langle \sigma_{ \nu \nu \to \vp \vp} v \rangle \to 0$
as $T \to 0$.
On the other hand the rate of $\nu$-pair creation,  
$\vp \vp \to \nu \nu$ has a non-zero limit as $T \to 0$ (see
the
right panel of fig.\ref{fig_cs_AB}).  

In this case the BEQs (\ref{beq2}) read
\bea
f_\vp^\prime &=&
\sigma  \left[f_{\vp}^2- f_{\vp}^{EQ} {}^2\right] 
+\sigma_B \left[
 f_{\vp}^2- 
\left(\frac{f_{\vp}^{EQ}}{f_{\nu}^{EQ}}\right)^2 f_{\nu}^2
\right] \label{BEB1}\\
f_\nu^\prime &=& 
\sigma_B \left[
\left(\frac{f_{\vp}^{EQ}}{f_{\nu}^{EQ}}\right)^2f_{\nu}^2 - f_{\vp}^2
\right]
\label{BEB2}
\eea
where $\sigma_B \equiv \sigma^B_0 + \sigma^B_1 T + \sigma^B_2 T^2$.
As it is shown in the right panel of fig.~\ref{fig_cs_AB},
for low temperatures $\sigma$ is well approximated by a constant
while for $\sigma_B$  we used low-temperature expansion
keeping linear and quadratic terms in $x^{-1}$. We have 
estimated the quality of the approximation by comparing the exact numerical integration of the BEQ's
for a number of points in the parameter space
with the one obtained with quadratic expansion of $\sigma_B$, resulting errors for $f_\vp(T_{\rm CMB})$ and $f_\nu(T_{\rm CMB})$
are the following $\delta^B_\vp =6.3\% $, $\delta^B_\nu = 2.6\% $.
It is also useful to notice that the ratio of equilibrium
distributions,
$\left(f_{\vp}^{EQ}/ f_{\nu}^{EQ}\right)^2 \propto
e^{-2(\mvp-\mdm)/T}$
vanishes as $T \rightarrow 0$ since $\mdm < \mvp$. 

Numerical solutions of (\ref{BEB1},\ref{BEB2}) are shown in the
right panel of fig.~\ref{fig_case_AB},
where the neutrino mass was fixed at $m_\nu=100\gev$ for three choices
of scalar mass: $\mvp=120, 400$ and $700\gev$.
Note that for parameters adopted in the figure ($\lx=0.1,\,g_\nu=2.5$), 
$\nu$ and $\vp$ decouple roughly simultaneously; we
have verified
numerically that this is typical throughout most of the 
relevant region of parameter space~\footnote{Neutrinos
decouple earlier for small neutrino Yukawa coupling $g_\nu\sim 0.1$, 
but in this case the DM relic abundance does not match the one derived
from the WMAP data, which requires larger Yukawa
couplings $g_\nu \gesim 1.8$.}.
Since $T_f^\nu\simeq T_f^\vp$ and $f_{\vp}^{EQ} < f_{\nu}^{EQ}$
for $\mdm < \mvp$, the asymptotic low-temperature density will be larger
for neutrinos, $f_{\vp}(T) < f_{\nu}(T)$. 
Therefore, in  case B, it is typical that the number
density of DM at low temperatures
is dominated by neutrinos. In fact, $f_\nu$ domination at
low temperatures can be understood intuitively
since neutrinos do no couple
directly to the SM, and in consequence,
they annihilate into SM particles slower than scalars.

Contrary to 
naive expectation, it is remarkable that in case B and for fixed
$\mdm$ the fermion freeze-out temperature is
strongly dependent on $\mvp $
(right panel of fig.~\ref{fig_case_AB}), it varies from $T_f^\nu \simeq 4 \gev$ for
$\mvp=120\gev$ to
$T_f^\nu \simeq 30\gev$ for $\mvp=700\gev$. 
Note that in this case $x_f$ differs from its standard value $20-30$,
for instance for $\mvp=700\gev$ and $m_\nu=100\gev$ corresponding
values are $x_\vp^f\simeq 23$ and $x_\nu^f\simeq 3$ for $\vp$ and $\nu$
respectively. This results
in a rapid grow of low-temperature $f_\nu$ with $\mvp$
at fixed $\mdm$:
$f_\nu(T_{\rm CMB})\sim 10^{-8}$ at $\mvp=120\gev$,
to $f_\nu(T_{\rm CMB})\sim 10^{-1}$ at $\mvp=700\gev$. On the
other hand,
the low-temperature $f_\vp(T_{\rm CMB})$ is 
roughly independent of $\mvp$, even though the scalar decoupling
temperature, $T_f^\vp$ varies with $\mvp$.
This case nicely illustrates the dramatic
influence of the presence and interaction among DM components
upon their thermal evolution.

Another comment is in order here. As one can clearly see
in first panel on the right of fig.~\ref{fig_case_AB}, there are
parameter ranges such that after decoupling from equilibrium,
scalars (black dashed line) do not freeze-out immediately (in
contrast to single-component DM or in case A):
$f_\vp $ deviates from equilibrium, but is still temperature
dependent and only later freezes out. 
This happens because even below the temperature
at which the $\nu $ and $ \vp $ decouple from the equilibrium with the SM, $\vp$ pairs can still
annihilate into $\nu$ pairs. This effect can be seen from the
BEQs (\ref{BEB1}-\ref{BEB2}).
After the $\nu$ decouple, we have $f_\nu \gg f_\nu^{EQ}$ and the BEQ
for scalars, eq.~\ref{BEB1} becomes
\bea
&& f'_{\vp} =
(\sigma + \sigma_B ) \left( f_{\vp}^2 - h_\vp^2 \right)\,, \quad
h_\vp^2 = 
f_{\vp}^{EQ\;2} \left[
\frac\sigma{\sigma+ \sigma_B} + \left(\frac{f_{\nu}}{f_{\nu}^{EQ}}\right)^2 \frac{\sigma_B}{\sigma + \sigma_B} 
\right]
\label{BEB11}
\eea

\begin{figure}
\centering
\includegraphics[height = 5.5 cm]{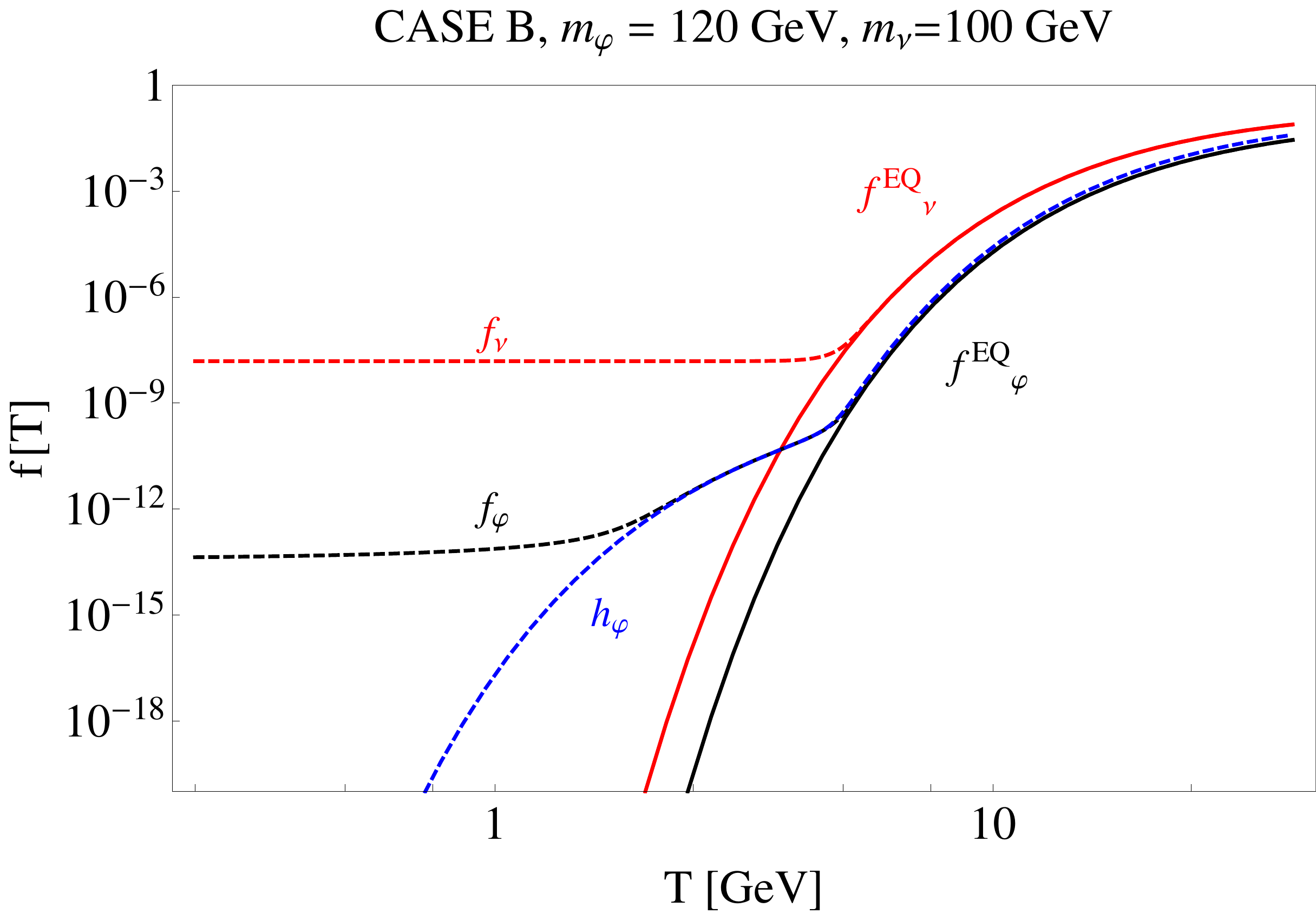}
\caption{Solution of the BEQs, case B  ($\mvp = 120$ GeV, $\mdm = 100$ GeV); for $\lx = 0.1$, $g_\nu = 2.5$. 
Solid black (red) line illustrates equilibrium distributions, $f_{\vp}^{EQ}$ ($f_{\nu}^{EQ}$) 
for scalars (neutrinos), dashed lines are the corresponding numerical solutions of the BEQs.  
Blue dashed line shows the distribution $h_\vp$ from (\ref{BEB11}).
}
\label{fig_case_B_bump}
\end{figure}

We interpret this as follows: after neutrinos decouple, scalars approach a modified ``equilibrium'' distribution $h_\vp$ shown as the
blue dashed curve in fig.~\ref{fig_case_B_bump}. As it is seen 
in the right panels of fig.\ref{fig_case_AB} and in fig.~\ref{fig_case_B_bump}, 
as $T$ decreases, $f_\vp$ will eventually
decouple also from $h_\vp$ and  freeze-out.
In order to illustrate the difference between the modified evolution of scalars after the decoupling 
from $f_{\vp}^{EQ}$ we plot in the right panels of fig.\ref{fig_case_AB} also the numerical solutions of a single 
BEQ for scalars without neutrinos present in the theory (green dashed lines). This behavior of
$ f_\vp $ between decoupling and freeze-out is only possible in multi-component 
and self interacting DM scenarios and, to the best of our knowledge, has not been 
previously discussed in the literature.

The disappearance of scalars into neutrinos is, of course, more efficient and faster
as the mass difference between $\vp$ and $\nu$ grows, this can also be observed in the right
panels of fig.~\ref{fig_case_AB}. It is also seen that a
large mass splittings results in very large neutrino low-temperature density, while scalar density 
remains very small, $f_\vp\sim 10^{-12}-10^{-13}$. It follows that upper limits on 
the total DM density (implied e.g. by the WMAP data) favor small mass splitting.

%


\bigskip

Following a strategy analogous to the one we used in case A one can also find an approximate analytical 
solution of the BEQs in case B. Unfortunately the accuracy of the approximation is much
worse in this case, because of the difficulties in including the intermediate state where the
scalars have decoupled but have not yet frozen-out. For this reason in case B we will use only
numerical solutions.


\section{Relic abundance}  
\label{relicabundance}

\begin{figure}
\centering
\includegraphics[height=4.5cm]{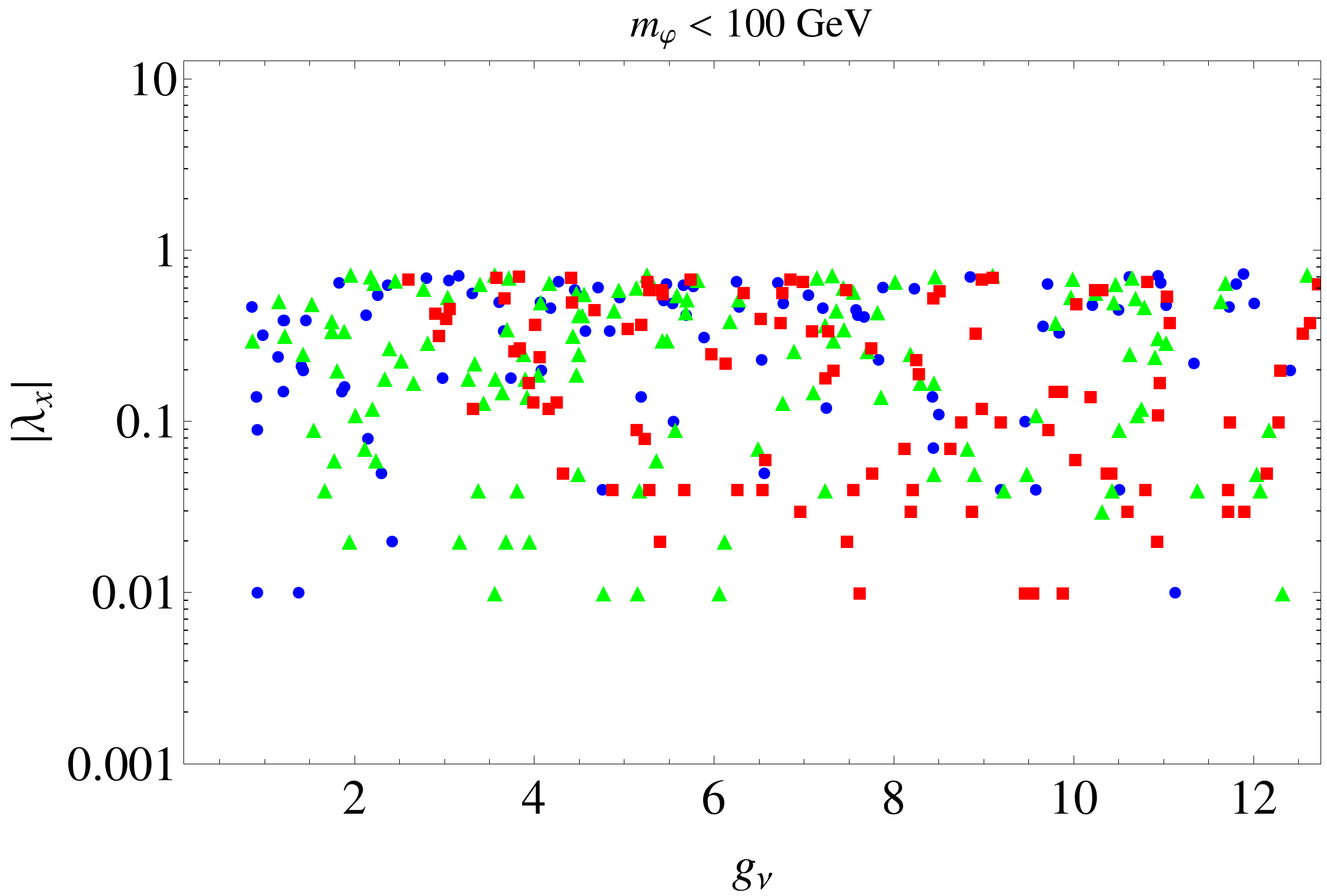}
\includegraphics[height=4.5cm]{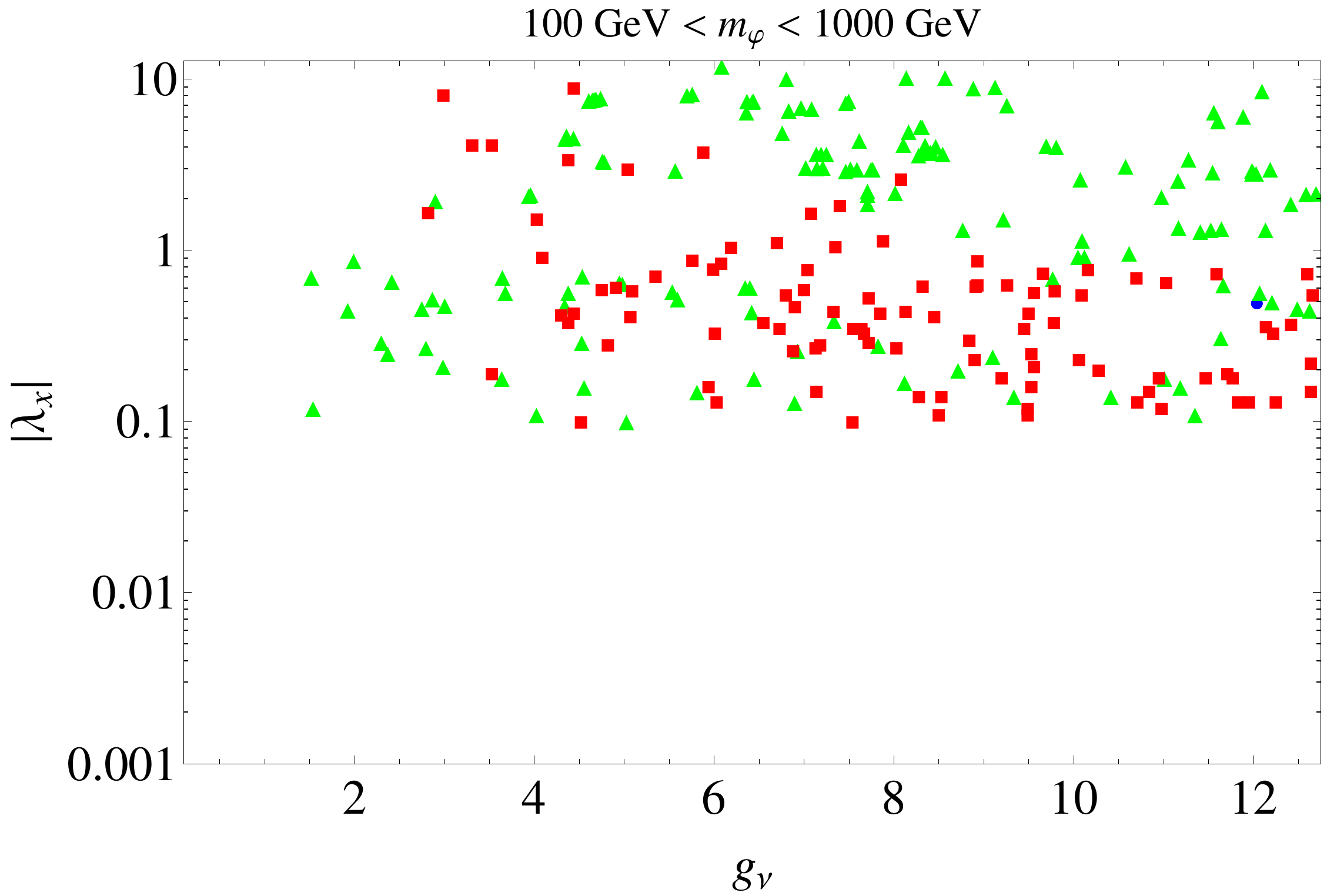}\\
\includegraphics[height=4.5cm]{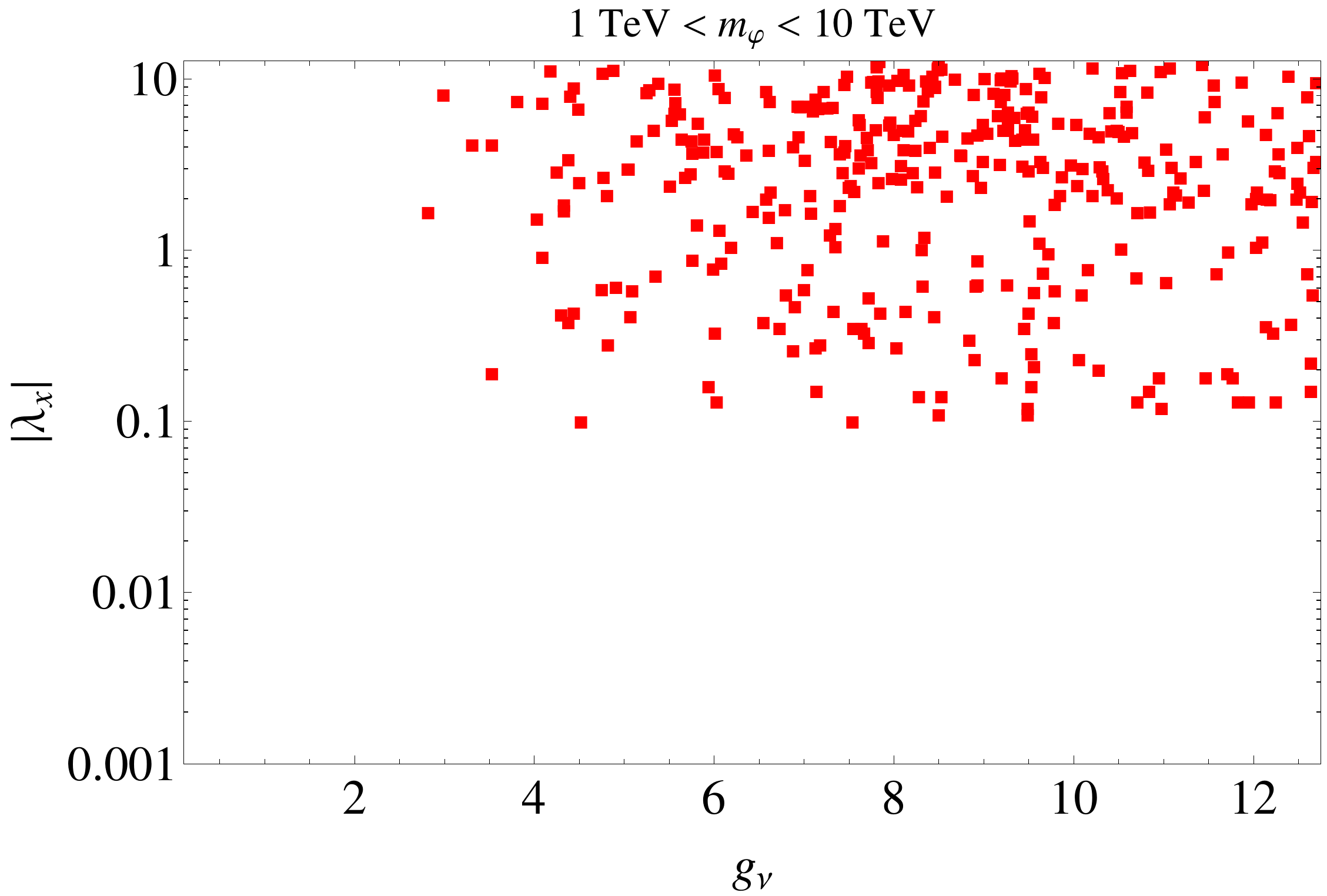}
\caption{Points (obtained by solving the BEQs numerically) that satisfy WMAP bound for cases A and B and
projected into the $(\lx, g_\nu)$ plane. Blue (circles): $\mdm<100\gev$, green (triangles): $100\gev<\mdm<1\tev$ 
red (squares): $1\tev<\mdm<2\tev$ and for scalar DM mass 
ranges as indicated in each panel.}
\label{wmap_g_lx}
\end{figure}

The total relic abundance of DM in our model is given by the sum of the neutrino and scalar abundances:
\beq
\Omega_{tot} = \Omega_\nu + \Omega_\vp = \frac{\mdm f_{\nu}+ \mvp f_{\vp} }{\rho_{crit}}T_{\gamma}^3
\eeq
The experimental data on the relic density measured at the $1\sigma$ level 
by WMAP \cite{Hinshaw:2012aka} shown in equation (\ref{eq:wmap-omega}).
In order to determine parameters of our model that satisfy the limit, we have performed a random 
scan over the 4-dimensional parameter space of our model ($\mvp, \mdm, \lx, g_\nu$) in a range: 
$1\gev < \mvp < 10 \tev$, $1\gev < \mdm < 2 \tev$,
$0.001 < \lx < 4\pi$ and $0.1 < g_\nu < 4\pi$. The results of the scan -- 
points satisfying the relic abundance constraint (within $3\sigma$) in the ($\lx,g_\nu$) plane, are shown in 
fig.~\ref{wmap_g_lx}.
\begin{figure}
\centering
\includegraphics[height=5.5 cm]{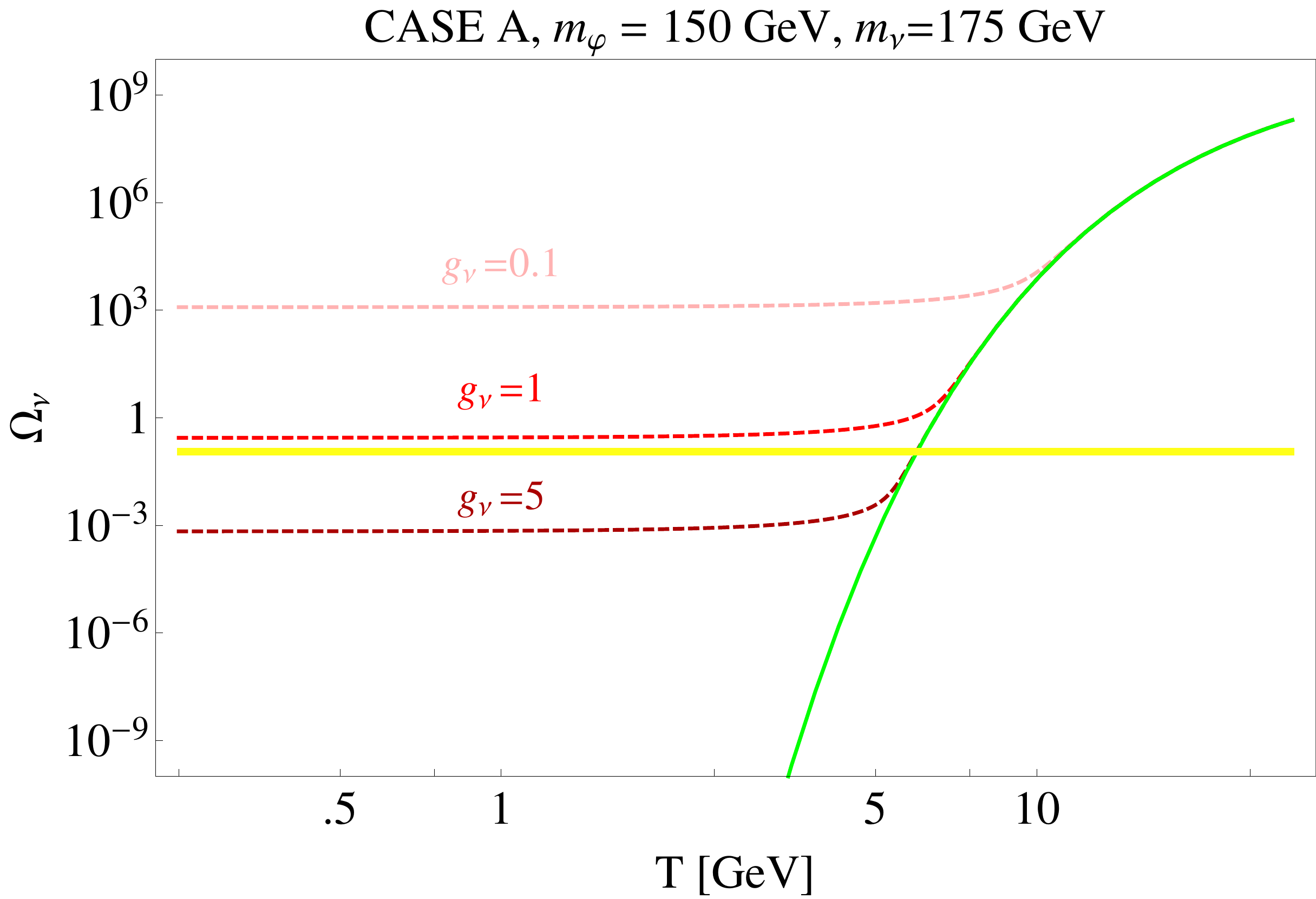}\\
\caption{Solutions of the BEQs for $m_\varphi = 150$ GeV, $m_\nu = 175$ GeV (case A), $\lambda_x = 1$.  
Pink, red, dark red lines: solutions for the neutrino abundance for   $g_\nu = 0.1, 1, 5$, respectively. 
Yellow lines: WMAP $6 \sigma$ limit on DM abundance. Green: equilibrium distribution for neutrinos at 175 GeV.}
\label{small_g_nu}
\end{figure}

\begin{figure}

\includegraphics[width=7.4 cm]{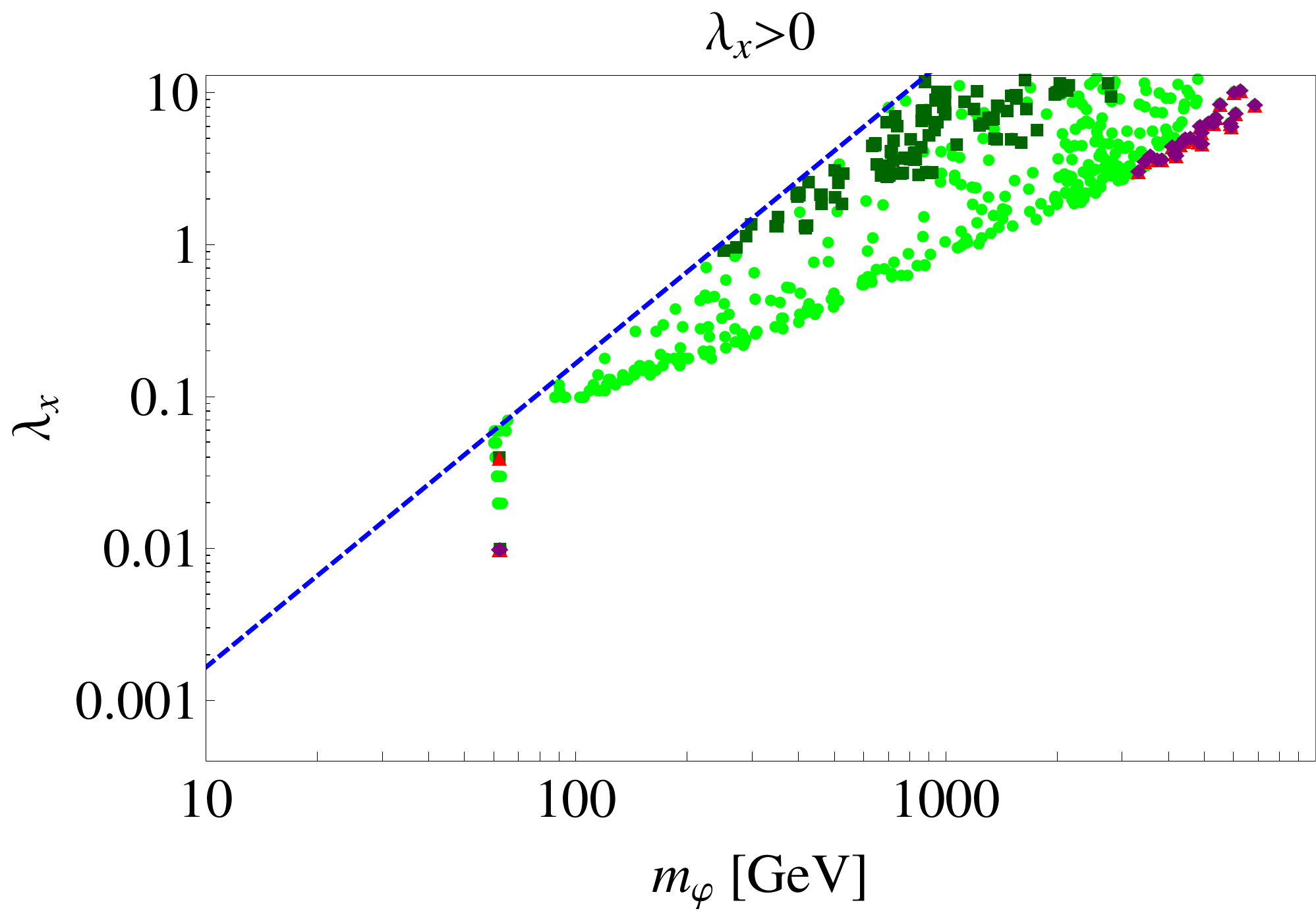}
\includegraphics[width=7.7 cm]{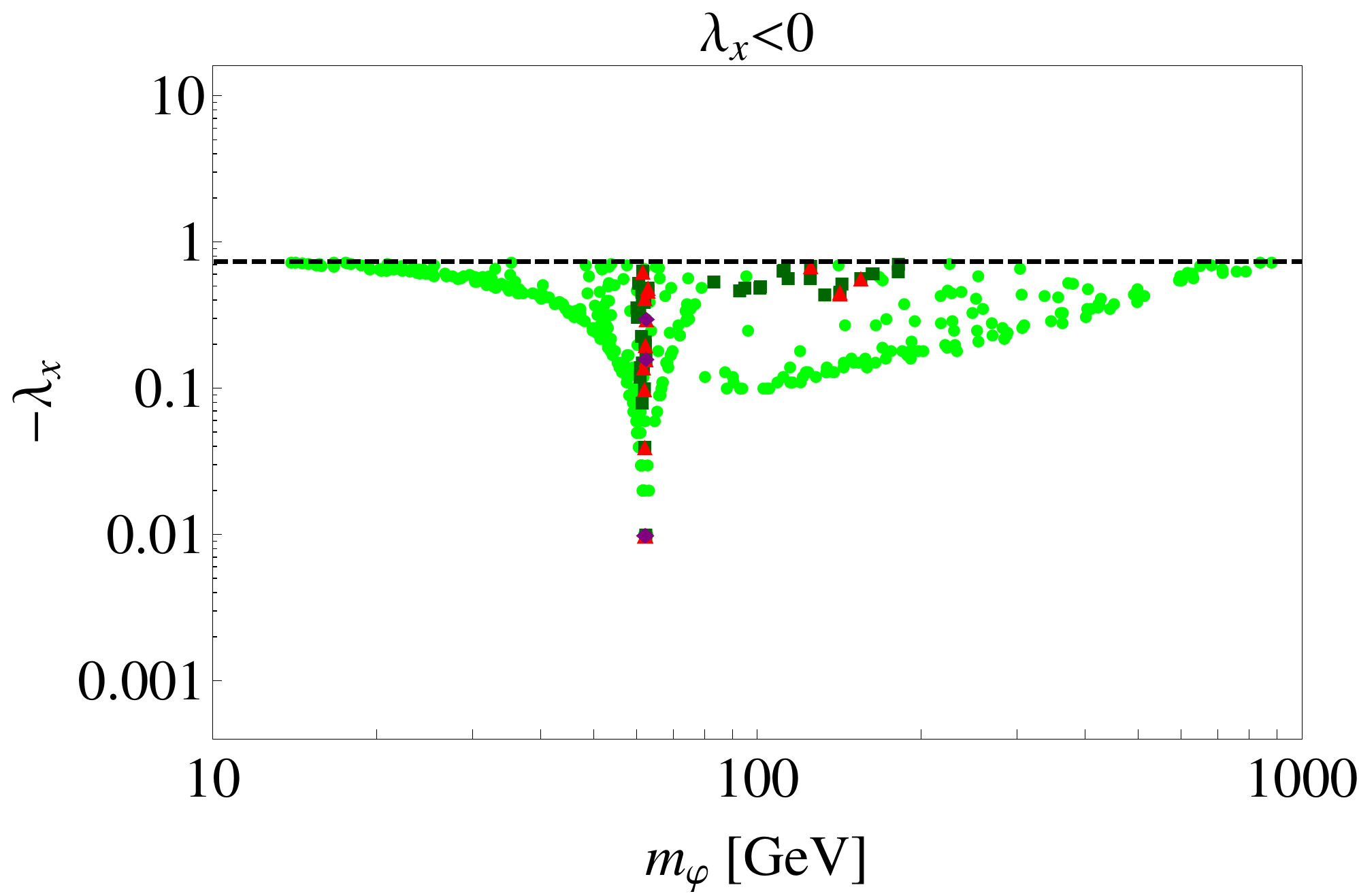}\\
\includegraphics[width=7.1 cm]{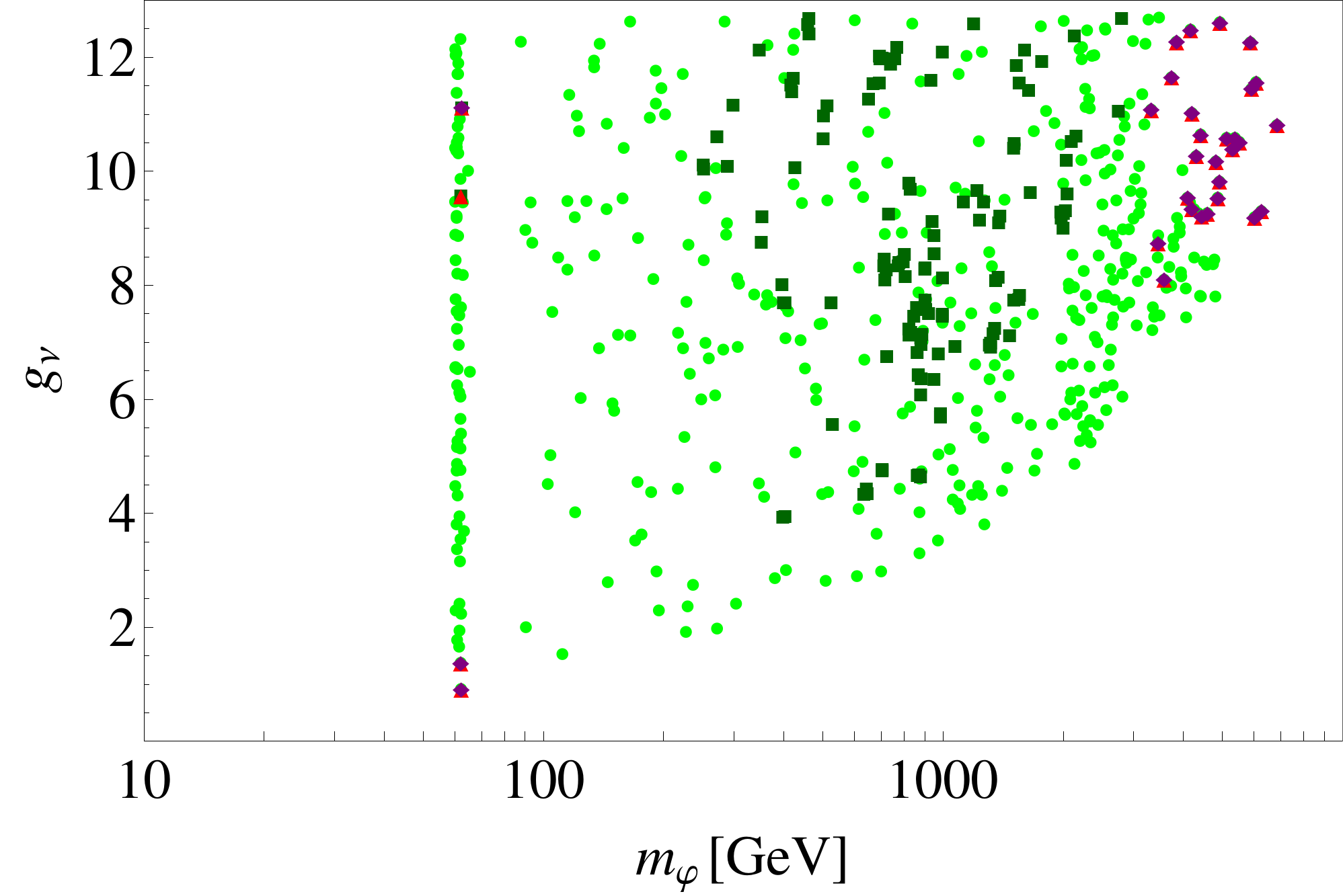}
\includegraphics[width=7.45 cm]{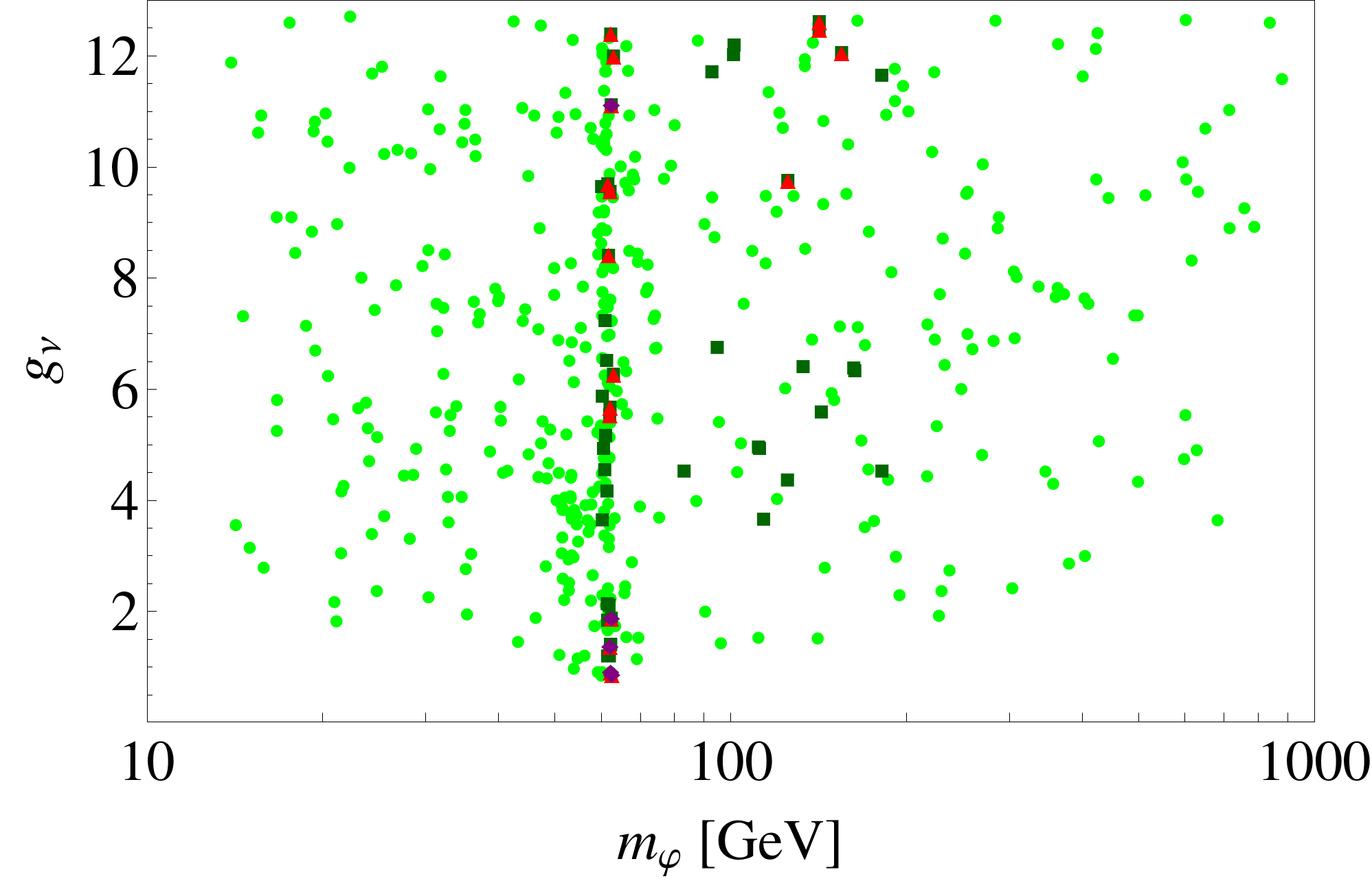}\\
\includegraphics[width=7.4 cm]{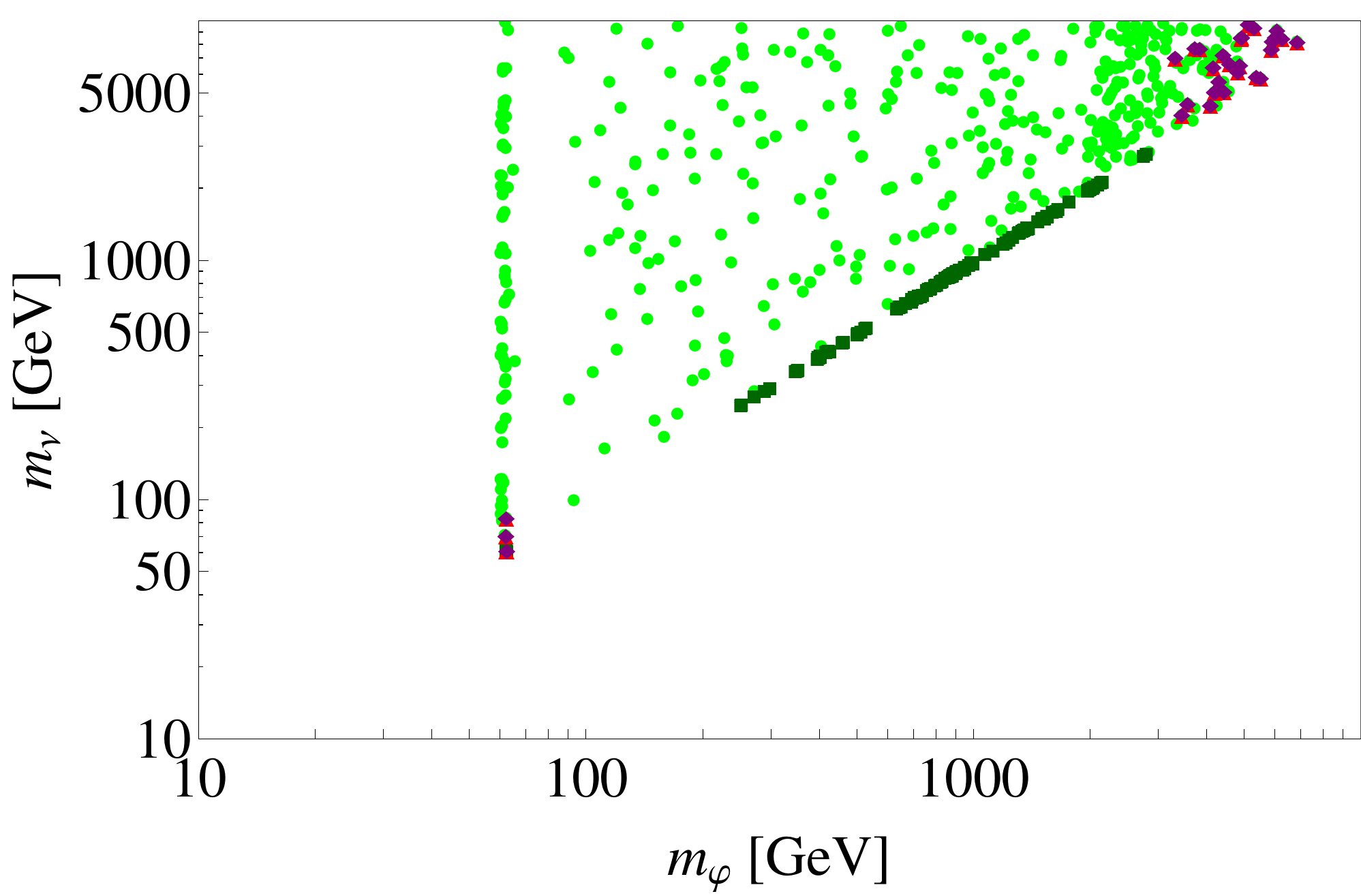}
\includegraphics[width=7.75 cm]{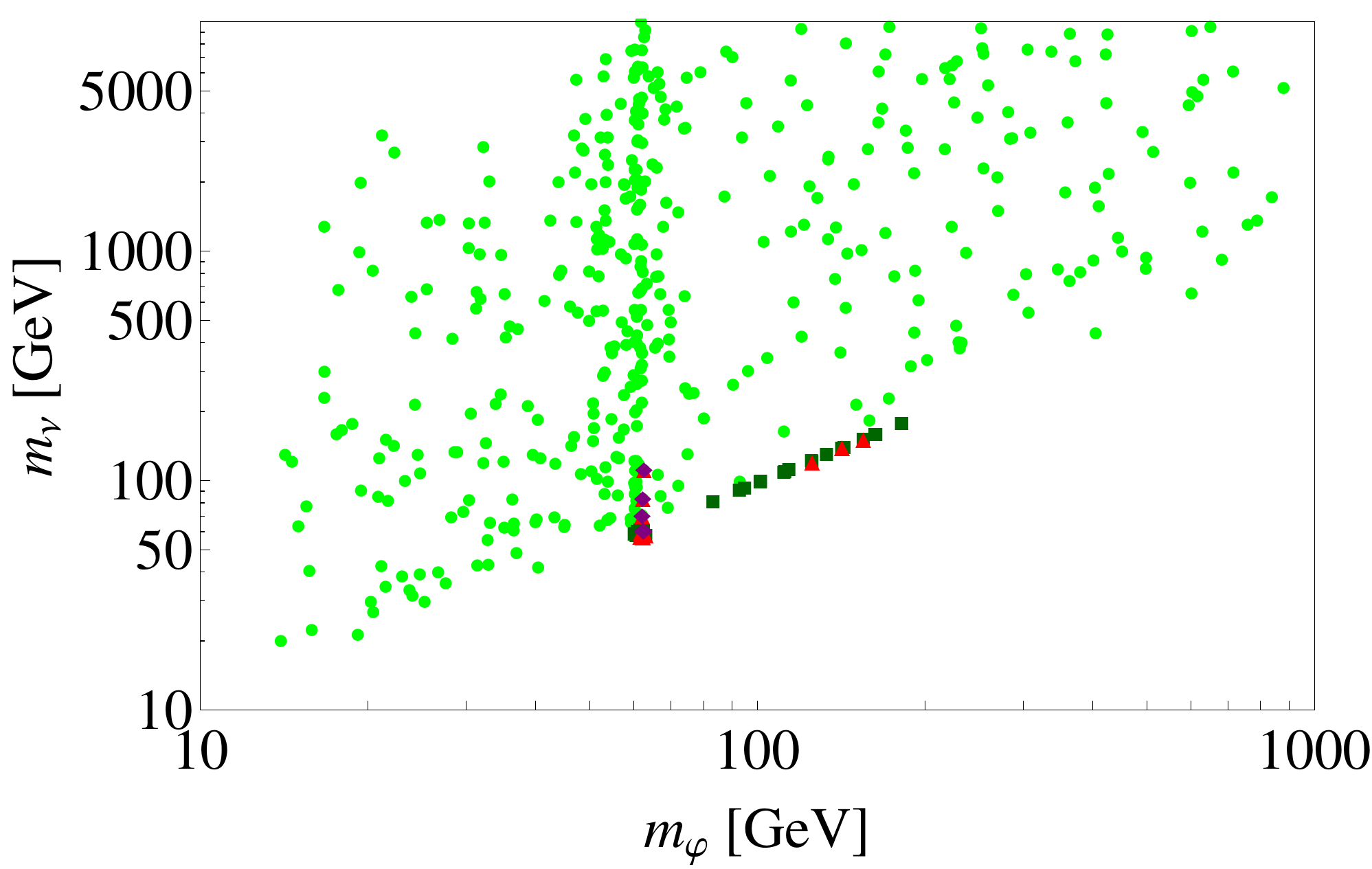}\\
\caption{Points that satisfy WMAP bound within $3 \sigma$ range
projected into
$(\lx,\mvp)$ (upper), $(g_\nu,\mvp)$ (middle) and $(m_\nu,\mvp)$
(lower) planes. 
Green circles - case A points, dark green squares - case
B points. Red triangles and purple diamonds - points for which the XENON100 limit
is separately satisfied, respectively by $\vp$ and $\nu$. 
The consistency limit on $\lx$ (\ref{lxregions1})
and the stability limit (\ref{lxregions2}) for $\lambda_\vp =
8\pi$ are satisfied.}
\label{wmap_1}
\end{figure}

\begin{figure}
\centering
\includegraphics[height=6 cm]{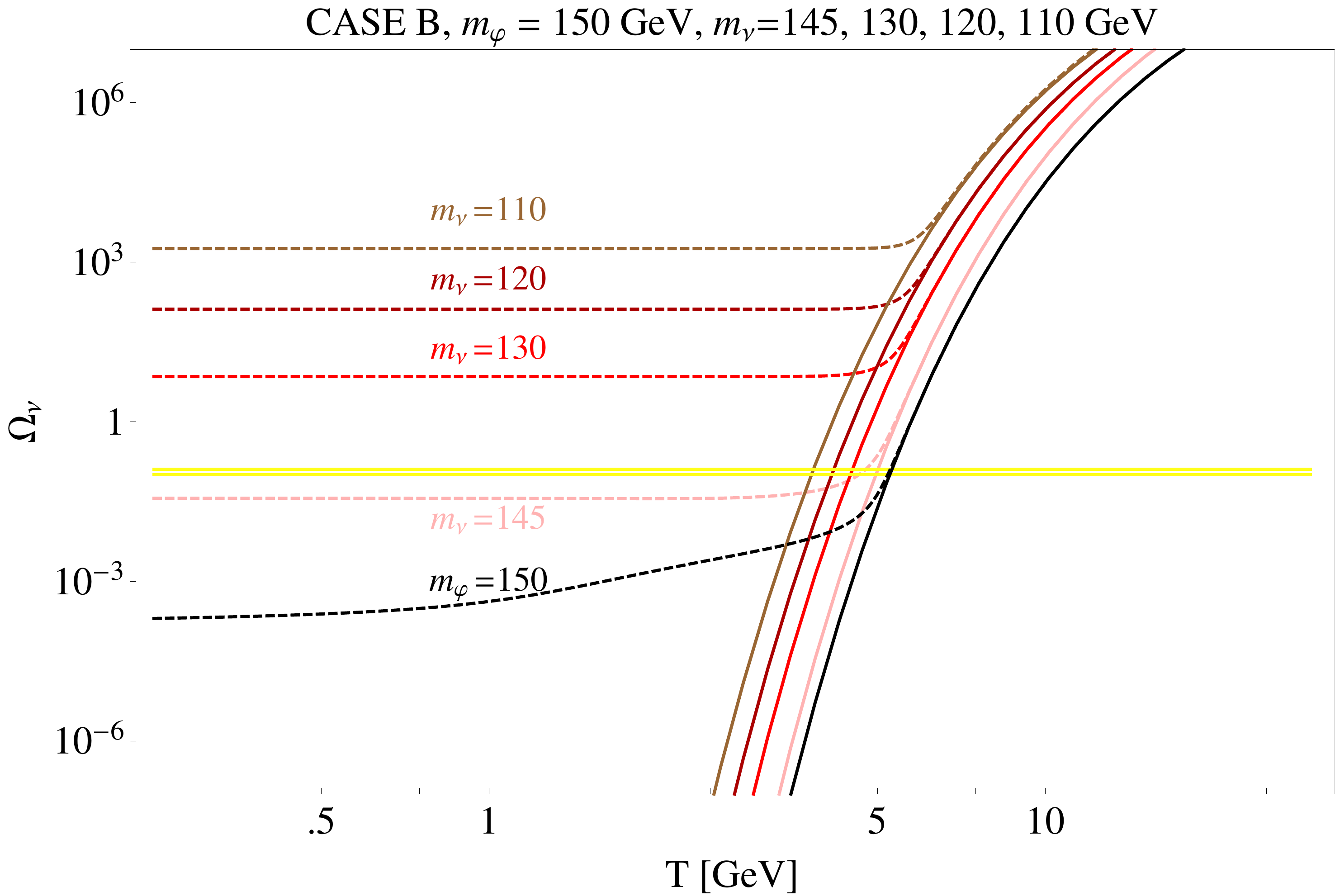}\\
\caption{Solutions to the BEQs: $f_\vp$ (dashed black line),
$f_\vp^{EQ}$ (solid black line) and $ f_\nu $ for
$ \mdm = 145,\,130,\,120,\,110\gev$ (light red, red, dark red and brown dashed lines,
respectively). In all cases we chose $\mvp = 150\gev$, $\lambda_x
= 1$, $g_\nu = 7.5$. Yellow lines are from the WMAP $6 \sigma$ allowed
region of DM abundance.}
\label{mass_split}
\end{figure}

It is seen from fig.~\ref{wmap_g_lx} that we did not find any
points satisfying the WMAP bound for $g_\nu < 0.92$. 
In fact, it is easy to understand why $g_\nu$ can not be very
small: as is seen from fig.~\ref{small_g_nu} the relic abundance of $\nu$
increases rapidly as  $g_\nu$ drops, since this suppresses annihilation into 
scalar pairs;  $g_\nu$ must be
large enough in order to avoid overabundance of neutrinos.
This reasoning is supported, in the case A, by our approximate
analytical solution (\ref{delnu1}) for which
$f_\nu(T_{\rm CMB}) \propto \sigma_A^{-1} \sim
g_\nu^{-4}$, so that an order of magnitude change in
$g_\nu$ implies 4 orders of magnitude change in the abundance of
neutrinos!
From fig.~\ref{wmap_g_lx} we also observe that 
the WMAP constraint requires that a growing $\mvp$ be
correlated with large $|\lx|$ and $g_\nu$,
so that with increasing $\mvp$ our points 
are more and more concentrated in the upper
right corner of the $g_\nu-\lx$ plane.
Note that in the lower panel only red squares survive, this is because
for heavier scalar DM masses, only
slightly heavier or degenerate neutrino DM masses, 
accompanied by large
values of $|\lx|$ and $g_\nu$ survive the relic
density constraint. This is also easy to understand: with increasing
scalar DM mass, scalar relic density increases for case A and
neutrino DM density increases for both case A and case B.
So, to bring the number density down within the observed
limit, we need  large couplings to increase the annihilation rates
and, in addition, the mass splitting has to be small 
in order to tame the neutrino DM
density; see, for example, fig.~\ref{fig_case_AB}. 
This is also understood from figs. \ref{wmap_1} and
\ref{mass_split} as discussed below.

It is instructive to look into various projections of the scan
points shown in fig.~\ref{wmap_1}.
The left panels are for $\lx > 0$, so the limits
(\ref{lxregions1}) are imposed, while
the right ones are for $\lx < 0$ in which case only the limit
(\ref{lxregions2}) applies.
In each case we show, as a function of $\mvp$, all the remaining
parameters, $\lx$, $g_\nu$
and $\mdm$ for which the WMAP bound is satisfied. The plots
in fig.~\ref{wmap_1} differentiate between case A (green circles)
and B (dark green squares) of which the former are much more plentiful
Note that the $m_\vp-m_\nu$ projections  (lower panels) show
that in case B the WMAP restriction can be satisfied only
for $\mvp \simeq \mdm$, as dark green squares are located just below
the diagonal line.
In fig.~\ref{mass_split} we illustrate the effects of the  $\mvp$-$\mdm$ 
splitting on the neutrino abundance $\Omega_\nu$ for  case B;
we can see that the WMAP bound can be met only
when the masses are close enough.
This can be understood from the right panels of
fig.~\ref{fig_case_AB} from which
it is clear that in order to reduce $ f_\nu $,
the dominant low-temperature component of DM,
the splitting between $\mvp$ and $\mdm$ must be small;
for large splittings the DM decouples and freezes-out
early  ($T_f^\nu\sim O(10)\gev$), and the neutrinos do not
have enough time
to disappear into SM particles, leading to an unacceptably large DM
relic abundance. When the mass splitting is small the neutrino
annihilation into scalars (followed by scalar annihilation
into SM particles) is still sufficiently efficient to yield
an acceptable relic abundance. Summarizing, in case B the
WMAP bound can be met only if
{\em i)} the neutrinos freeze-out relatively late, 
and {\em ii)} $\mdm \simeq \mvp$.

We also include in fig.~\ref{wmap_1} points that satisfy direct
detection limits
from XENON100 (red triangles) and CREST-II  (blue diamonds)
(direct detection of DM will be discussed in detail in
sec.~\ref{dirdet}).
It is important to note already at this point that there
exist three regions of $\vp$ mass which are consistent with XENON100:
$\mvp\simeq m_h/2$, $\mvp\simeq 130-140\gev$ and heavy
mass region $\mvp\gesim 3\tev$.

It is also worth discussing more quantitatively
the degeneracy $\mdm \simeq \mvp$
required for the case B.
As it is seen in the upper and middle panels of
fig.~\ref{fig_case_AB},  if the mass
splitting $\Delta m = \mvp - \mdm$ is not too large, then the
decoupling from equilibrium occurs in a range of temperatures
where the ``distance'' between
distributions, $\Delta f(T)\equiv
[\log_{10}f^{EQ}_\vp(T)-\log_{10}f^{EQ}_\nu(T)]$,
is approximately $T$-independent, and depends mainly on $\Delta m $.
Changing the coupling constants alters the
decoupling temperature of both DM particles,
but $\Delta f(T)$ remains unaltered.  
Since scalars and neutrinos decouple roughly simultaneously
$\Delta f(T_{\rm CMB})$ is also a function of $\Delta m$ only. 
It follows that, if $\mvp \simeq m_\nu$, the
difference between the $\vp$ and $\nu$ contributions to
$\Omega_{DM}$,
$\Delta\Omega = (\log_{10}\Omega_\vp-\log_{10}\Omega_\nu)$, is
roughly a function of $\Delta m$ only.
The minimal abundance found within the numerical scans is $\Omega\sim 10^{-8}$. In order to reach the WMAP range of abundance ($\Omega \sim 0.1$), the maximal value of
$\Delta\Omega$ should be $\sim 7$. From fig.~\ref{mass_split} we can estimate that this value of $\Delta\Omega$ corresponds to $\Delta m \lesim 40$ GeV. This very rough estimate agrees with our numerical scans where we find that (in case B) the maximal allowed splitting found is $\Delta m^{\rm MAX}_{\rm NUM}=29.8$ GeV.

\begin{figure}
\centering
\includegraphics[height=4.7 cm]{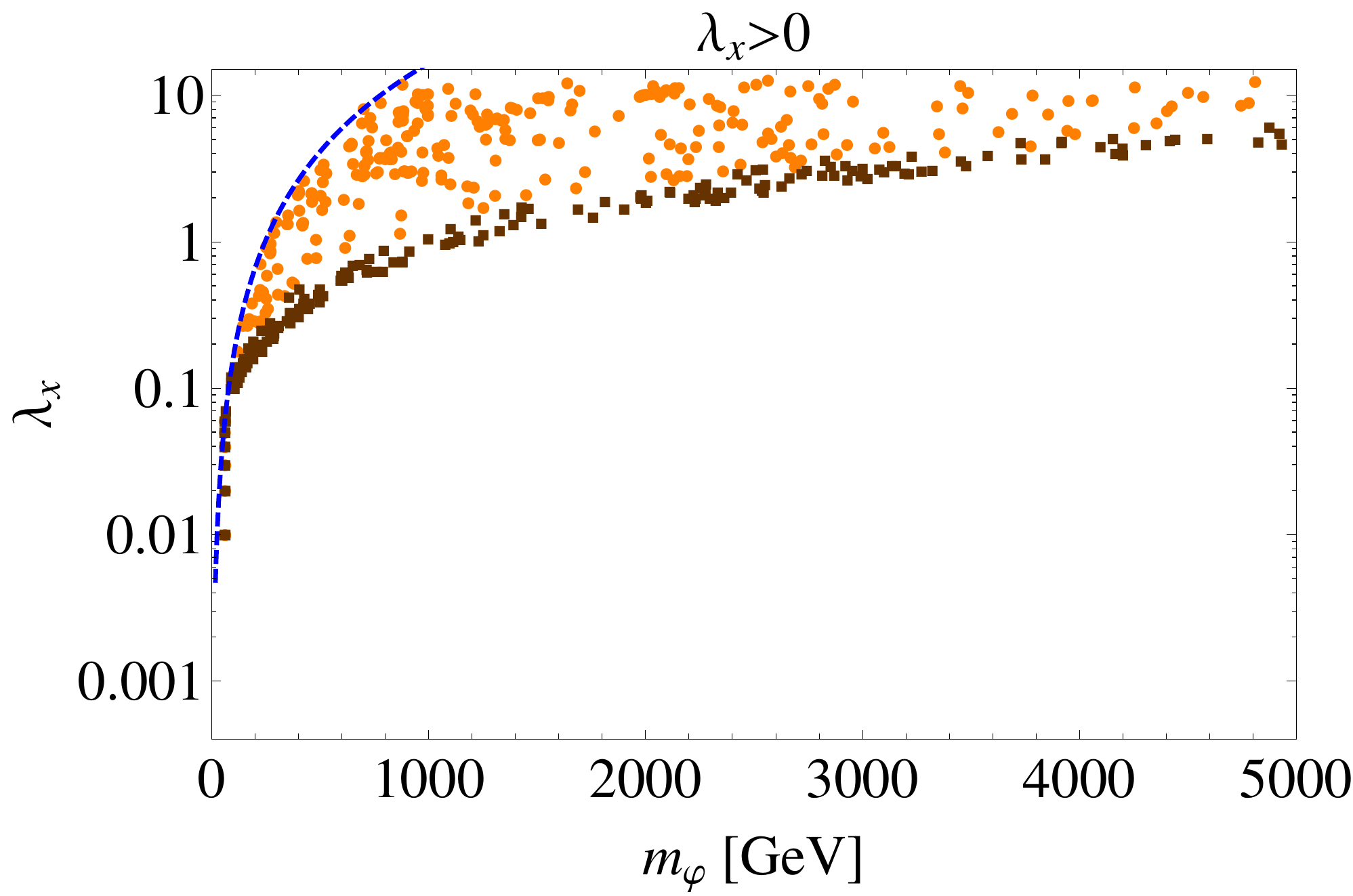}
\includegraphics[height=4.7 cm]{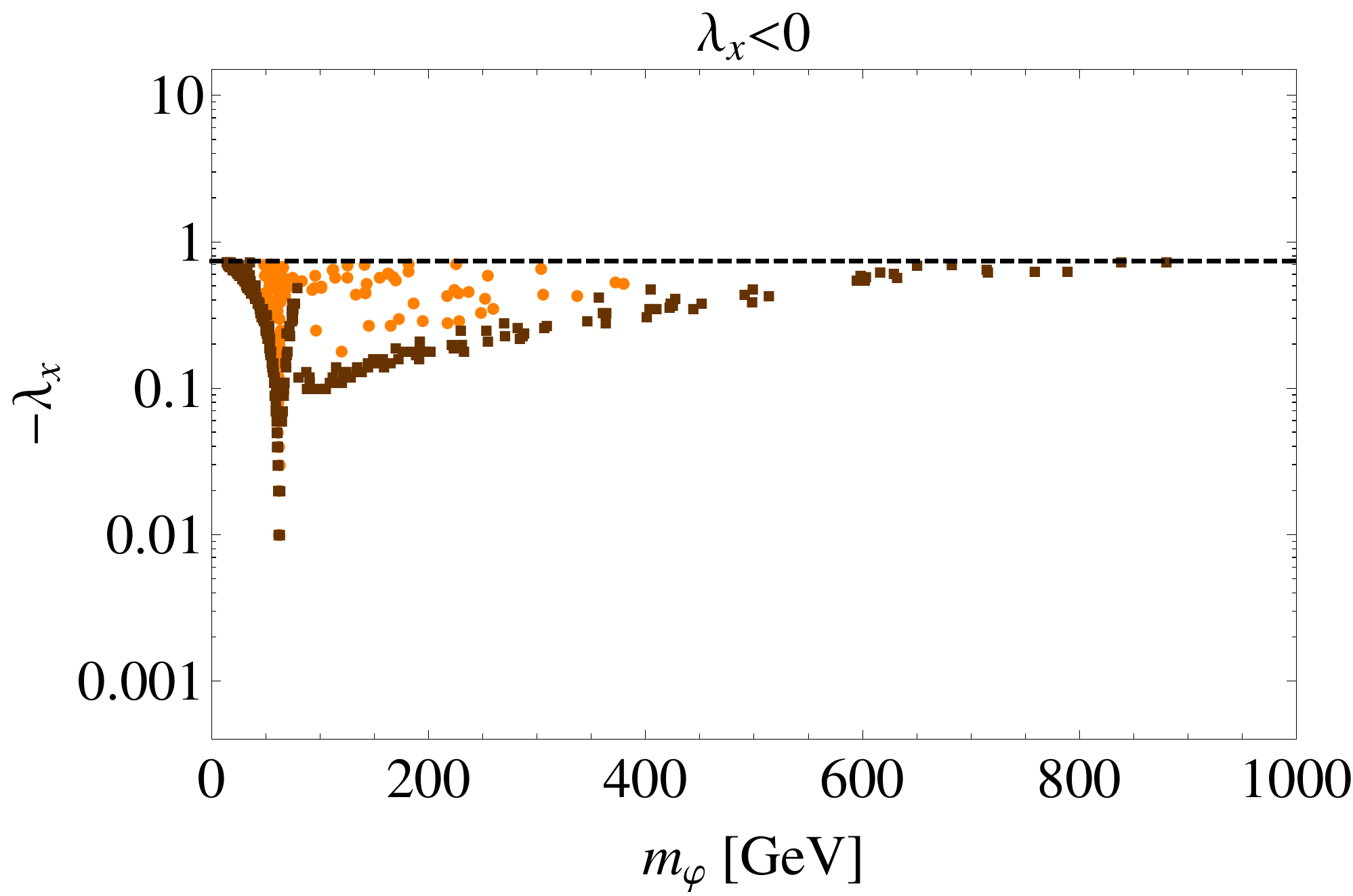}
\caption{Points that satisfy WMAP bound within $3 \sigma$ range
projected into
$(\lx,\mvp)$ plane. Orange circles - points where $\Omega_\vp <
\Omega_\nu$,
dark orange squares - points where $\Omega_\vp > \Omega_\nu$.
The left panel corresponds to the solutions for positive $\lx$,
while the right panel is for negative $\lx$. Blue dashed line is
the consistency limit on $\lx$ (\ref{lxregions1}), while the black horizontal
dashed line is the stability limit  $\lambda_\vp = 8\pi$.
}
\label{wmap_3}
\end{figure}

The top panel of fig.~\ref{wmap_1} clearly shows the 
resonance region $\mvp \simeq \mh/2 \sim 62.5 \gev$
in the $(\lx,\mvp)$ plane; $\lx$
must be small otherwise the resonant graph with a 
Higgs boson in the s-channel yields too large annihilation 
rate and consequently too small $\vp$ abundance.
The intermediate mass consistent with XENON100, 
$\mvp\simeq 130-140\gev$ requires $\lx<0$ that
causes a destructive interference between diagrams contributing
to the annihilation rate so that the annihilation rate could be suppressed 
even with substantial $\lx$. The high scalar mass region 
consistent with XENON100 requires large $\lx$.
In the middle panel of fig.~\ref{wmap_1} we 
again observe that usually large values of $g_\nu$ are allowed by
the WMAP data. In fig.~\ref{wmap_3} we present allowed region 
in the $(\lx,\mvp)$ plane for both $\Omega_\vp > \Omega_\nu$ 
and $\Omega_\vp < \Omega_\nu$; it is worth noting
that points that are close to the lower edge of the
WMAP allowed region generally
correspond to $\Omega_\vp > \Omega_\nu$ (dark orange squares).
When $\Omega_\vp$ dominates 
$\vp$ annihilation rate must
be sufficiently suppressed in order to keep the $\vp$ abundance
at the WMAP level. The edge
corresponds to the result for $\lx=\lx(\mvp)$ obtained for one
singlet DM case investigated in \cite{Drozd:2011aa},
(see fig.~7 in that reference). 
\begin{figure}
\centering
\includegraphics[width=7 cm]{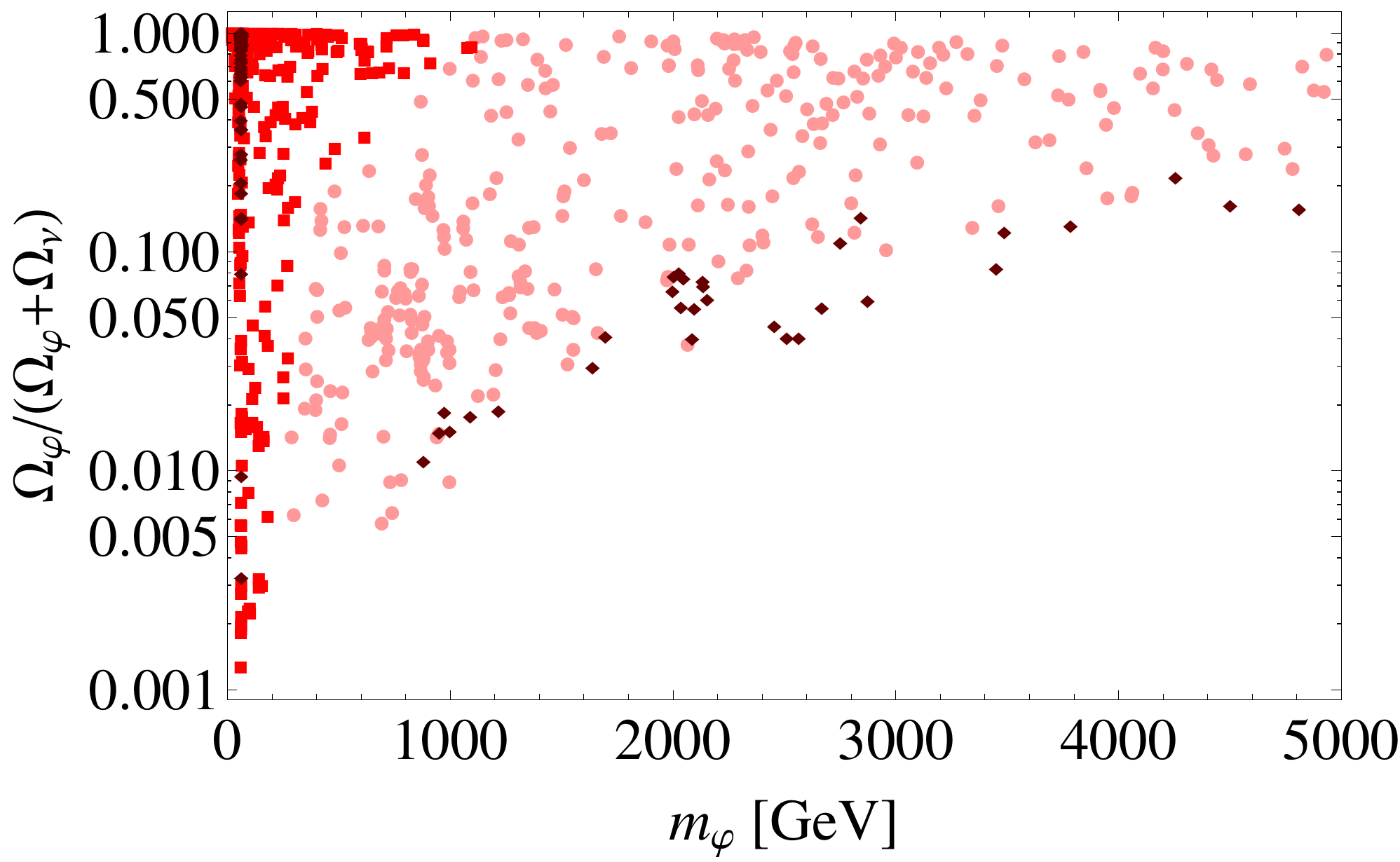}
\includegraphics[width=7 cm]{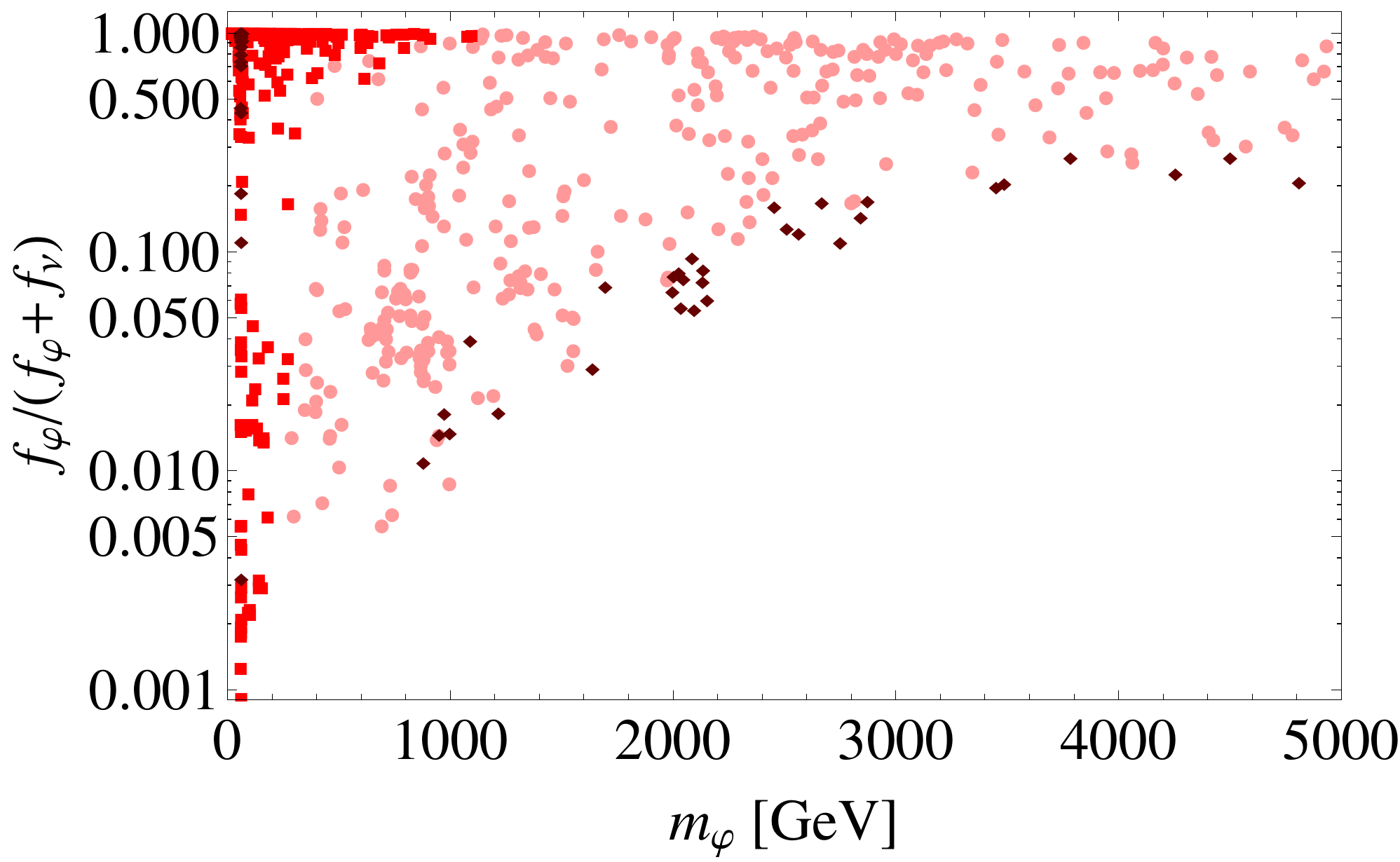}
\caption{Relative abundance of $\vp$ (left panel) and relative number density of $\vp$
(right panel) 
as a function of $\mvp$ for points that satisfy WMAP bound within $6 \sigma$. 
Light red points: $1< \lx < 10$;   red  points: $0.1< \lx < 1$;
dark red  points: $\lx <0.1$ }
\label{wmap_3_a}
\end{figure}

Fig.~\ref{wmap_3_a} illustrates the manner in which the DM abundance is split
between $\vp$ and $\nu$.

\section{Direct Detection}
\label{dirdet}

In this section we discuss constraints imposed on the model by
searches
for direct signals of DM particles scattering off a nuclei. We
focus here
on constraints obtained by the XENON100
experiment~\cite{Aprile:2012nq}
as they impose strongest limits on DM - nucleon scattering
cross-section $\sigma_{\rm DM-N}$
in the mass range of our interest. 
We will also comment on results obtained by the CREST-II
experiment \cite{Angloher:2011uu}.
\begin{figure}
\includegraphics[height = 3 cm, width = 3 cm]{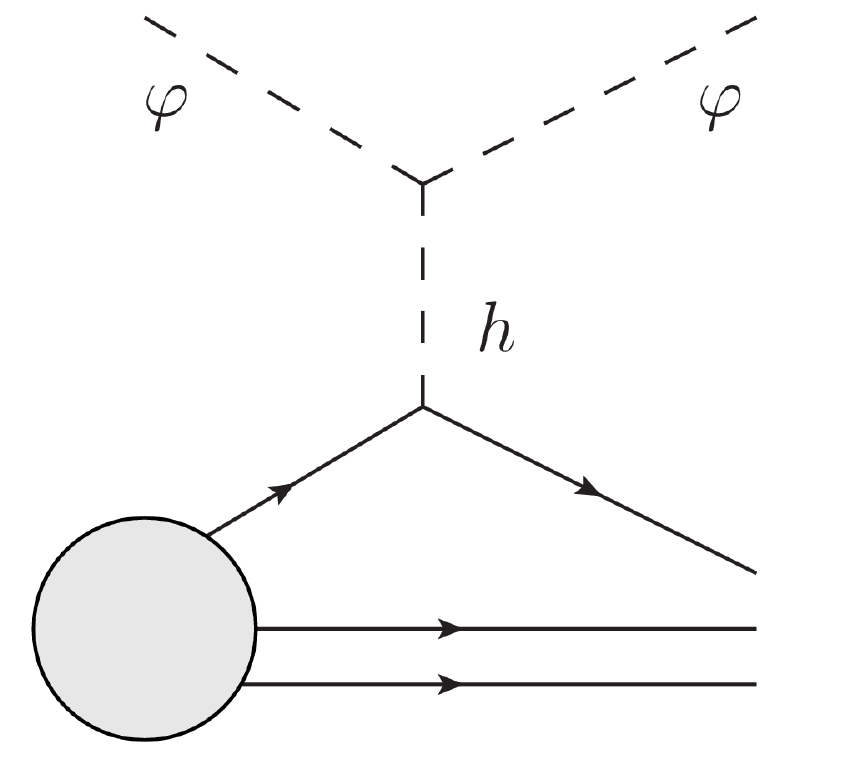}
\hspace{1cm}
\includegraphics[height = 3 cm, width = 3 cm]{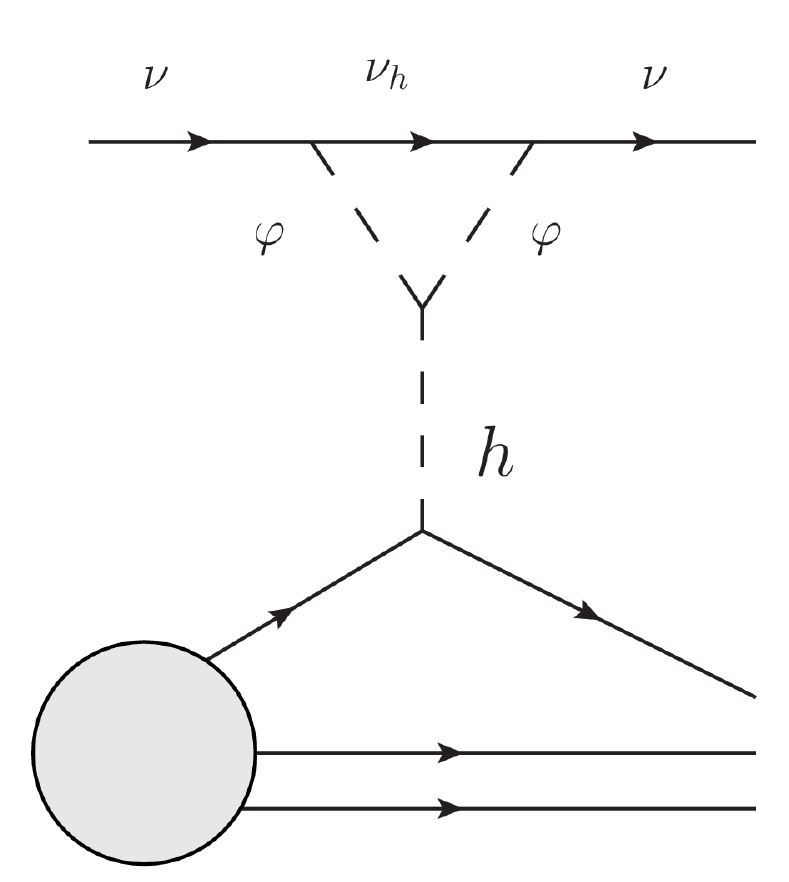}
\caption{The Feynman diagram for the elastic scattering of DM ($\vp$ and $\nu$) off a nucleon.}
\label{dir_det}
\end{figure}

In our model, at the tree level, scattering of DM
off nuclei originates from the interaction with the scalar DM component. 
Neutrino's leading contribution to the scattering appears at the one-loop level. 
However, as it has been multiply illustrated, the DM is often dominated by dark neutrinos.
Therefore, even though $\nu$ nucleon scattering is loop induced, it might be relevant. 
Therefore, the dominant contributions to the
scattering of DM are described by the two Feynman diagrams in
fig.~\ref{dir_det}.
We start with $\vp$ nucleon scattering, the corresponding cross section is the following
\beq
\sigma_{\vp N} = \frac{4 \mu^2}{\pi} 
\left( \frac{\lx m_N}{2 \mvp m_h^2}  \sum_{q} f^{N}_{q} \right)^2
\label{dir_vp_xs}
\eeq
where the sum runs over all quark flavors $q$, $m_n$ is the
nucleon mass and $f^{N}_{q}$ are
the nucleon form factors as defined in~\cite{Belanger:2008sj} and $\mu \equiv \mvp m_N/(\mvp+m_N)$.

\begin{figure}
\centering
\includegraphics[width=7.8 cm]{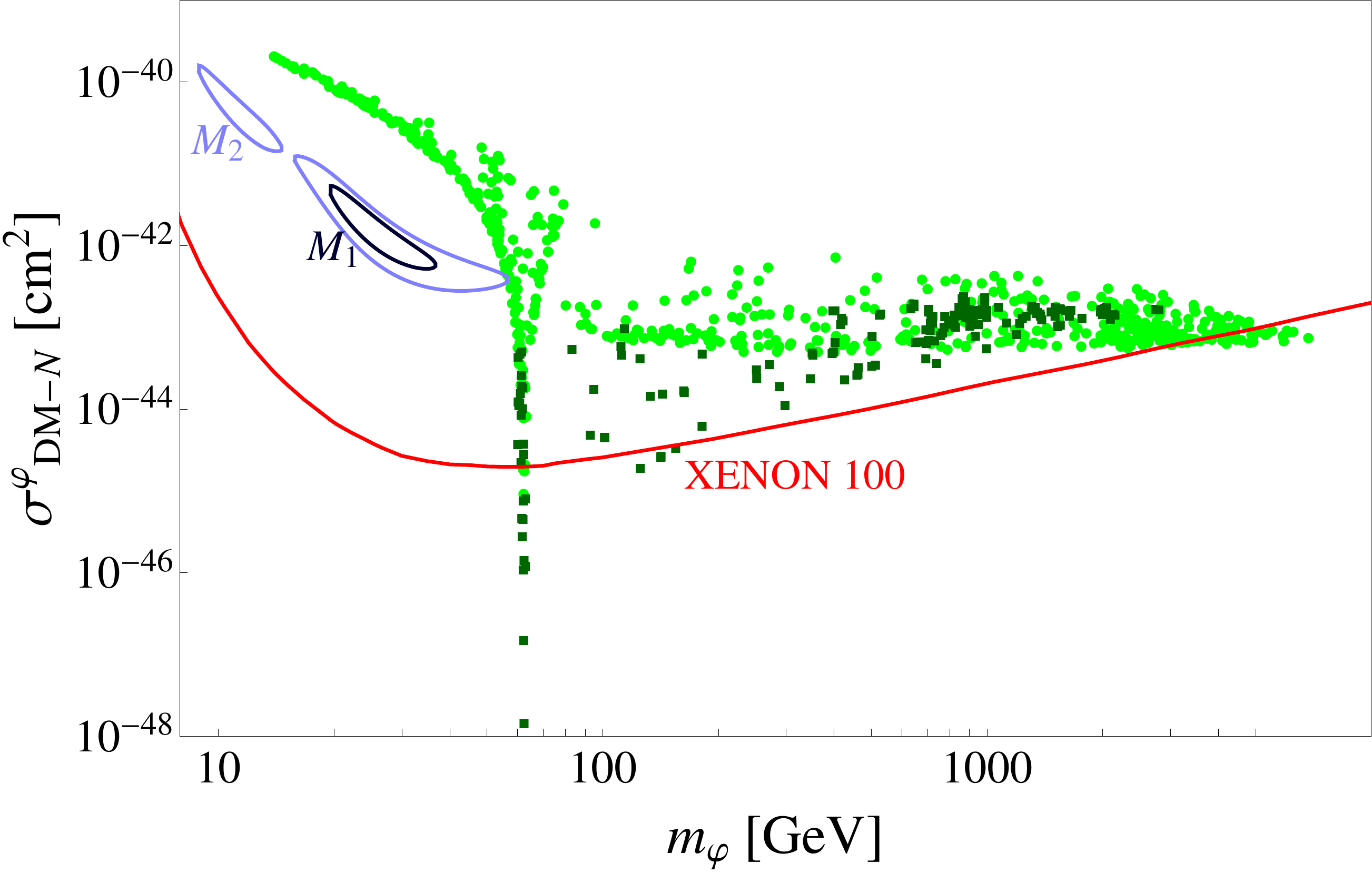}
\includegraphics[width=7.8 cm]{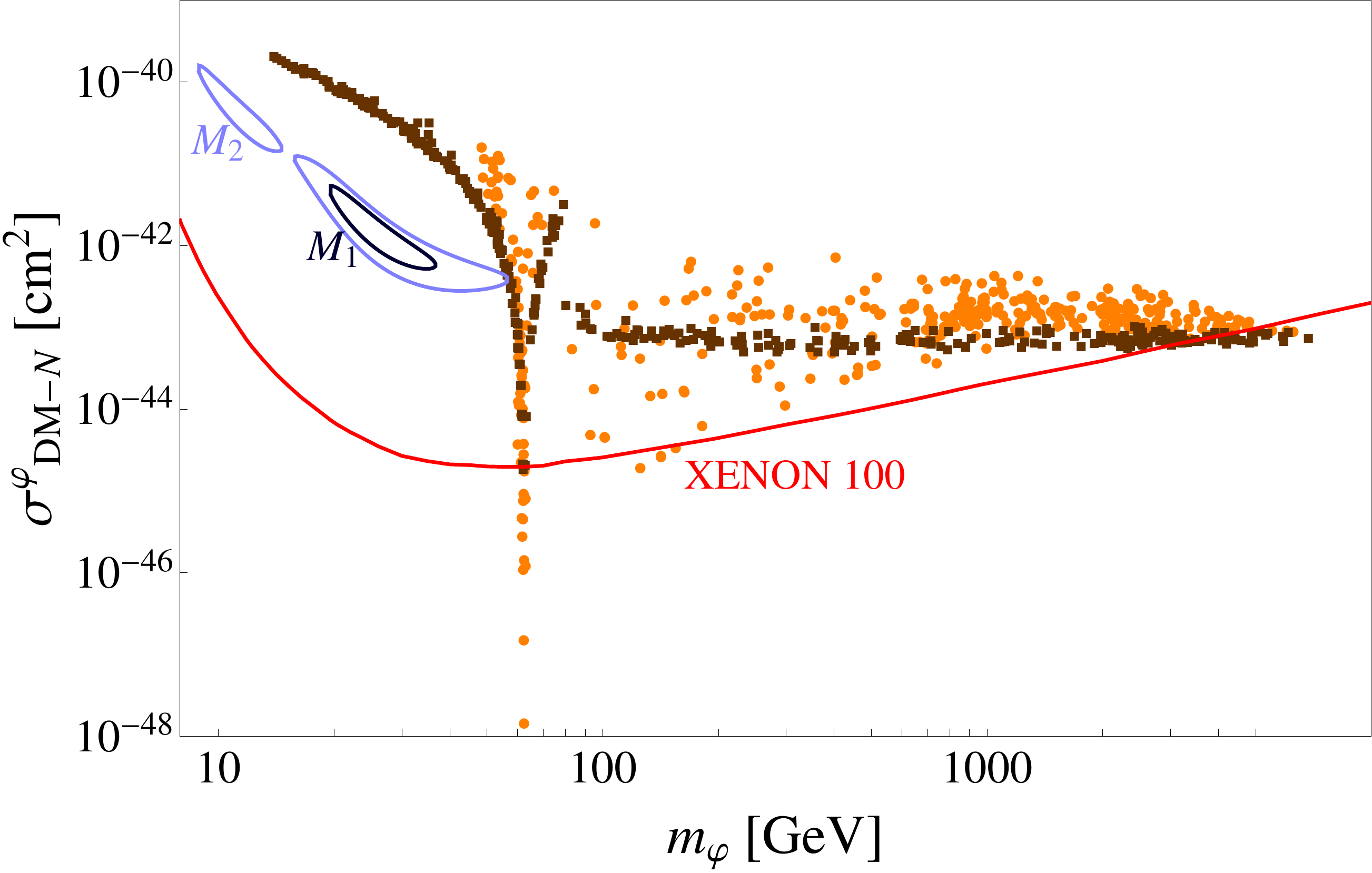}
\caption{
Plot of the cross section $\sigma_{\rm DM-N}^\vp$ 
as a function of $\mvp$ for points satisfying the WMAP 
data within $3 \sigma$; the other parameters are randomly chosen in the ranges defined in the text
(including both signs of $\lx$). Left panel: green circles (dark green squares) 
correspond to case A (case B) solutions. Right panel: orange circles (dark orange squares) 
correspond to $\Omega_\vp < \Omega_\nu$ ( $\Omega_\vp > \Omega_\nu$).
The red line shows the XENON100 data, and the two islands in blue indicate 
1 and $2\sigma$ CRESST-II results.}
\label{DD_vp_res}
\end{figure}

To compare the prediction for the direct detection cross section
obtained within our 2-component DM scenario
with experimental results from the XENON collaboration one has
to take into account that the standard
limits on DM direct detection assume all DM particles to be
interacting with SM with the same rate.
In our case, this is not true as we have two components of DM
and their number densities are in general different.
Therefore, to compare with the data, we need to rescale the
$\vp N$ cross section 
by a factor that accounts for the fact that two DM
components are present:
\beq
\sigma_{\rm DM-N}^\vp = \frac{n_{\vp}}{n_{\vp}+n_{\nu}} \sigma_{\vp  N}.
\label{dir_DM_vp}
\eeq
In fig.~\ref{DD_vp_res} we plot the rescaled cross section
$\sigma_{\rm DM-N}^\vp$ as a function of $\mvp$ 
calculated for points satisfying the WMAP bounds for cases A and 
B. It follows from this figure that in the resonance region
$\mvp\simeq m_h/2$ and in the middle mass region $\mvp \simeq 130-140\gev$ 
direct detection constraints favor $\mdm < \mvp$ (case B). However for the heavy
scalars solution $\mvp \gesim 3\tev$ it turns out that $\mdm > \mvp$ (case A)
is required.
\begin{figure}
\centering
\includegraphics[width=7.8 cm]{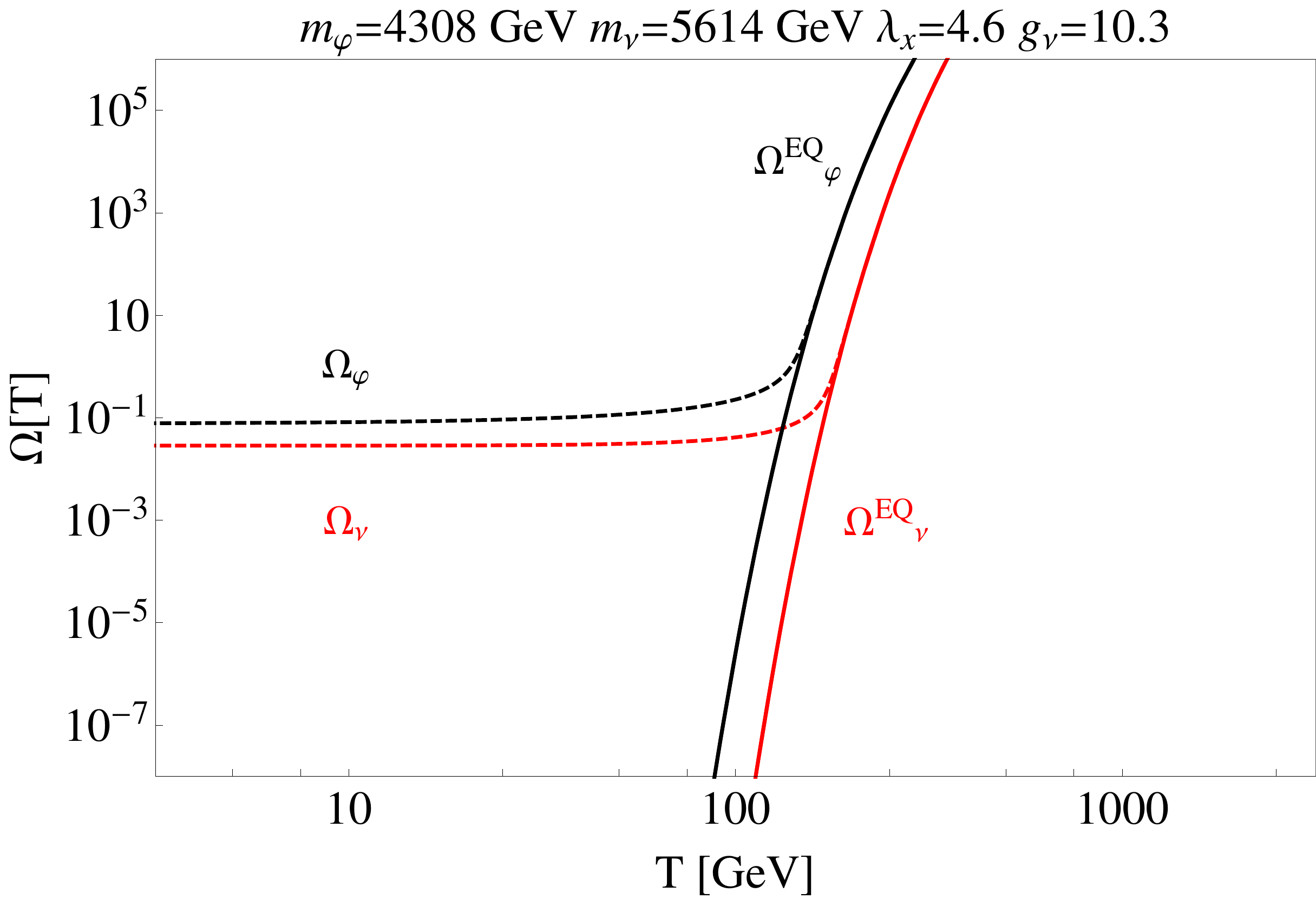}
\includegraphics[width=7.8 cm]{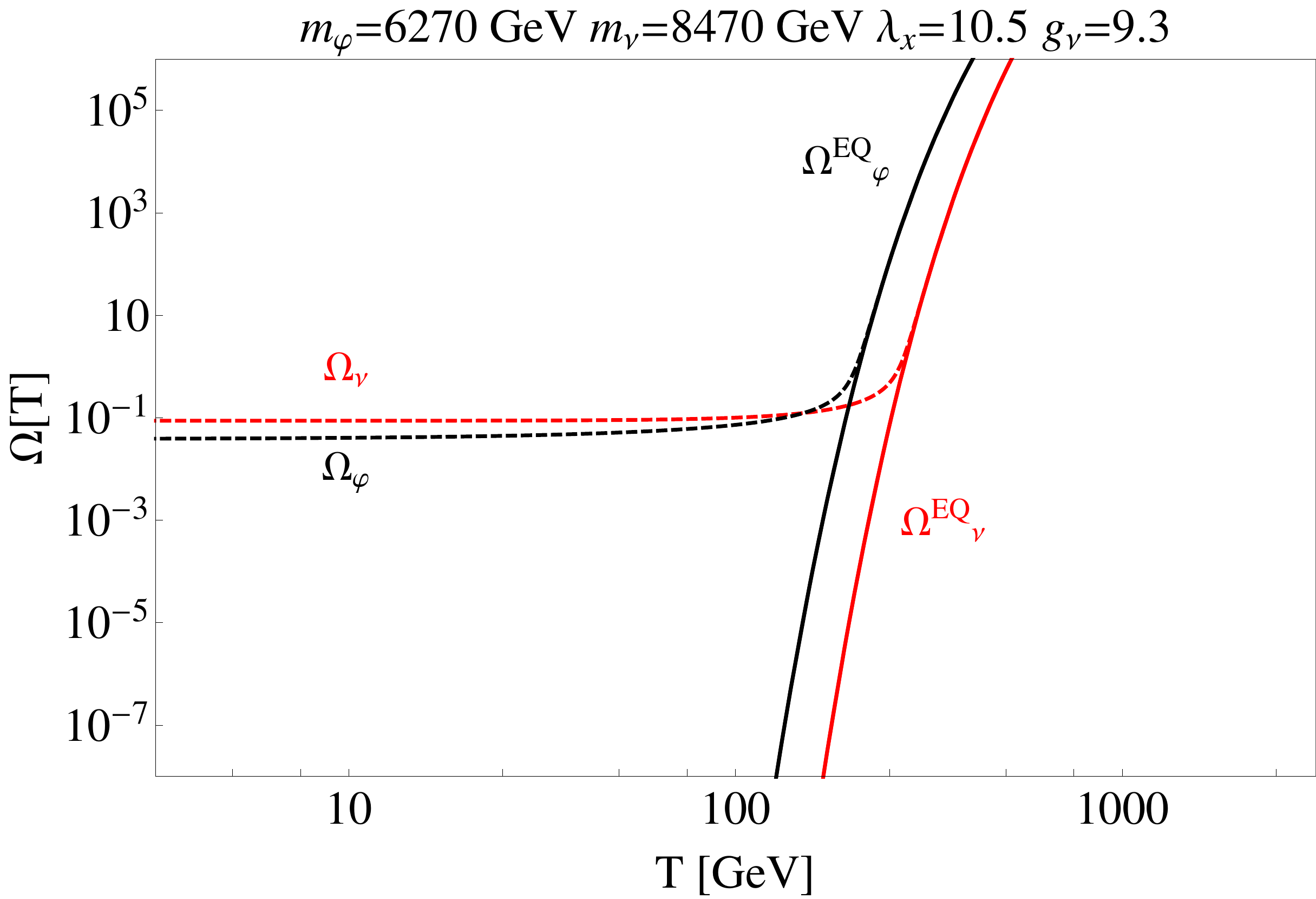}
\includegraphics[width=7.8 cm]{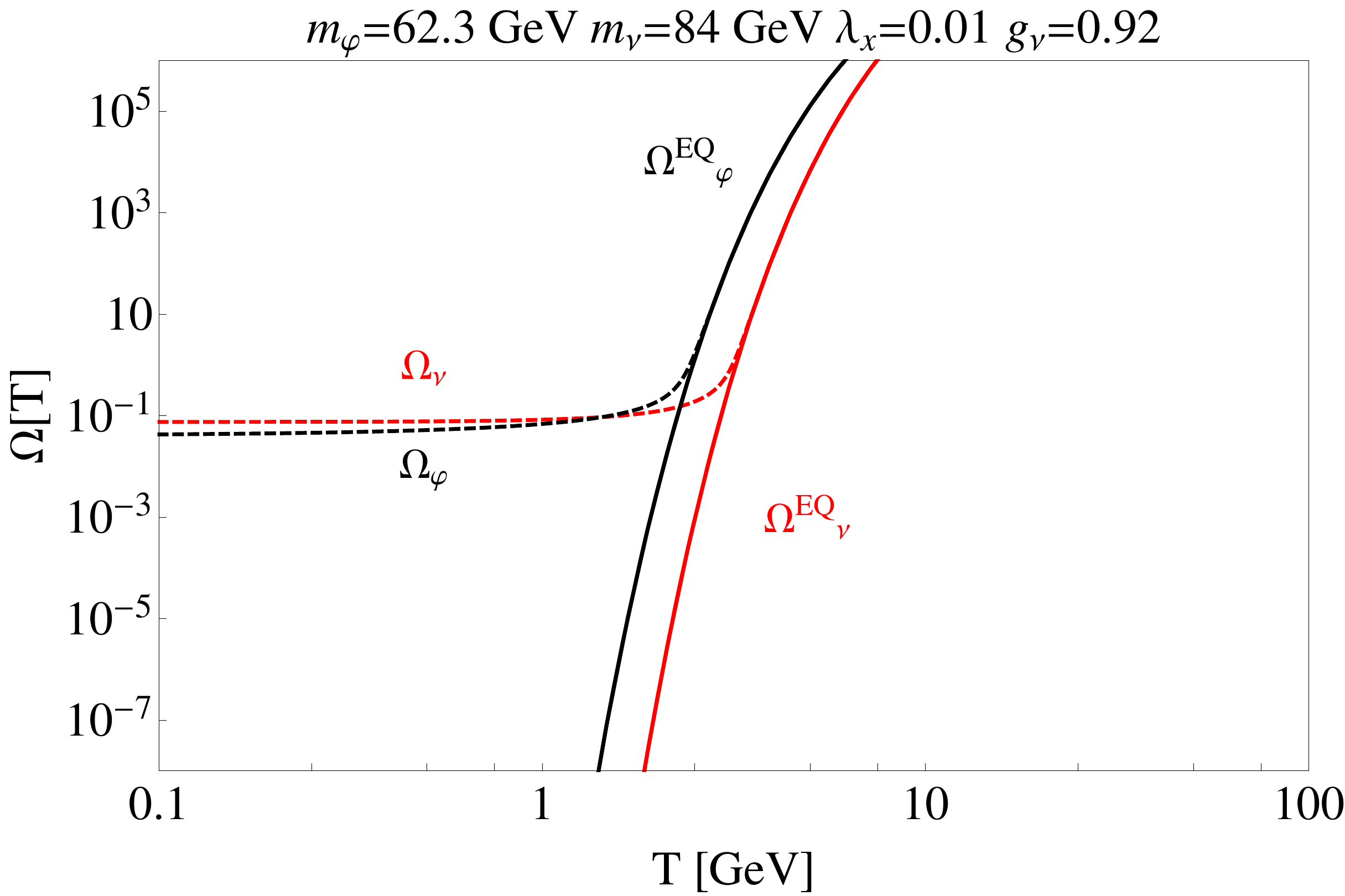}
\includegraphics[width=7.8 cm]{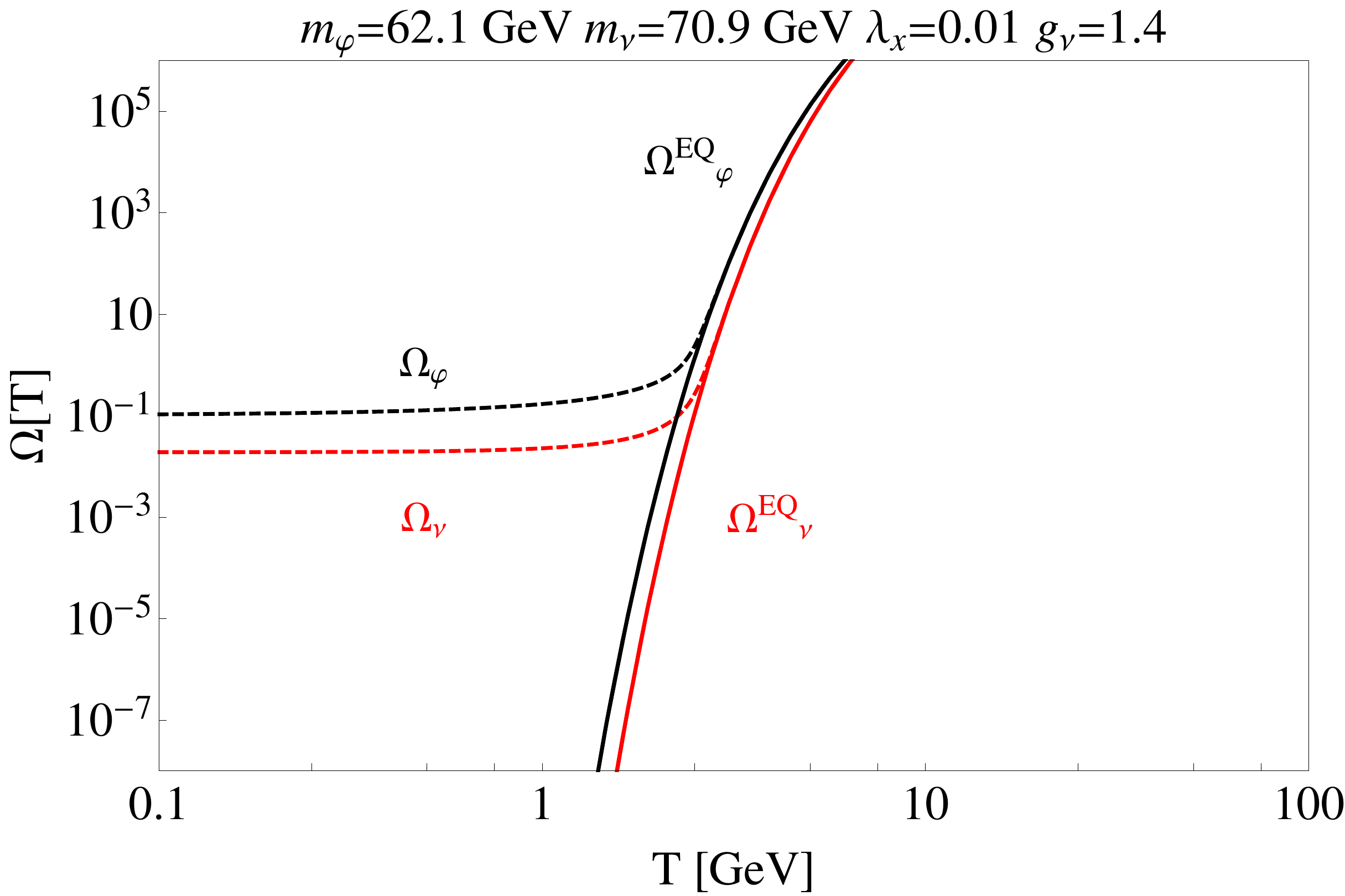}
\caption{Selected solutions of the Boltzmann equation for parameters that satisfy both WMAP and XENON constraints.}
\label{DD_res2}
\end{figure}
The right panel of fig.~\ref{DD_vp_res} illustrates the correlation between 
$\sigma_{\rm DM-N}^\vp$ and the relative abundance 
of $\vp$ and $\nu$. 
We observe that in the resonance region and in middle mass region
$\Omega_\vp < \Omega_\nu$ (more neutrinos) while for
the large mass $\Omega_\vp > \Omega_\nu$ (more scalars).

As seen from fig.~\ref{DD_vp_res} the majority of points lie above (i.e. 
are excluded by) the XENON100 lower limit.
This  is easy to understand. 
Using (\ref{dir_vp_xs}) and (\ref{dir_DM_vp}) we find that
\beq
\sigma_{\rm DM-N}^\vp \propto \frac{\lx^2}{\mvp^2} \frac{f_\vp(T_{\rm CMB})}{f_\vp(T_{\rm CMB})+f_\nu(T_{\rm CMB})}.
\label{dir_DM_xs2}
\eeq
In order to minimize $\sigma_{\rm DM-N}^\vp$ one should (for a given
$\mvp$) choose $\lx$ and 
$f_\vp(T_{\rm CMB})/(f_\vp(T_{\rm CMB})+f_\nu(T_{\rm CMB}))$
as small as possible.
These factors, however, are correlated.
For a conservative estimate of the $\mvp$ dependence 
we choose the lower edge of the allowed $(\lx,\mvp)$ region
from the upper panel of
fig.~\ref{wmap_1}, and the lower edge of the $f_\vp(T_{\rm
CMB})/(f_\vp(T_{\rm CMB})+f_\nu(T_{\rm CMB}))$
region found in fig.~\ref{wmap_3_a}.
From fig.~\ref{wmap_1} we  find that for $100\gev < \mvp <
1000\gev$
\beq
\log_{10}(\lambda_{x\;{\rm min}}) \simeq 
\log_{10}\left( \frac{\mvp}{1\gev} \right) -3\,,
\label{lxap}
\eeq
while from fig.~\ref{wmap_3_a} we obtain
\beq
\log_{10}\left[\left.\frac{f_\vp(T_{\rm CMB})}{f_\vp(T_{\rm CMB})+f_\nu(T_{\rm CMB})}\right|_{\rm min}\right]\simeq 
0.4 \cdot 10^{-3} \frac{\mvp}{1\gev} - 2.4
\label{frac}
\eeq
Combining (\ref{dir_DM_xs2}-\ref{frac}) we find that
\beq 
\log_{10}\left[\left.\sigma_{\rm DM-N}^\vp\right|_{\rm min}\right] \simeq
-43 + 0.4\cdot 10^{-3} \frac{\mvp}{1\gev}
\label{sigmin}
\eeq
where the constant is such that around $\mvp\sim 100\gev$ the
scan points are above the XENON100 limit as shown in fig~\ref{DD_vp_res}.
The linearly growing part is a reminiscent of the $\mvp$ dependence
present in (\ref{frac}), as the mass dependence of $\lambda_{x\, \rm{min}}$ and
$\mvp$ in (\ref{lxap}) cancel. Note however that the remaining mass dependence is very weak
and in fact disappears after saturating (\ref{frac}) around $5\tev$, see fig.~\ref{wmap_3_a}.

Since there exist solutions in the resonance region it is 
important to calculate
the Higgs-boson-decay branching ratio to $\vp\vp$, 
as those points could be excluded by measurements of the invisible Higgs-boson width.
It turns out that for most of those solutions the $BR(h\to
\vp\vp$) is typically small and in agreement
with the present data \cite{Espinosa:2012vu}.
\begin{figure}
\centering
\includegraphics[width=7.8 cm]{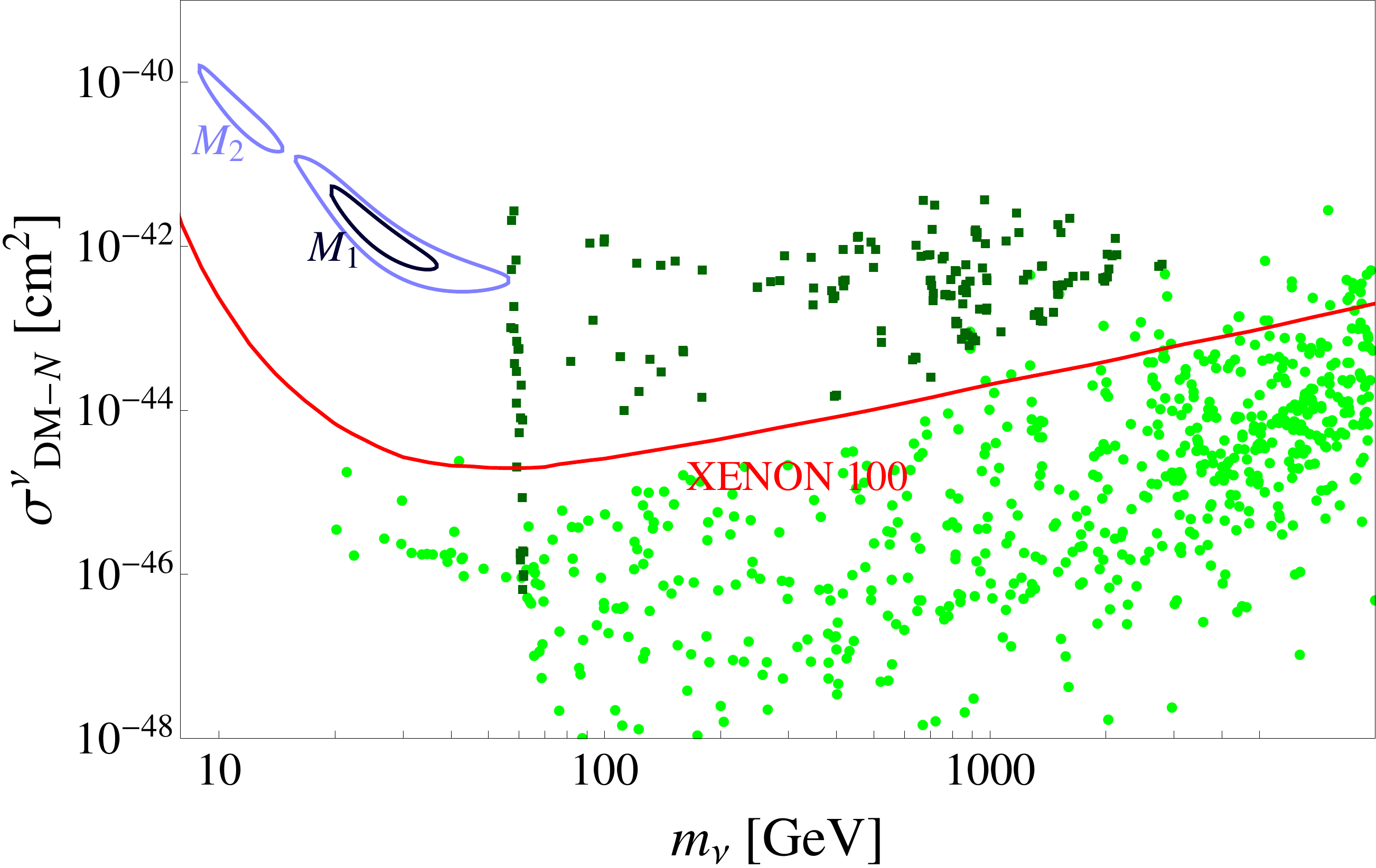}
\includegraphics[width=7.8 cm]{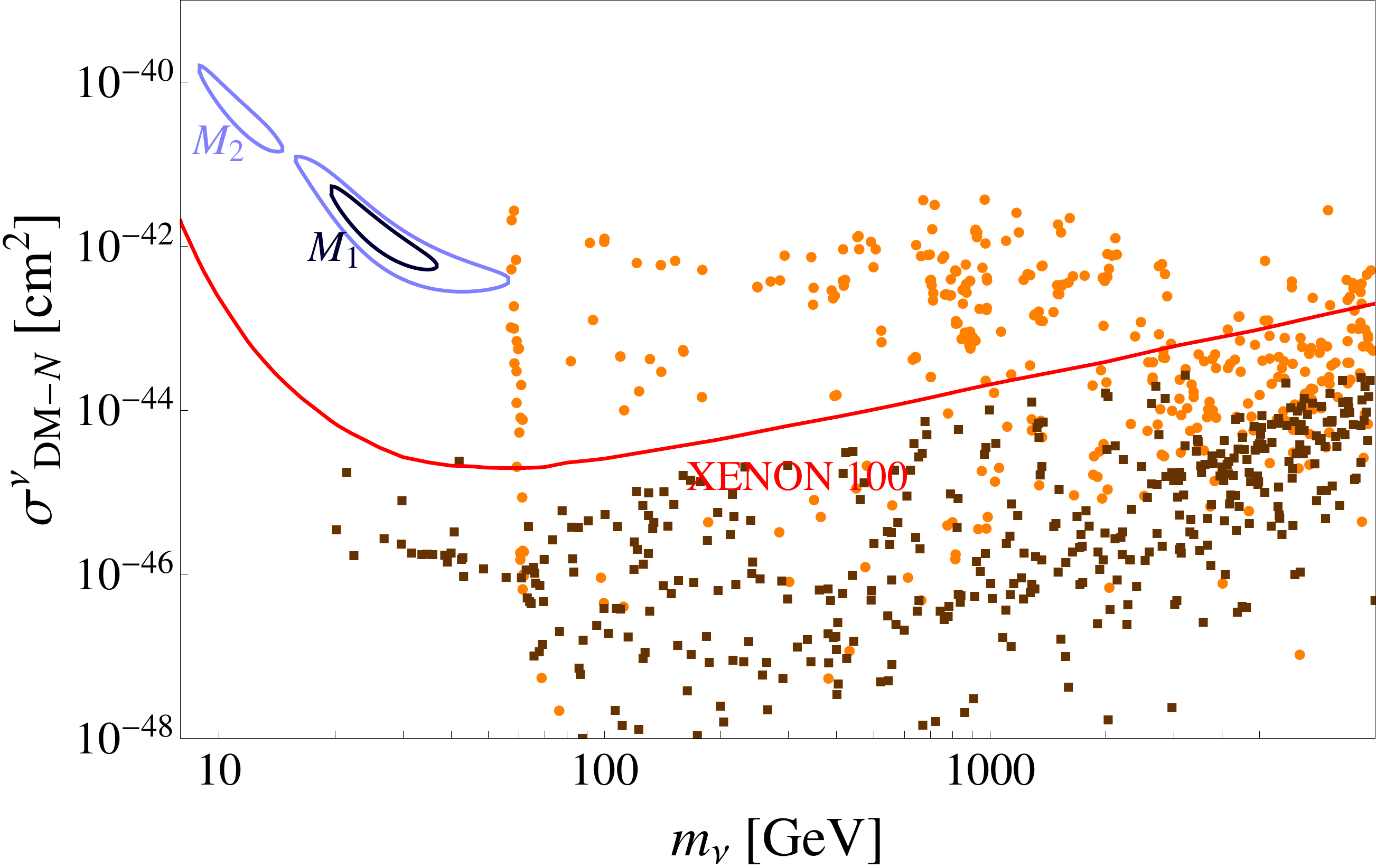}
\caption{
Plot of the cross section $\sigma_{\rm DM-N}^\nu$ 
as a function of $\nu$ for points satisfying the WMAP 
data within $3 \sigma$; the other parameters are randomly chosen in the ranges defined in the text
(including both signs of $\lx$). Left panel: green circles (dark green squares) 
correspond to case A (case B) solutions. Right panel: orange circles (dark orange squares) 
correspond to $\Omega_\vp < \Omega_\nu$ ( $\Omega_\vp > \Omega_\nu$).
The red line shows the XENON100 data, and the two islands in blue indicate 
1 and $2\sigma$ CRESST-II results.}
\label{DD_nu_res}
\end{figure}
It is worth noting that even though the
XENON100 data excludes $\sigma_{\rm DM-N} \gesim
10^{-40}-10^{-44}\cm^2$, other experiments, e.g. 
CREST-II ~\cite{Angloher:2011uu}, claim an observation of DM
scattering with cross sections $\sigma_{\rm DM-N} 
\sim 10^{-40} - 5 \cdot 10^{-43}\cm^2$
(significantly above the XENON100
limits) and for DM mass range $10-60\gev$. 
It is not our intention here to fit our model parameters to the
CREST-II  data,
however few remarks are here in order. First, we have
verified that our model could accommodate CREST-II  $2\sigma$ data,
though in that
region of $\sigma_{\rm DM-N}^\vp$,large $\lx$ is necessary; and since
$\mvp < \mh/2$, the solutions that agree with CREST-II  inevitably 
imply $BR(h\to \vp\vp)\sim 1$, which 
is in conflict with the present collider
data. 
Note however that, since the CREST-II  $2\sigma$ region is 
close to the threshold for $h\to \vp\vp$, therefore a modest ($
\sim 3\sigma$) extension of the region towards
the threshold allows us to find acceptable points above the
threshold for which $BR(h\to \vp\vp) = 0$
since the decay is kinematically forbidden; a sample
of those is shown in tab.~\ref{table2}. It is also worth
noticing from the middle right panel of
fig.~\ref{wmap_1} that the corresponding Yukawa
couplings could be smaller,
$g_\nu \gesim 4$, than those that are needed to satisfy the XENON100
limit (red triangles) in the resonance
region.   
\begin{table}[h]
\begin{tabular}{||c||c|c|c|c|c|c|c|c|c|c|c|c|c|c|c|c|c|c|c|c|c|c|c|c|c|c|c|c|c|c|c|c|c|c||}
\hline\hline
\input{BRaCRESSTwide.dat}
\end{tabular}
\caption{
Points with $BR(h\to\vp \vp)=0$ that satisfy WMAP bound within
$3 \sigma$ range
and for which the cross section $\sigma^\vp_{\rm DM-N}$ 
is within a $3 \sigma $
range of the CREST-II  region $M_1$ and with a
$\mvp$ that is not more than $10\gev$ above the maximal ($2
\sigma$) mass range for CREST-II.
}
\label{table2}
\end{table}

In fig.~\ref{DD_res2} we illustrate temperature evolution of number densities 
(normalized such that at $T_{\rm CMB}$ they coincide with relict abundances) 
for a sample of points that are below XENON100 limit in fig.~\ref{DD_vp_res}.

As it has already been mentioned large abundance of dark neutrinos $\nu$ may imply 
that their contribution, although suppressed at the level of an amplitude, may be relevant 
after taking into account their relative number density:
\beq
\sigma_{\rm DM-N}^\nu = \frac{n_{\nu}}{n_{\vp}+n_{\nu}} \sigma_{\nu  N}
\label{dir_DM_nu}
\eeq
Results for the cross section $\sigma_{\rm DM-N}^\nu$ 
as a function of $\nu$ for points satisfying the WMAP are
confronted with the XENON100 bound in fig.~\ref{DD_nu_res}.
It is seen that the case A points are mostly in agreement with
the bound, while the case B points leads to too large cross section.
It is also worth to notice that points below the XENON100 limit
correspond to $\Omega_\vp > \Omega_\nu$ in agreement with our intuition.

\begin{figure}
\centering
\includegraphics[width=7.8 cm]{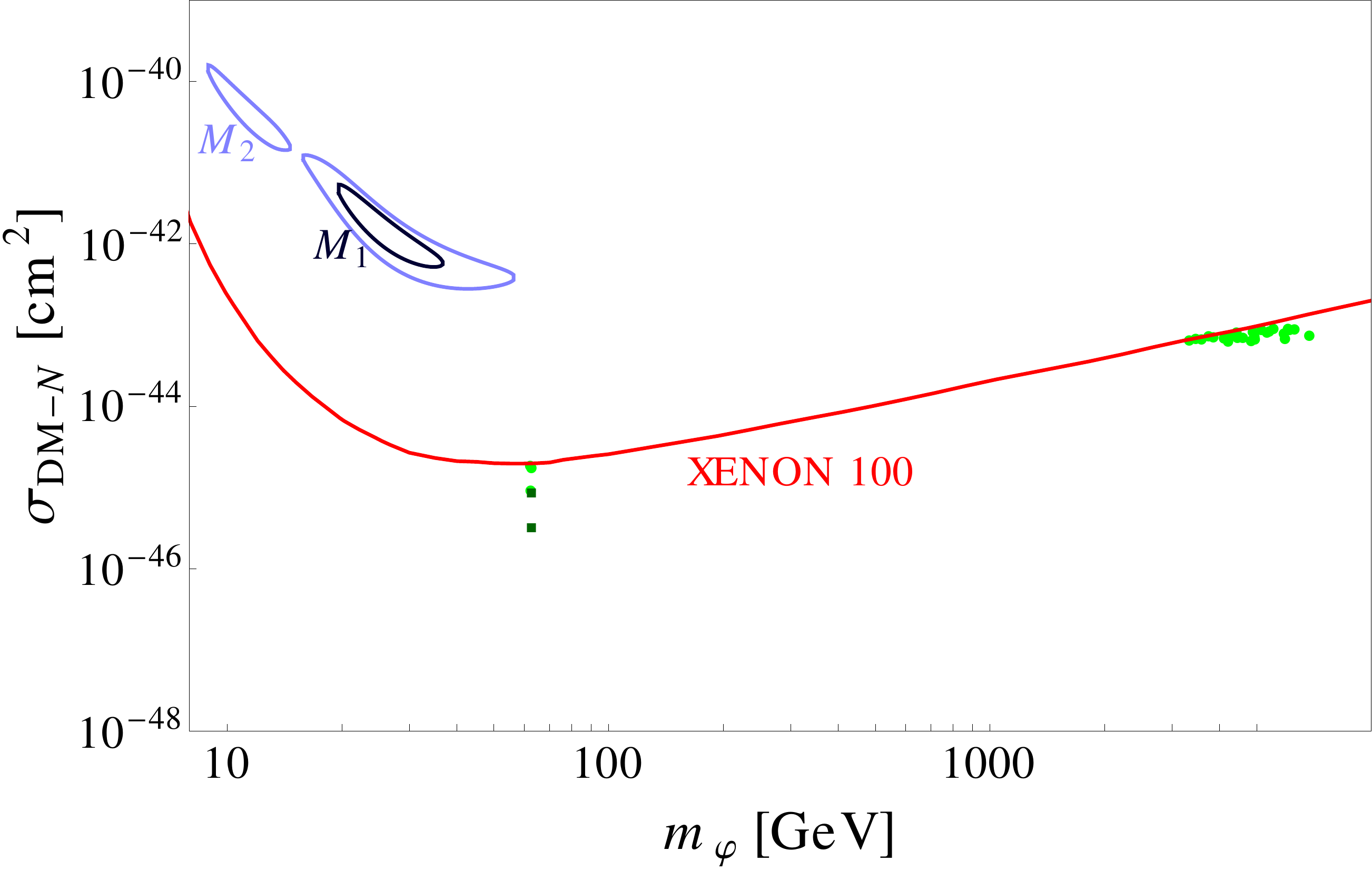}
\caption{
Plot of the cross section $\sigma_{\rm DM-N} = \sigma_{\rm DM-N}^\vp + \sigma_{\rm DM-N}^\nu$  
as a function of $\vp$ for points satisfying the WMAP 
data within $3 \sigma$; the other parameters are randomly chosen in the ranges defined in the text
(including both signs of $\lx$). Green circles (dark green squares) 
correspond to case A (case B) solutions. 
The red line shows the XENON100 data, and the two islands in blue indicate 
1 and $2\sigma$ CRESST-II results.}
\label{DD_nu_vp_res}
\end{figure}

So far we have been comparing separately $\vp$ and $\nu$ cross sections with
experimental data. However one should take into account the fact that we do have
{\it two component DM}. That is not quite straightforward if masses of the
two components are different or their contributions are of the same order.
Fortunately, it turns out that in almost all cases of interest it is meaningful
to compare $\sigma_{\rm DM-N} \equiv \sigma_{\rm DM-N}^\vp + \sigma_{\rm DM-N}^\nu$ with the experimental
limits. 
The reason is that for all points of interest either $ \vp$ and $\nu$ are almost
degenerate, or the abundance is dominated by $ \vp $, and both these cases  are well 
described by plotting $\sigma_{\rm DM-N}$ vs. $ \mvp $. These results are presented
in fig.~\ref{DD_nu_vp_res}.
The dark green squares stand for the case B points, so with , therefore
in the first approximation we may compare $\sigma_{\rm DM-N}$ for those points with the limits. 
On the other hand, it turns out that light green circles correspond to points for which
the cross section is dominated by scalars, so again those points might be compared with
single-component DM limits.

%

\section{Conclusions}
\label{conc}

We have discussed the main features of a
two-component cold Dark Matter model  composed of a neutral Majorana
fermion ($\nu$) and
a neutral real singlet ($\vp$). The Boltzmann
equations for number
densities of $\nu$ and $\vp$ were solved numerically and, for
the case $\mdm > \mvp$,
an approximate analytical solution for the present DM abundance
of both components
was found. In order to determine a region of parameter space
that is consistent both with
WMAP and XENON100 data a scan over 4-dim parameter space was
performed.

It has been shown that the agreement with the WMAP data requires
that neutrinos cannot be
substantially lighter than scalars, i.e. consistent solutions
are found only for
$\mdm \gesim \mvp$. In the region where $\mdm \simeq \mvp$
we observe interesting and strong implications of
the presence (and interactions)
of {\em two components} of DM, in particular, the thermal evolution
of their number densities could be dramatically altered.

It has been shown that in a majority of the parameter space
$\nu$ constitute
the dominant component of the present energy density of DM. This
observation agrees
with a naive intuition: since $\nu$'s do not interact with the
SM directly therefore
they annihilate slower 
than $\vp$'s that couple directly to
Higgs bosons.
In order to enhance the annihilation rate for $\nu$, large
values of the $\nu - \vp$ coupling $g_\nu \simeq 1 - 12$ are
favored by the WMAP data.
One could safely generalize the above observations and conclude
that in the multi-component DM
models the generic difficulty is an overabundance of the DM
components that have no direct couplings to the SM.
Another remark is that when the scalars $ \vp $ are relatively heavy 
($100 \lesim \mvp \lesim 1000\gev$) their annihilation
rate into SM particles must be amplified
in order to maintain agreement with the WMAP data, that implies
the Higgs portal ($\propto H^\dagger H\vp^2$)
coupling $\lx$ must grow with $\mvp$.  

The XENON100 upper limit in DM-nucleon cross section,
$\sigma_{\rm DM-N}$, turns out to be a very
restrictive condition on the model. Let's first focus on
the case with $\sigma_{\rm DM-N}$ dominated by the $\vp$-N
scattering. 
Naively one could expect
that the prediction for $\sigma_{\rm DM-N}^\vp$
could be reduced below the XENON100 limit by increasing $\mvp$.
However there exist two obstacles that prohibit suppression of
$\sigma_{\rm DM-N}^\vp$ by enlarging the scalar mass
(in the range $\mvp\gesim 100\gev$): {\em (i)} 
in order to meet the WMAP constraint data on
the present DM abundance the minimal value of the Higgs portal coupling
constant $\lx$ must grow rapidly with $\mvp$, and {\em(ii)} the
minimal relative scalar
density $f_\vp(T_{\rm CMB})/[f_\vp(T_{\rm CMB})+f_\nu(T_{\rm
CMB})]$ also increases rapidly with $\mvp$. The two factors imply that
the WMAP constraint restrict parameters to those for which
$\sigma_{\rm DM-N}^\vp$ is a approximately a constant 
function of $\mvp$; in particular, a large $\mvp$ does not
help to suppress $\sigma_{{\rm DM-N}}^\vp$. Nevertheless for $\mvp \gesim 3 \tev$
$\sigma_{\rm DM-N}^\vp$ starts to be consistent with the XENON100 data
since the bound becomes weaker at large $\mvp$. For those points 
$\Omega_\vp > \Omega_\nu$. We have also found consistent solutions for
$\mvp \simeq m_h/2$ and $\mvp\simeq 130-140\gev$ corresponding
to $\Omega_\vp < \Omega_\nu$.

The $\nu$-DM cross section, $\sigma_{{\rm DM-N}}^\nu$,
that appears at the 1-loop level was also calculated
and its contribution was confronted with the XENON100 data.
It has been shown that in the case A ($m_\nu>\mvp$) points that
satisfy the WMAP constraint are mostly in agreement with
the XENON100 bound, while in the case B ($m_\nu<\mvp$) the cross section
is usually too large. It is also worth to notice that points below the XENON100 limit
correspond to $\Omega_\vp > \Omega_\nu$.  

When both $\nu$-N and $\vp$-N cross sections are taken into account, it turns out
only solutions with $\mvp \simeq m_h/2$ and $\mvp \gesim 3 \tev$ survive.

It has been noticed that, since the CREST-II  $2\sigma$ region is 
close to the threshold for $h \to \vp\vp$, therefore a moderate
($ \sim 3\sigma$) extension of the region towards
the $h\to \vp\vp$ threshold allowed us to find points consistent with
the WMAP data with vanishing invisible decay width.

As a final remark we note that
such a model is difficult to test at the Large Hadron
Collider (LHC). The leading new effect
would be production of scalar DM pairs, with
a signature of missing energy associated with one or more jets.
Such an analyses lie beyond the scope of this work.

\section*{Acknowledgments}

This work has been supported in part by the National Science
Centre
(Poland) as a research project, decision no
DEC-2011/01/B/ST2/00438
and by the Foundation
for Polish Science International PhD Projects Programme
co-financed by
the EU European Regional Development Fund. 
 The work of SB is supported by the U.S. Department of Energy under 
Grant No. DE-SC0008541.


\appendix

\section{Dark Matter annihilation}
\label{vpvpXsec}

\begin{figure}
\centering
\includegraphics[height=2.7 cm]{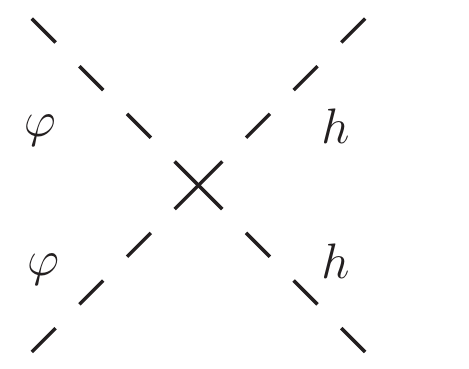}
\includegraphics[height=2.7 cm]{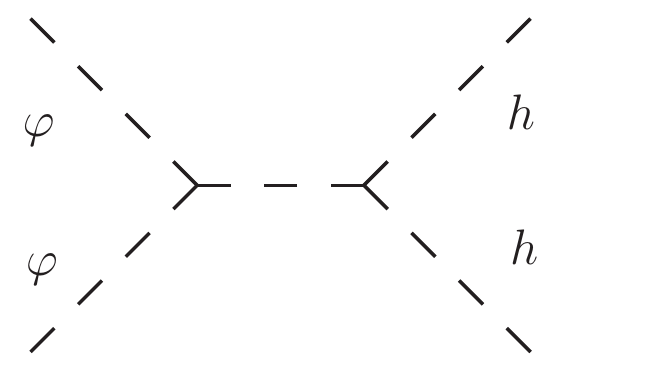}
\includegraphics[height=3.1 cm]{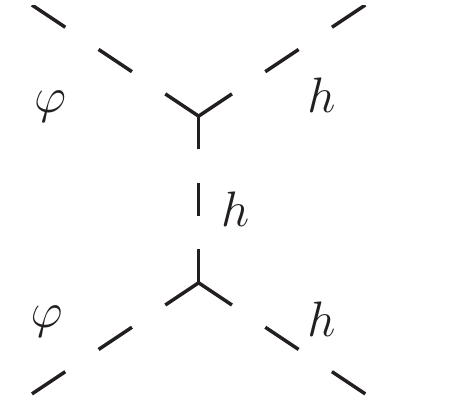}\\
\includegraphics[height=2.7 cm]{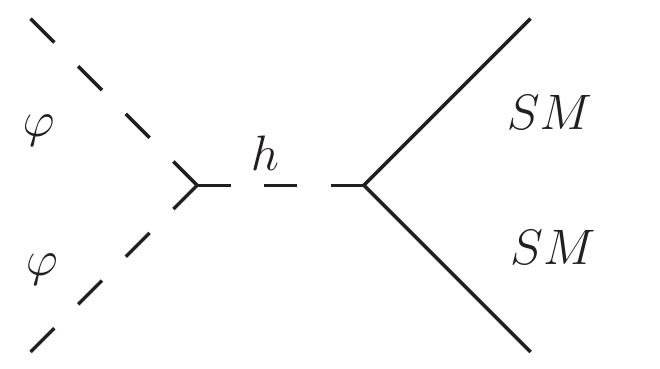}
\includegraphics[height=2.7 cm]{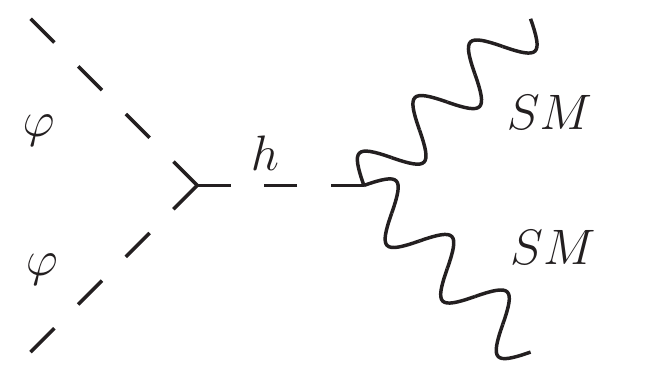}
\caption{Diagrams contributing to the scalar $\vp \vp$ annihilation into SM particles.}
\label{ann_diag}
\end{figure}

The diagrams contributing to the scalar $\vp \vp$ annihilation into SM particles are shown in fig.~\ref{ann_diag}. The corresponding cross sections 
are available in the literature(e.g. \cite{Burgess:2000yq} and \cite{Guo:2010hq}); we have verified the results of \cite{Guo:2010hq}:
\begin{eqnarray}
\hat{\sigma}_{WW}(s) &=& \frac{\lambda_{x}^2 }{2\pi} \sqrt{1 - \frac{4 M_{W}^2}{s}}   \frac{ s^2 }{(s-\mh^2)^2+\mh^2 \Gamma_{h}^2} 
\left( \frac{12 M_W^4}{s^2} - \frac{4 M_W^2}{s} + 1 \right) \cr
\hat{\sigma}_{ZZ}(s) &=& \frac{\lambda_{x}^2 }{4\pi} \sqrt{1 - \frac{4 M_{Z}^2}{s}}   \frac{ s^2 }{(s-\mh^2)^2+\mh^2 \Gamma_{h}^2} 
\left( \frac{12 M_Z^4}{s^2} - \frac{4 M_Z^2}{s} + 1 \right) \cr
\hat{\sigma}_{\overline{f}f}(s) &=& \frac{\lambda_{x}^2}{\pi} \left(\sqrt{1 - \frac{4  m_{f}^2}{s}} \, \right)^3
 \frac{ m_{f}^2 \, s }{(s-m_{h}^2)^2+\mh^2 \Gamma_{h}^2} \cr
\hat{\sigma}_{hh}(s) &=&
\frac{\lambda_{x}^2 }{4 \pi}\sqrt{1 - \frac{4 \mh^2}{s}} \left(
 \frac{(s + 2m_h^2)^2}{(s-m_h^2)^2}
+ \frac{32 v^4 \lambda_{x}^2}{(s-2m_h^2)^2} \left(\frac{1}{1-\xi^2} + F(\xi)
\right)
-\frac{16 v^2 \lambda_{x}(s+2m_h^2)}{(s-2m_h^2)(s-m_h^2)} F(\xi)
\right) \cr &&
\label{xsec1}
\end{eqnarray}
where $F(\xi) = \mathrm{ArcTanh}(\xi)/\xi$, $\xi = \sqrt{(s-4\mh^2)(s-4\mvp^2)}/(s-2\mh^2)$. 
The total cross section is then
\beq
\hat{\sigma}_{\vp\vp \to {\rm SM \,SM}} (s) = \hat{\sigma}_{WW}(s) + \hat{\sigma}_{ZZ}(s)  + \sum_{f} \hat{\sigma}_{\overline{f}f}(s) +\hat{\sigma}_{hh}(s) \label{xsec}\\
\eeq
where the sum runs over all fermions $f$. The remaining DM$\leftrightarrow$DM cross sections are
\bea
\hat\sigma_{\vp\vp \to \ndm\ndm}(s) &=& \int d\Pi_{\nu} d\Pi'_{\nu} |M_{\vp \vp \nu \nu}|^2 (2\pi)^4 \delta^4(P-p_{\nu}-p'_{\nu}) \cr
\hat\sigma_{\ndm\ndm \to \vp\vp }(s) &=& \int d\Pi_{\vp} d\Pi'_{\vp} |M_{\vp \vp \nu \nu}|^2 (2\pi)^4 \delta^4(P-p_{\vp}-p'_{\vp})
\eea
where $d\Pi_X = \gt_X d^3p_X/[(2\pi)^3 2E_X] $,
$P$ is the incoming momenta and the matrix element is given by
\bea
|M_{\vp \vp \nu \nu}|^2 (s) &=& -
\frac{
g^4 \left(4 \mdm^2-s\right) 
A(s)
}{
2 B(s)^2 C(s)^2}\\
A(s)&=&
64\left(
2(\mdm-M_h)^2(\mdm+M_h)^4+4(\mdm-M_h)M_h(\mdm+M_h)^2\mvp^2+(\mdm^2+2M_v)^2\mvp^4
\right)
\nonumber\\
&&
16\left(
4(\mdm-M_h)(\mdm+M_h)^2(\mdm+2M_h)+2(\mdm^2+2\mdm M_h+4M_h^2)\mvp^2-\mvp^4
\right)s
\nonumber\\
&&
+4\left(3\mdm^2+8\mdm M_h+8M_h^2-2\mvp^2\right)s^2 +s^3
+
\left(4 \mdm^2-s\right) \left(-4 \mvp^2+s\right)^2 \text{Cos}4\theta
\nonumber\\
&&
-16 \mdm^2  \left(4 \mvp^2-s\right)\left(4 (\mdm - M_h)(\mdm + M_h)^2+4M_h \mvp^2 - (\mdm + 2M_h)s\right) \text{Cos}2\theta
\nonumber\\
B(s)&=&
\left(2 \mdm^2-2 M_h^2+2 \mvp^2-s+\sqrt{-4 \mdm^2+s} \sqrt{-4 \mvp^2+s} \text{Cos}[\theta ]\right)
\nonumber\\
C(s)&=&
\left(-2 \mdm^2+2 M_h^2-2 \mvp^2+s+\sqrt{-4 \mdm^2+s} \sqrt{-4 \mvp^2+s} \text{Cos}[\theta ]\right)
\nonumber
\eea

When $\mh > 2 \mvp $ the decay $h \to \vp \vp $ is allowed and one has to modify the $h$ width accordingly:
\begin{eqnarray}
\Gamma_{h} = \Gamma_{h \rightarrow SM} +  \Gamma_{h \rightarrow \varphi \varphi} 
\label{h_width}
\end{eqnarray}
\begin{eqnarray}
 \Gamma_{h \rightarrow \vp \vp} = \frac{v^2 \lx^2 }{8 \pi \mh^2} \sqrt{\mh^2-4 \mvp^2} 
\, \theta_{H}(\mh - 2 \mvp)
\label{eq:h.inv}
\end{eqnarray}
where $\theta_{H}$ is the Heaviside step function (we also ignore 1-loop corrections
to $ \Gamma_h $ that might include a contribution from $ h \to \nu_h\nu_h$).

\begin{figure}
\centering
\includegraphics[height=3.2 cm]{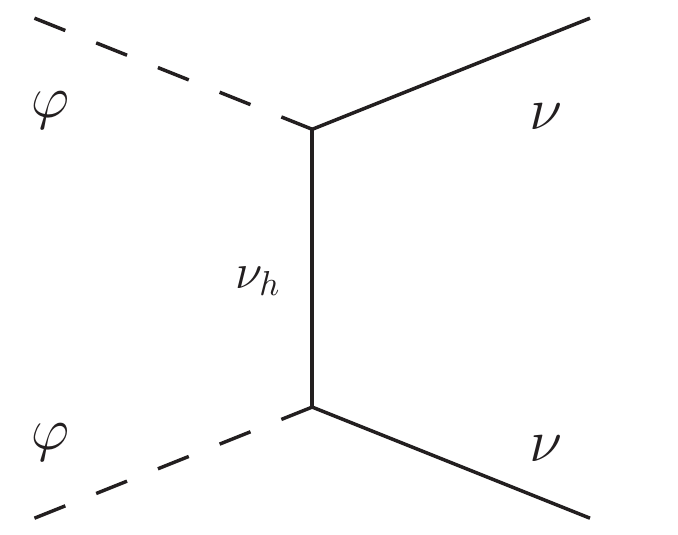} \includegraphics[height=3.1 cm]{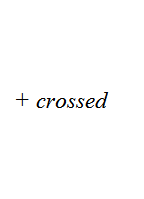}\\
\caption{Diagrams contributing to the scalar $\vp \vp$ annihilation into DM neutrinos.}
\label{ann_diag_phi_nu}
\end{figure}
%

\section{Neutrino Scattering}
\label{nunuScattXsec}

\begin{figure}
\centering
\includegraphics[height=3.75 cm]{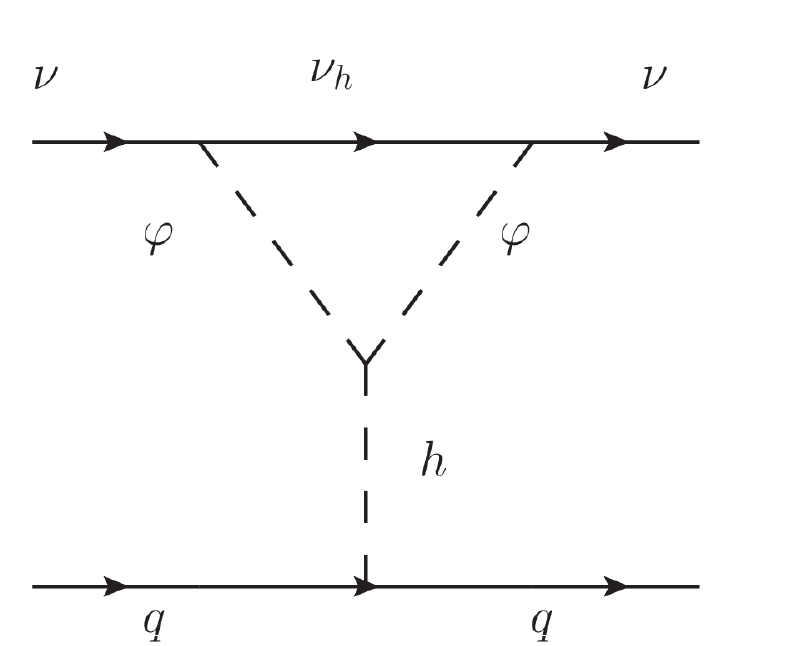}
\caption{Diagram of the neutrino $\nu \nu$ scattering off a nucleon.}
\label{scatt_diag_nu}
\end{figure}

The diagram for neutrino scattering off a nucleon at 1-loop level is shown in fig.~\ref{scatt_diag_nu}. The amplitude modulus squared for a scattering $\nu  q \rightarrow \nu q$ is the following:
\beq
|M_{\nu\nu q q}|^2 = 
\left( \frac{g_\nu^2 \lambda_x m_q}{32 \pi^2 M_h}
\right)
\left( \frac{ \xi(m_\nu/M_h,\mvp/ M_h, \sqrt{t}/M_h)}{t-m_h^2} \right)
\left( 4m_\nu^2-t \right)
\left( 4m_q^2-t \right)
\eeq
where 
\beq
\xi(a,b,c)=\int_0^1 dz \frac{z(1+az)}{1-z-z(1-z)a^2+zb^2-z^2u(1-u)tc^2}
\eeq
We are interested at the cross section at zero momentum transfer

\beq
\sigma_{\nu N} = \int_0^{4\mu ^2 w^2}\frac{d \sigma (t=0)}{dt} dt = 
\frac{\mu^2}{\pi} 
\left( \frac{g_\nu^2 \lambda_x \xi(m_\nu/M_h,\mvp/ M_h)m_N F}{32 \pi^2 M_h m_h^2}
\right)
\eeq
where $w$ is the relative velocity of dark matter to the nucleon, $F=\left( \sum_{q} f^{N}_{q} \right)$ (see \cite{Belanger:2008sj}) and $\mu$ is defined as in \ref{dir_vp_xs}.

\end{document}

%% file: mylatex.tex
\usepackage{bbm}                                                  

\newcommand{\nc}{\newcommand}
\nc{\beq}{\begin{equation}}  \nc{\eeq}{\end{equation}}
\nc{\bea}{\begin{eqnarray}}  \nc{\eea}{\end{eqnarray}}
\nc{\baa}{\begin{array}}     \nc{\eaa}{\end{array}}
\nc{\bit}{\begin{itemize}}   \nc{\eit}{\end{itemize}}
\nc{\ben}{\begin{enumerate}} \nc{\een}{\end{enumerate}}
\nc{\bce}{\begin{center}}    \nc{\ece}{\end{center}}
\nc{\bpm}{\begin{pmatrix}}   \nc{\epm}{\end{pmatrix}}
\nc{\bvt}{\begin{verbatim}}  \nc{\evt}{\end{verbatim}}
\nc{\non}{\nonumber} 

\def\half{\frac12}	

\def\to{\rightarrow}

\def\gesim{\gtrsim}
\def\lesim{\lesssim}
\def\boldoverdot{\,{\raise6pt\hbox{\bf.}\!\!\!\!\>}}
\def\lsp{\;\;\;\;}

\def\then{{\quad\Rightarrow\quad}}

\def\gcal{{\cal G}}

\def\lcal{{\cal L}}

\def\ocal{{\cal O}}

\def\zBB{{\mathbbm Z}}
\usepackage{bm}
\def\lhs{left hand side\ }
\def\rhs{right hand side\ }

\def\diag{\hbox{\diag}}

\def\gev{\;\hbox{GeV}}
\def\tev{\;\hbox{TeV}}

\def\cm{\;\hbox{cm}}

\def\doubleundertext#1{
{\undertext{\vphantom{y}#1}}\par\nobreak\vskip-\the\baselineskip\vskip4pt%
\undertext{\hbox to 2in{}}}
\def\inbox#1{\vbox{\hrule\hbox{\vrule\kern5pt
     \vbox{\kern5pt#1\kern5pt}\kern5pt\vrule}\hrule}}
\def\sqr#1#2{{\vcenter{\hrule height.#2pt
      \hbox{\vrule width.#2pt height#1pt \kern#1pt
         \vrule width.#2pt}
      \hrule height.#2pt}}}

\def\today{\ifcase\month\or
  January\or February\or March\or April\or May\or June\or
  July\or August\or September\or October\or November\or December\fi
  \space\number\day, \number\year}
\def\pmb#1{\setbox0=\hbox{#1}%
  \kern-.025em\copy0\kern-\wd0
  \kern.05em\copy0\kern-\wd0
  \kern-.025em\raise.0433em\box0 }

\def\pmbb#1{\setbox0=\hbox{#1}%
  \kern-.02em\copy0\kern-\wd0
  \kern.04em\copy0\kern-\wd0
  \kern-.02em\raise.03464em\box0 }
\def\sumprime_#1{\setbox0=\hbox{$\scriptstyle{#1}$}
  \setbox2=\hbox{$\displaystyle{\sum}$}
  \setbox4=\hbox{${}'\mathsurround=0pt$}
  \dimen0=.5\wd0 \advance\dimen0 by-.5\wd2
  \ifdim\dimen0>0pt
  \ifdim\dimen0>\wd4 \kern\wd4 \else\kern\dimen0\fi\fi
\mathop{{\sum}'}_{\kern-\wd4 #1}}

%% file: BRaCRESSTwide.dat
 $\mvp$  & 65.82 & 66.73 & 66.94 & 67.05 & 67.08 \cr\hline 
 $\mdm$  & 776 & 5373 & 654 & 85 & 4713 \cr\hline 
 $\lx $  & 0.18 & 0.1 & 0.11 & 0.11 & 0.11 \cr\hline 
 $g_\nu$  & 2.3 & 12. & 9.6 & 8.5 & 11. \cr\hline 
 $\mathrm{Log}_{10} \sigma_{DM-N}^{\vp}$  & -42.3 & -42.7 & -42.6 & -42.6 & -42.6 \cr\hline 